\documentclass[%
reprint,
superscriptaddress,
onecolumn,
amsmath,amssymb,
aps,
]{revtex4-2}

\usepackage{graphicx}% Include figure files
\usepackage{dcolumn}% Align table columns on decimal point
\usepackage{bm}% bold math
\graphicspath{{Pictures/}}
\usepackage[detect-none]{siunitx}
\usepackage{float}
\usepackage{multirow}
\usepackage{tikz}
\usepackage{wasysym}
\usepackage{MnSymbol}

\definecolor{mycolor1}{RGB}{0, 114, 189}
\definecolor{mycolor2}{RGB}{217.6000   83.2000   25.0880}
\definecolor{mycolor3}{RGB}{237.8240  177.6640   32.0000}
\definecolor{mycolor4}{RGB}{126.4640   47.1040  142.3360}
\definecolor{mycolor5}{RGB}{119.2960  172.5440   48.1280}
\definecolor{mycolor6}{RGB}{77.0560  190.7200  238.8480}
\definecolor{mycolor7}{RGB}{162.5600   19.9680   47.1040}
\definecolor{mycolor8}{RGB}{64    64    64}
\definecolor{mycolor9}{RGB}{0 0 255}
\definecolor{mycolor10}{RGB}{0 0 0}

\begin{document}

\preprint{APS/123-QED}

%\title{Transient growth in compressible boundary-layer flow with supercritical fluids }
\title{Transient growth in diabatic boundary layers with fluids at supercritical pressure}
% Force line breaks with \\
\author{Pietro Carlo Boldini}
   \email{pietro.c.boldini@gmail.com}
    \affiliation{Process and Energy Department, Delft University of Technology, Leeghwaterstraat 39, 2628 CB Delft, The Netherlands}
\author{Benjamin Bugeat}%
    \affiliation{School of Engineering, University of Leicester, University Road, Leicester, LE1 7RH, United Kingdom}
    \author{Jurriaan W.~R.~Peeters} 
    \affiliation{Process and Energy Department, Delft University of Technology, Leeghwaterstraat 39, 2628 CB Delft, The Netherlands}
\author{Markus Kloker} 
    \affiliation{Institute of Aerodynamics and Gas Dynamics, University of Stuttgart, Pfaffenwaldring 21, 70569 Stuttgart, Germany}
\author{Rene Pecnik} 
   \email{r.pecnik@tudelft.nl}
    \affiliation{Process and Energy Department, Delft University of Technology, Leeghwaterstraat 39, 2628 CB Delft, The Netherlands}

\date{\today}% It is always \today, today,

\begin{abstract}

In the region close to the thermodynamic critical point and in the proximity of the pseudo-boiling (Widom) line, strong property variations substantially alter the growth of modal instabilities, as revealed in Ren et al.~(\textit{J.~Fluid Mech.}, vol.~871, 2019, pp.~831–864). Here, we study non-modal disturbances in the spatial framework using an eigenvector decomposition of the linearized Navier-Stokes equations under the assumption of locally parallel flow. To account for non-ideality, a new energy norm is derived. Several heat transfer scenarios at supercritical pressure are investigated, which are of practical relevance in technical applications. The boundary layers with the fluid at supercritical pressure are heated or cooled by prescribing the wall and free-stream temperatures so that the temperature profile is either entirely subcritical (liquid-like), supercritical (gas-like), or transcritical (across the Widom line). The free-stream Mach number is set to $10^{-3}$. In the non-transcritical regimes, the resulting streamwise-independent streaks originate from the lift-up effect. Wall cooling enhances the energy amplification for both subcritical and supercritical regimes. When the temperature profile is increased beyond the Widom line, a strong sub-optimal growth is observed over very short streamwise distances due to the Orr mechanism. Due to the additional presence of transcritical Mode II, the optimal energy growth at large distances is found to arise from an interplay between lift-up and Orr mechanism. As a result, optimal disturbances are streamwise-modulated streaks with strong thermal components and with a propagation angle inversely proportional to the local Reynolds number.
The non-modal growth is put in perspective with modal growth by means of an N-factor comparison. In the non-transcritical regimes, modal stability dominates regardless of a wall-temperature variation. In contrast, in the transcritical regime, non-modal $N$-factors are found to resemble the imposition of an adverse pressure gradient in the ideal-gas regime. When cooling beyond the Widom line, optimal growth is greatly enhanced, yet strong inviscid instability prevails. When heating beyond the Widom line, optimal growth could be sufficiently large to favor transition, particularly with a high free-stream turbulence level.

\end{abstract}

\maketitle

\section{\label{sec:1}Introduction}

Supercritical fluids have recently gained interest in many industrial applications to enhance the thermodynamic efficiency of energy conversion systems. These applications have been thoroughly investigated for various heat sources, including nuclear energy, solar energy, fuel cell-power cycles, and industrial waste heat \cite{Liao1}. At supercritical pressures, close to the thermodynamic critical point, an ideal-gas law is inadequate to model the thermodynamic state. Therefore, higher-order non-ideal equation of states are necessary. Several studies have been conducted to explore how these non-ideal gas effects impact turbulence \cite{Kawai1,Peeters1}, heat transfer \cite{Yoo1,Nemati1}, and more recently, modal instability mechanisms \cite{Gloerfelt1,Ren1,Ren2,Ly1,Bugeat1,Bugeat2,Ren3}. Amongst these studies, Ren et al.~\cite{Ren2} conducted linear stability analyses of adiabatic flat-plate boundary layers with a highly non-ideal fluid, namely carbon dioxide at supercritical pressure. The flow undergoes stabilization for both the subcritical (liquid-like) and supercritical (gas-like) temperature regimes, specifically when the flow temperature approaches the Widom or pseudo-boiling line. The latter is defined as $\max \{c_{p}(T)\}$, representing the maximum specific heat at constant pressure \cite{Banuti1}. Crossing the Widom line from a liquid-like free stream, defined hereafter as transcritical heating temperature regime, induces sharp gradients in thermo-physical-properties, leading to the presence of two modal instabilities. The second unstable mode (Mode II) is inviscid according to the generalized-inflection-point (GIP) criterion and exhibits growth rates an order of magnitude larger than the other conventional mode (Mode I), linked to Tollmien-Schlichting waves. Recently, in Ref.~\cite{Bugeat1}, the presence of Mode II was associated with the appearance of a minimum of the kinematic-viscosity base-flow profile at the Widom line. Since all non-polar fluids at supercritical pressure and transcritical temperature exhibit such a minimum, their corresponding boundary layers are inviscidly unstable with a GIP in the proximity of the Widom line. A physical mechanism of this instability was proposed by Ref.~\cite{Bugeat2}, based on the interaction of shear and baroclinic waves. 

In the case of compressible boundary-layer flows under the ideal-gas assumption, it is well-known that various transition scenarios involve instabilities preceding or bypassing the aforementioned exponential modal growth \cite{Schmid2}. Non-modal growth, also known as transient growth, has emerged as a potential mechanism for explaining transition scenarios over a wide range of parameter values, such as in the blunt-body paradox \cite{Reshotko1,Reshotko3}. Hanifi et al.~\cite{Hanifi1} conducted the first transient growth analysis in compressible boundary layers in a temporal formulation. Similar to incompressible flows \cite{Andersson1}, initial optimal perturbations took the form of local streamwise vorticity disks, which evolve into linearly growing streamwise streaks driven by the lift-up effect \cite{Landahl1,Luchini1}. Simultaneously, for compressible flows, density and temperature fluctuations also increase linearly with time \cite{Hanifi2}. Analogous conclusions were drawn for a spatial framework \cite{Tumin1}. Optimal output perturbations, in the form of streamwise velocity streaks, were found in supersonic and hypersonic flows under the ideal-gas assumption \cite{Bitter1,Paredes1}. Here, wall cooling below the adiabatic temperature reduces the level of non-modal energy amplification. Conversely, at low Mach numbers (e.g.,~$M_\infty=0.5$) and with highly cooled walls (e.g.,~$T_w/T_\infty=0.3$, where $T_w$ is the wall temperature), no modal instabilities are found, and thus, transition to turbulence is likely to be driven only by transient growth \cite{Bitter1}. 

Another example, where large thermodynamic and transport property gradients play a major role in flow instability, is in stratified flows \cite{Govindarajan1}. The impact of viscosity stratification on transient growth has been studied both for both incompressible and compressible Couette \cite{Malik1,Saikia1} and channel \cite{Chikkadi1,Sameen1,Sameen2} flows. In the latter, it was discovered that any viscosity stratification does not significantly affect transient growth, whereas an increase in Prandtl number can lead to a total-energy amplification rise by over an order of magnitude. Recently, Jose et al.~\cite{Jose1} studied the role of buoyancy on stratified viscosity profiles in water and air. When the bottom layer is denser than the top one, transient-growth stabilization occurs, and vice versa. The optimal perturbation energy is found to be constantly localized in low-viscosity regions, leading to the formation of strong localized streaks. As the Richardson number is increased, these optimal perturbations are no longer streamwise independent. A similar behavior was found in stably stratified boundary layers \cite{Parente1} under the Boussinesq assumption, where the optimal energy growth is a combination of the lift-up and the Orr mechanism. Furthermore, as buoyancy effects become larger than shear production, the optimal energy gain and optimal time decrease, while the optimal streamwise wavenumber increases, confirming the existence of oblique optimal perturbations. 
The first study on non-modal growth with supercritical carbon dioxide at $p=\SI{80}{bar}$ was performed in a Poiseuille flow \cite{Ren1}. Regardless of the temperature flow regime (subcritical, supercritical, or transcritical with respect to the pseudo-boiling temperature), non-ideality always promoted transient growth at a constant $T_w/T_{center}$, where $T_{center}$ is the centerline temperature. The same behavior was discovered when the level of viscous heating, controlled by the product of the Prandtl $Pr_\infty$ and Eckert $Ec_\infty$ number, was increased in contrast to the ideal-gas assumption. Optimal initial perturbations were found to be streamwise-independent vortices, while optimal output perturbations were represented by velocity streaks. In the transcritical temperature case, either when $Pr_\infty Ec_\infty$ is sufficiently large or when the Widom line is crossed, the most non-modally unstable condition appear, leading to the additional formation of large thermal streaks in the dense near-wall region close to the Widom line.

The main focus of this work is to investigate the non-modal growth of two-dimensional zero-pressure-gradient (ZPG) flat-plate boundary layers with fluids close to the critical point. Specifically, we aim to elucidate the role of Mode II in transient growth, as first discussed by Ref.~\cite{Robinet1}. In order to include non-ideal gas effects in the non-modal stability analysis, a new energy norm has been derived, differently from Ref.~\cite{Ren1}. The transient growth calculations reported in this work are conducted in the spatial framework, enabling the comparison with the spatial exponential amplification of unstable modes. The competition between modal and non-modal growth mechanisms remains a fundamental research question, independently of the gas behavior. For instance, $N$-factor correlations provide valuable insights into the classical $e^N$ method for transition prediction (see \cite{Levin1,Tempelmann1}). Moreover, understanding the conditions under which the increased level of transient growth could be critical allows for a shift from transition models based solely on linear growth towards amplitude-based transition prediction. This methodology is applied for the first time to fluids at supercritical pressure, opening new perspectives in the study of instability and transition mechanisms in non-ideal fluids.

The work is organized as follows: in Sec.~\ref{sec:2}, the governing equations for base-flow calculation and modal/non-modal stability analysis are unveiled. The derivation of the new energy norm is then undertaken. Subsequently (Sec.~\ref{sec:3}), base-flow properties of eight flow cases at a constant supercritical pressure of $\SI{80}{bar}$ for carbon dioxide ($p^*/p^*_c=1.084$, with $p^*_c$ being the critical pressure) are introduced. Given a diabatic wall, different temperature profiles are considered, namely below, above, and crossing the Widom line. These regimes are defined as subcritical, supercritical, and transcritical (pseudoboiling, \cite{Banuti1}) with respect to the pseudo-critical temperature. Note that critical here refers to the thermodynamic critical point, not to be confused with subcritical growth below critical Reynolds numbers in hydrodynamic stability theory \cite{Schmid3}. Non-modal stability analysis results are then reported in Sec.~\ref{sec:4}. Optimal amplifications and perturbations are investigated for all regimes. Effects of initial Reynolds number and wall temperature, with a special focus on the transcritical wall-heating and -cooling cases, are considered. The final section, Sec.~\ref{sec:44}, is dedicated to the comparison between transient growth of perturbation energy and exponential growth of unstable discrete eigenvalues by means of $N$-factor.

\section{\label{sec:2}Methodology}

\subsection{\label{subsec:2a}Flow-conservation equations}
We consider a single-phase flow of a supercritical fluid governed by the fully compressible Navier-Stokes equations (conservation of mass, momentum, and energy) in differential and dimensionless form as
\begin{subequations}
\label{eq:NS}
\begin{equation}
\dfrac{D\rho} {Dt }+\rho\dfrac{\partial  u_j}{\partial x_j}=0,
\end{equation}
\begin{equation}
\rho \dfrac{D u_i}{D t}= -\dfrac{\partial p}{\partial x_i} +\dfrac{1}{Re_L}\dfrac{\partial \tau_{ij}}{\partial x_j},
\end{equation}
\begin{equation}
\rho \dfrac{D e}{D t}=-Ec_\infty p  \dfrac{\partial u_j}{\partial x_j}  +  \dfrac{Ec_\infty}{Re_L} \dfrac{\partial \left(u_i \tau_{ij} \right)}{\partial x_j} -   \dfrac{1}{Re_L Pr_\infty}\dfrac{\partial q_j }{\partial x_j},
\end{equation}
 \end{subequations}
where $x_i=(x,y,z)$ are the Cartesian coordinates in the streamwise, wall-normal, and spanwise directions, $t$ is the time, $\rho$ is the fluid density, $u_i=(u,v,w)$ are the velocity components, $p$ is the pressure, and $E=e+u_i u_j/2$ is the specific total energy with $e$ as the specific internal energy. Under the assumption that the fluid is Newtonian, the viscous stress tensor $\tau_{ij}$ is given as
\begin{subequations}
\begin{gather}
	\tau_{ij}=\lambda \delta_{ij} \dfrac{\partial u_k}{\partial x_k} + \mu\left( \dfrac{\partial u_i}{\partial x_j} + \dfrac{\partial u_j}{\partial x_i} \right), \quad  
\lambda=\mu_b -\dfrac{2}{3} \mu,  \tag{2a,b}
\end{gather}
 \end{subequations}
where $\mu$ is the dynamic viscosity, $\lambda=\mu_b -2/3 \mu $ is the second viscosity coefficient, and $\delta_{ij}$ is the Kronecker delta. The bulk viscosity $\mu_b$ has proven to have a very limited influence on the stability of Poiseuille flows at supercritical pressure (see Ref.~\cite{Ren1}), and therefore it is set to $0$ in agreement with Ref.~\cite{Ren2}. An additional assumption is that buoyancy effects are not considered here. The convective heat flux vector $q_j$ is modeled according to Fourier's law as $q_j=-\kappa \ \partial T/\partial x_j$, where $\kappa$ is the thermal conductivity, and $T$ the fluid temperature. The above conservation equations have been non-dimensionalized by the following reference values
\begin{subequations}
\begin{gather}
t=\frac{t^* U^*_{\infty}}{L^*},	\quad x_i=\frac{x^*_i}{L^*}, \quad u_i=\frac{u^*_i}{U^*_{\infty}}, \quad \rho=\frac{\rho^*}{\rho^*_{\infty}}, \nonumber\\
p=\frac{p^*}{\rho^*_{\infty} {U^{*^{2}}_{\infty}}}, \quad T=\frac{T^*}{T^*_{\infty}}, \quad e=\frac{e^*}{c^*_{p,\infty} T^{*}_{\infty}},	\quad 	\mu=\frac{\mu^*}{\mu^*_{\infty}},  \tag{3a--k} \\
\mu_b=\frac{\mu^*_b}{\mu^*_{\infty}}, \quad \kappa=\frac{\kappa^*}{\kappa^*_{\infty}}, \quad \nu=\frac{\nu^*}{\nu^*_{\infty}}, \nonumber
  \label{eq:variables}
\end{gather}
\end{subequations}
where $(\cdot)^*$ denotes dimensional quantities, and $(\cdot)_\infty$ corresponds to free-stream flow conditions. The corresponding non-dimensional characteristic parameters are defined as
\begin{subequations}
\begin{gather}
Re_L=\frac{\rho^*_{\infty}U^*_{\infty}L^*}{\mu^*_{\infty}}, \quad 	Ec_\infty=\frac{{U^{*2}_\infty}}{c^*_{p,\infty}T^*_\infty}, \quad 
Pr_\infty=\frac{c^*_{p,\infty} \mu^*_{\infty}}{\kappa^*_\infty}, \tag{4a--c}
\label{eq:nondimnumbers}
\end{gather}
\end{subequations}
 where $c^*_{p,\infty}$ is specific heat at constant pressure, and $Re_L$ is the Reynolds number based on a chosen length scale $L^*$. We opt for the local Blasius length scale $\delta^*$ such that
\begin{subequations}
\begin{gather}
Re_L=Re_\delta=\frac{\rho^*_{\infty}U^*_{\infty}\delta^*}{\mu^*_{\infty}}=\sqrt{Re_x}, \quad \delta^*=\sqrt{\dfrac{\mu^*_\infty x^*}{\rho^*_\infty U^*_\infty}}. \tag{5a,b}
	\label{eq:blasius} 
\end{gather}
\end{subequations}
In Eq.~\eqref{eq:nondimnumbers}, $Ec_\infty$ is the Eckert number, and $Pr_\infty$ is the Prandtl number (all based on free-stream conditions). The Mach number $M_\infty=U^*_\infty/a^*_\infty$, with $a^*_\infty$ as the speed of sound, can be obtained from $Ec_\infty$. In order to close the conservation equations in Eqs.~\eqref{eq:NS}, an equation of state (EoS) and transport properties need to be defined. We choose $\SI{80}{bar}$ as the reference pressure of supercritical $\text{CO}_2$ based on previous studies (see, for instance, Ref.~\cite{Ren2}). Its properties are summarized in Table~\ref{tab:tableCO2}. Both thermodynamic and transport properties are obtained from the NIST REFPROP library (see Ref.~\cite{Lemmon1}), which is transformed into two-dimensional (2-D) look-up tables as functions of $p$ and $T$. These properties are needed both for the laminar base-flow calculation and the stability analysis.
 
\subsection{\label{subsec:2b}Base-flow calculation}
The calculation of the base-flow profiles is based on the compressible boundary-layer equations for self-similar flow \cite{Ren2}, combined with the tabulated NIST REFPROP library and under the assumption of zero-pressure gradient. After the coordinate transformation based on the Lees-Dorodnitsyn variables \cite{White1}, with $d\xi=\rho^*_\infty \mu^*_\infty U^*_\infty \, dx^*$ and $d\eta=\rho^* U^*_\infty/\sqrt{2\xi} \,dy^*$, the transformed ordinary differential equations become \begin{equation}
  \left.
      \begin{array}{cc}
			\left( C   f_{\eta\eta}\right)_\eta + f  f_{\eta \eta}= 0,  \\ [3ex]
		f  g_\eta  + \left( \dfrac{C }{ Pr } g_\eta \right)_\eta + C \dfrac{ U^{*2}_\infty}{ h^*_\infty} \left(  f_{\eta \eta}\right)^2 = 	0,
	    \end{array}
	  \right\}
	    \label{eq:BL2}
\end{equation}
where $(\cdot)_\eta$ denotes a partial derivative with respect to $\eta$ and
\begin{subequations}
\begin{gather}
f_\eta=\dfrac{u^{*}}{U^{*}_\infty}, \quad  g=\dfrac{h^{*}}{h^{*}_\infty}, \quad  C=\dfrac{\rho^* \mu^*}{\rho^*_\infty \mu^*_\infty}=\rho \mu, \quad Pr=\dfrac{c^*_{p} \, \mu^*}{\kappa^*}.  \tag{7a--d}
  \label{eq:BL3}
\end{gather}
\end{subequations}
In Eqs.~\eqref{eq:BL2} and \eqref{eq:BL3}, $f$ is related to the stream function, $g$ is the dimensionless static specific enthalpy, $C$ is the Chapman-Rubesin parameter, and $Pr$ is the local Prandtl number. Moreover, derivatives of the transport and thermodynamic properties are numerically evaluated using a second-order finite differences method as a function of $p$ and $T$ (see \cite{Ren1}). In this context, isothermal boundary conditions (BC) are considered as
\begin{equation}
  \left.
      \begin{array}{cc}
\label{eq:BF_BC}
    f(0)=0, \quad f_\eta(0)=0, \quad f_\eta(\infty)=1, \\
    g(0)=g_w, \quad g(\infty)=1,
	    \end{array}
	  \right\}
\end{equation}
where $g_w(p,T_w)$ denotes the prescribed enthalpy at the wall. The system of boundary-layer equations is solved numerically with a 4th-order Runge-Kutta scheme, together with the Newton-Raphson method to iteratively satisfy the BCs at the wall. Grid-independent results are achieved with a wall-normal resolution of at least 10000 points to accurately capture the strong property gradients around the Widom line, and a domain size approximately equal to 10 times the boundary-layer thickness.

\subsection{\label{subsec:2c}Stability calculation}

The stability analysis is performed under the framework of linear stability theory (LST). Thus, the flow field $\mathbf{q}=[p,u,v,w,T]^\mathrm{T}$ is initially decomposed into a steady laminar part $\bar{\mathbf{q}}$, obtained from the boundary-layer equations, and a fluctuating component $\mathbf{q}'$, infinitesimally small compared to $\bar{\mathbf{q}}$. This decomposition results in
\begin{equation}
    \mathbf{q}(x,y,z,t)=\bar{\mathbf{q}}(y)+\epsilon\mathbf{q}'(x,y,z,t),
    \label{eq:flowdec}
\end{equation}
where $\epsilon \ll 1$. The base flow is assumed to be 1-D and locally parallel in the streamwise direction. The fluctuating thermodynamic and transport properties (e.g.,~$\rho',\mu',\kappa'$) are determined as functions of the two independent thermodynamic properties $p$ and $T$. For instance, the viscosity perturbation $\mu'$ is expressed as a first-order Taylor series in terms of the base-flow properties:
\begin{eqnarray}
    \mu'=&&\left.\dfrac{\partial \bar{\mu}}{\partial \bar{p}}\right\vert_{\bar{T}}p'+\left.\dfrac{\partial \bar{\mu}}{\partial \bar{T}}\right\vert_{\bar{p}}T'.
    \label{eq:pT}
\end{eqnarray}
The perturbation $\mathbf{q}'$ is assumed to depend solely on the wall-normal direction, and be periodic in all other directions. Therefore, using the classical Fourier ansatz, $\mathbf{q}'$ is written as
\begin{equation}
\mathbf{q}'(x,y,z,t)=\hat{\mathbf{q}}(y)\exp[\mathrm{i}(\alpha x+\beta z-\omega t)] + \text{c.c.},
\label{eq:qLST}
\end{equation}
where $\hat{\mathbf{q}}(y)$ is the perturbation eigenfunction, $\alpha$ is the non-dimensional streamwise wavenumber, $\beta \equiv \beta_r$ is the real spanwise wavenumber, $\omega \equiv \omega_r$ is the real angular frequency, and c.c.~stands for the complex conjugate. In this work, the spatial problem is considered by prescribing $\beta_r$ and $\omega_r$. The streamwise wavenumber in Eq.~\eqref{eq:qLST} is set to be complex ($\alpha=\alpha_r+\mathrm{i} \alpha_i$), where $\alpha_i$ represents the local spatial growth rate. Hence, modal amplification occurs for $\alpha_i<0$.
\subsubsection{\label{subsubsec:2c1}Modal analysis}

According to LST, Eq.~\eqref{eq:flowdec} is substituted into Eq.~\eqref{eq:NS}, and non-linear terms of order $O(\epsilon^2)$ are neglected. Subsequently, after subtracting the steady base flow, the linearized stability equations can be expressed in a compact matrix form as follows:
\begin{eqnarray}
 && \mathbf{L_t}\frac{\partial \mathbf{q}'}{\partial t} + \mathbf{L_x}\frac{\partial \mathbf{q}'}{\partial x} + \mathbf{L_y}\frac{\partial \mathbf{q}'}{\partial y} + \mathbf{L_z}\frac{\partial \mathbf{q}'}{\partial z} + \mathbf{L_{q'}}\mathbf{q}' \nonumber \\
&& + \mathbf{V_{xx}}\frac{\partial^2 \mathbf{q}'}{\partial x^2} + \mathbf{V_{yy}}\frac{\partial^2 \mathbf{q}'}{\partial y^2} + \mathbf{V_{zz}}\frac{\partial^2 \mathbf{q}'}{\partial z^2} + \mathbf{V_{xy}}\frac{\partial^2 \mathbf{q}'}{\partial x \partial y} + \mathbf{V_{xz}}\frac{\partial^2 \mathbf{q}'}{\partial x \partial z} + \mathbf{V_{yz}}\frac{\partial^2 \mathbf{q}'}{\partial y \partial z} = 0.
\label{eq:LSTmatrices}
\end{eqnarray}
Here, $\mathbf{q}'=\left[p', u', v', w', T'\right]^\mathrm{T}$ is the perturbation vector, and matrices $\mathbf{L_t}$, $\mathbf{L_x}$, $\mathbf{L_y}$, $\mathbf{L_z}$, $\mathbf{L_{q'}}$, $\mathbf{V_{xx}}$, $\mathbf{V_{yy}}$, $\mathbf{V_{zz}}$, $\mathbf{V_{xy}}$, $\mathbf{V_{xz}}$ and $\mathbf{V_{yz}}$ are functions of the base-flow properties and dimensionless parameters in Eq.~\eqref{eq:nondimnumbers}. The expressions for these matrices are documented in Ref.~\cite{Ren1} for $\bar{\chi}=f(\bar{\rho},\bar{T})$, where $\chi$ is an arbitrary thermodynamic or transport property. However, in this study, $\bar{\chi}$ is a function of $\bar{p}$ and $\bar{T}$. This modification results in new base-flow matrices, detailed in Appendix \ref{sec:app1b}. The use of pressure, instead of density, is motivated by the low Mach number in this study. At $M_\infty=10^{-3}$ (see Sec.~\ref{sec:3}), the density $\rho'$ is effectively decoupled from pressure $p'$, as the latter is only of hydrodynamic nature rather than of acoustic origin \cite{Nemati2}. Consequently, we choose $p'$ and $T'$ as the two independent variables as in Eq.~\eqref{eq:pT}. A pseudo-spectral collocation method based on $N$-Chebyshev collocation points is employed with near-wall grid clustering in agreement with Ref.~\cite{Hanifi1}. The boundary conditions for $\hat{\mathbf{q}}$ are given as:
\begin{equation}
  \left.
      \begin{array}{cc}
\hat{u}=\hat{v}=\hat{w}=\hat{T}=0, \quad \text{at} \ \ y=0, \\
\hat{u}=\hat{v}=\hat{w}=\hat{T}=0, \quad \text{at} \ \  y=y_{max}.
	    \end{array}
	  \right\}
\end{equation}
Additionally, the pressure $\hat{p}$ is not prescribed at the wall. Finally, the system in Eq.~\eqref{eq:LSTmatrices} is written as 
\begin{equation}
\mathbb{A}_0\hat{\mathbf{Q}} = \alpha \mathbb{A}_{\alpha} \hat{\mathbf{Q}}+\alpha^2\mathbb{A}_{\alpha^2}\hat{\mathbf{Q}},
\label{eq:EVP}
\end{equation}
where $\mathbb{A}_0$, $\mathbb{A}_{\alpha}$, $\mathbb{A}_{\alpha^2}$ are $5N \times 5N$ and $\hat{\mathbf{Q}}=(\hat{\mathbf{q}}_1,\ldots,\hat{\mathbf{q}}_N)^\mathrm{T}$ is a $5N$-column vector containing all discretized perturbations. Eq.~\eqref{eq:EVP} is a non-linear eigenvalue problem, which is recast as a linear eigenvalue problem (Ref.~\cite{Malik2}) as 
\begin{equation}
\begin{bmatrix} 
\mathbb{A}_0 & \mathbf{0}   \\
\mathbf{0} & \mathbf{I}  \\
\end{bmatrix}
\begin{bmatrix} 
\hat{\mathbf{Q}} \\
\alpha \hat{\mathbf{Q}} \\
\end{bmatrix}
= \alpha
\begin{bmatrix} 
\mathbb{A}_\alpha   & \mathbb{A}_{\alpha^2}  \\
 \mathbf{I} & \mathbf{0}  \\
\end{bmatrix}
\begin{bmatrix} 
\hat{\mathbf{Q}}\\
\alpha \hat{\mathbf{Q}} \\
\end{bmatrix},
\label{eq:EVP2}
\end{equation}
where $\mathbf{I}$ is the identity matrix. In this procedure, the size of the state vector (hence the size of the system) is increased. The eigenvalue problem is solved using the LAPACK implementation of the QZ algorithm.

\subsubsection{\label{subsec:2c3}Energy norm}

Before addressing transient growth analysis, it is necessary to define an inner product and an associated energy norm. In Ref.~\cite{Ren1}, the energy norm for non-ideal gas flows was chosen to ensure convergence with respect to energy amplification. Nevertheless, pressure-related energy terms persisted. Their contribution to the total disturbance energy must vanish, as compression work is conservative \cite{Hanifi1}. Thus, following Mack's norm for ideal gas in Ref.~\cite{Hanifi1}, a rigid definition of the norm for non-ideal gas flows must be formulated. This norm will allow us to consider the characteristic non-orthogonality of the eigenfunctions and to quantify the magnitude of transient-growth energy amplification. First, a scalar product based on the energy density, as proposed by Ref.~\cite{Tumin1} is defined as
\begin{equation}
    (\mathbf{\hat{q}}_k,\mathbf{\hat{q}}_l)_E=\int_0^\infty \mathbf{\hat{q}}^H_k \mathbf{M} \mathbf{\hat{q}}_l \ dy,
    \label{eq:innerproduct}
\end{equation}
where the superscript $H$ denotes the complex conjugate transpose, and the matrix $\mathbf{M}$ is the energy matrix. The associated norm is expressed as:
\begin{equation}
    2E=\| \mathbf{\hat{q}} \|^2_{E}=(\mathbf{\hat{q}},\mathbf{\hat{q}})_E.
     \label{eq:associatednorm}
\end{equation}
For compressible flows under the ideal-gas assumption \cite{Chu1,Mack1,Hanifi1}, Eq.~\eqref{eq:associatednorm} results in
\begin{equation}
     2E=\int_\Omega m_{u_i} u'_i u'^\dagger_i + m_{\rho} \rho' \rho'^\dagger + m_{T} T' T'^\dagger  d\Omega, 
     \label{eq:generalE}
\end{equation}
where $(\cdot)^\dagger$ and $\Omega$ denote the complex conjugate and the spatial domain, respectively. The coefficients $m_{u_{1,2,3}}$, $m_{\rho}$, and $m_{T}$ of the symmetric positive-definite matrix $\mathbf{M}$ were derived by Ref.~\cite{Hanifi1} and take the form of
\begin{subequations}
\begin{gather}
    m_{\rho}=\dfrac{R_g \bar{T}}{\bar{\rho}}, \quad m_{u}=\bar{\rho}, \quad m_{v}=\bar{\rho}, \quad m_{w}=\bar{\rho}, \quad m_{T}=\dfrac{\bar{\rho} c_v}{Ec_\infty\bar{T}}, \tag{19a--e}
    \label{eq:normIG}
\end{gather}
\end{subequations}
with $R_g$ and $c_v$ being the gas constant and the specific heat at constant volume, respectively. However, for non-ideal gas conditions in Ref.~\cite{Ren1}, the coefficients $m_\rho$ and $m_T$ were assumed to be unity, revealing a robust convergence behavior for energy amplification. Here, instead, we follow the same procedure as outlined in Ref.~\cite{Hanifi1}. As will be shown below, the coefficients in Eq.~\eqref{eq:generalE} are modified so that the pressure-related term, appearing in the form $ \nabla \cdot \left( p'  \mathbf{u}' \right)$, can be eliminated after spatial integration. By multiplying each disturbance equation in Eq.~\eqref{eq:LSTmatrices} with $m_{\rho}\rho'$, $m_{u_i}u'_i$, $m_{T}T'$, respectively, and combining the equations, the following result is obtained
\begin{eqnarray}
\label{eq:RHS}
&&\dfrac{\partial E}{\partial t}=\dfrac{1}{2} \dfrac{\partial}{\partial t} \left( m_\rho  \rho'^2 + m_{u_{i}} u'^2_i + m_T  \dfrac{\partial \bar{e}}{\partial \bar{T}}_{\bar{\rho}} T'^2 + 2  m_T  \dfrac{\partial \bar{e}}{\partial \bar{\rho}}_{\bar{T}} \rho' T' \right) \nonumber \\
&&=  -m_\rho \bar{\rho} \rho' \frac{\partial u'_i}{\partial x_i} -\dfrac{m_{u_i}}{\bar{\rho} }u'_i\frac{\partial p'}{\partial x_i} - m_T \dfrac{ Ec_\infty \bar{p}}{\bar{\rho} }  T' \frac{\partial u'_i}{\partial x_i}  \\
&& - m_T  \dfrac{\partial \bar{e}}{\partial \bar{\rho}}_{\bar{T}} \rho'  \frac{\partial T'}{\partial t} + \text{visc.}, \nonumber
\end{eqnarray}
where visc.~represents the viscous terms. When integrating over the spatial domain, the pressure-work term in the RHS of Eq.~\eqref{eq:RHS} disappears, as both of the following properties are satisfied \cite{Bitter1,Chen1}: (1) the disturbances are periodic in space, and (2) the wall-normal velocity is zero. Consequently, using integration by parts yields:
\begin{eqnarray}
  &&\int_\Omega \nabla \cdot \left( p'  \mathbf{u}' \right) d\Omega = \int_\Omega  m_\rho \bar{\rho} \rho' \frac{\partial u'_i}{\partial x_i} d\Omega   -\int_\Omega m_{u_i} \dfrac{1}{\bar{\rho}}p'\frac{\partial u'_i}{\partial x_i} d\Omega  + \int_\Omega m_T \dfrac{ Ec_\infty \bar{p}}{\bar{\rho} }  T' \frac{\partial u'_i}{\partial x_i}  d\Omega.
\end{eqnarray}
The LHS of Eq.~\eqref{eq:RHS} represents the time rate of change of the disturbance energy as:
\begin{equation}
\dfrac{\partial E}{\partial t}=	\dfrac{1}{2} \dfrac{\partial}{\partial t} \int_\Omega  \left( m_\rho  \rho'^2 + m_{\mathbf{u}} u'^2_i + m_T  \dfrac{\partial \bar{e}}{\partial \bar{T}}_{\bar{\rho}} T'^2 +  2 m_T \dfrac{\partial \bar{e}}{\partial \bar{\rho}}_{\bar{T}} \rho' T' \right) d\Omega.
\end{equation}
By choosing $m_{\mathbf{u}}=\bar{\rho}$ similar to the ideal-gas norm \cite{Hanifi1}, and 
\begin{subequations}
\begin{gather}
    m_\rho=\dfrac{1}{ \bar{\rho}}\dfrac{\partial \bar{p}}{\partial \bar{\rho}}_{\bar{T}}, \quad m_T=\dfrac{\bar{\rho}}{Ec_\infty \bar{p}}\dfrac{\partial \bar{p}}{\partial \bar{T}}_{\bar{\rho}}, \tag{23a,b}
\end{gather}
\end{subequations}
the pressure-work term is eliminated. Thus, the resulting energy norm for non-ideal gas is written as:
\begin{eqnarray}
	2E= &&\int_\Omega   \dfrac{1}{ \bar{\rho}}\dfrac{\partial \bar{p}}{\partial \bar{\rho}}_{\bar{T}}  \rho'^2 + \bar{\rho} u'^{2}_i  + \dfrac{\bar{\rho}}{ Ec_\infty \bar{p}}\dfrac{\partial \bar{p}}{\partial \bar{T}}_{\bar{\rho}} \dfrac{\partial \bar{e}}{\partial \bar{T}}_{\bar{\rho}} T'^2 +2 \dfrac{\bar{\rho}}{ Ec_\infty \bar{p}}\dfrac{\partial \bar{p}}{\partial \bar{T}}_{\bar{\rho}}   \dfrac{\partial \bar{e}}{\partial \bar{\rho}}_{\bar{T}} \rho' T' d\Omega.
 \label{eq:realE}
\end{eqnarray}
The kinetic energy of a perturbation is therefore
\begin{eqnarray}
	2E_{kin.}= &&\int_\Omega  \bar{\rho} u'^{2}_i d\Omega,
 \label{eq:realE_ekin}
\end{eqnarray}
whereas its internal energy is given as:
\begin{eqnarray}
	2E_{int.}=2\left( E_{\rho'}+E_{T'}+E_{\rho'T'}\right)= &&\int_\Omega   \dfrac{1}{ \bar{\rho}}\dfrac{\partial \bar{p}}{\partial \bar{\rho}}_{\bar{T}}  \rho'^2  + \dfrac{\bar{\rho}}{ Ec_\infty \bar{p}}\dfrac{\partial \bar{p}}{\partial \bar{T}}_{\bar{\rho}} \dfrac{\partial \bar{e}}{\partial \bar{T}}_{\bar{\rho}} T'^2 \nonumber \\
 +2 \dfrac{\bar{\rho}}{ Ec_\infty \bar{p}}\dfrac{\partial \bar{p}}{\partial \bar{T}}_{\bar{\rho}}   \dfrac{\partial \bar{e}}{\partial \bar{\rho}}_{\bar{T}} \rho' T' d\Omega.
 \label{eq:realE_eint}
\end{eqnarray}

The inner product and the associated norm can then be expressed in symbolic form as in Eqs.~\eqref{eq:innerproduct} and \eqref{eq:associatednorm}, respectively. The newly proposed energy matrix $\mathbf{M}$ for a non-ideal gas is then:
\begin{equation}
	\mathbf{M}= \begin{bmatrix}
		\dfrac{1}{ \bar{\rho}}\dfrac{\partial \bar{p}}{\partial \bar{\rho}}_{\bar{T}} & 0 & 0 & 0 &  \dfrac{\bar{\rho}}{ Ec_\infty \bar{p}}\dfrac{\partial \bar{p}}{\partial \bar{T}}_{\bar{\rho}}   \dfrac{\partial \bar{e}}{\partial \bar{\rho}}_{\bar{T}} \\
		0 & \bar{\rho} & 0 & 0 & 0 \\
		0 & 0 & \bar{\rho} & 0 & 0 \\
		0 & 0 & 0 & \bar{\rho} & 0 \\
		\dfrac{\bar{\rho}}{ Ec_\infty \bar{p}}\dfrac{\partial \bar{p}}{\partial \bar{T}}_{\bar{\rho}}   \dfrac{\partial \bar{e}}{\partial \bar{\rho}}_{\bar{T}} & 0 & 0 & 0 &  \dfrac{\bar{\rho}}{ Ec_\infty \bar{p}}\dfrac{\partial \bar{p}}{\partial \bar{T}}_{\bar{\rho}} \dfrac{\partial \bar{e}}{\partial \bar{T}}_{\bar{\rho}} \\
	\end{bmatrix}.
 \label{eq:energymatrix}
\end{equation}
The matrix $\mathbf{M}$ is positive definite, meeting the requirement of the energy-norm definition \cite{Chu1}. For an ideal gas, Eq.~\eqref{eq:energymatrix} simplifies to
\begin{equation}
	\mathbf{M}= \operatorname{diag} \left( \dfrac{R_g \bar{T}}{\bar{\rho}}, \bar{\rho}, \bar{\rho}, \bar{\rho},  \dfrac{\bar{\rho} \, c_v}{ Ec_\infty \bar{T}}  \right),
\end{equation}
recovering Chu's \cite{Chu1} and Mack's norm \cite{Hanifi1}.

\subsubsection{\label{subsec:2c4}Non-modal analysis}
Once the eigenmode spectrum is obtained, the spatial optimization method of Ref.~\cite{Tumin1} is applied. The disturbance vector $\mathbf{q}'$ is projected onto the truncated eigenvector space spanned by the first $K$ spatial eigenvalues $\alpha_k$ and eigenfunctions $\hat{\mathbf{q}}_k$. Consequently, the optimal transient growth, or optimal energy amplification, at the specific downstream coordinate is then calculated as
	\begin{equation}
G=G(\omega,\beta,Re_\delta,x)=\max_{\hat{\mathbf{q}}(0)}\dfrac{E(\hat{\mathbf{q}}(x))}{E(\hat{\mathbf{q}}(0))}=\max_{\hat{\mathbf{q}}(0)}\dfrac{\| \hat{\mathbf{q}}(x) \|^2_E }{\| \hat{\mathbf{q}}(0) \|^2_E},
	\end{equation}
where $E(\hat{\mathbf{q}}(x))$ is the total perturbation energy defined in Eq.~\eqref{eq:realE}, and $E(\hat{\mathbf{q}}(0))$ is its value at the initial location $x=0$. The optimal transient growth can be expressed in terms of the $L_2$-(Euclidean)-norm as
\begin{equation}
		G=\| \mathbf{F} \bm{\Lambda} \mathbf{F}^{-1}\|^2_2,
  \label{eq:trans_opt}
	\end{equation}
with 
\begin{subequations}
\begin{gather}
\mathbf{\Lambda}=\operatorname{diag} (\exp(\mathrm{i} \alpha_1 x),...,\exp(\mathrm{i} \alpha_K x)), \quad \mathbf{A}=\mathbf{F}^H\mathbf{F}, \quad 
		\mathbf{A}_{kl} = (\hat{\mathbf{q}}_k,\hat{\mathbf{q}}_l) = \int_{0}^{\infty} \hat{\mathbf{q}}^H_k \mathbf{M} \hat{\mathbf{q}}_l \ dy,  \tag{31a--c}
  \end{gather}
\end{subequations}
and $\mathbf{M}$ as in Eq.~\eqref{eq:energymatrix}. The $L_2$-norm of matrix $\mathbf{F} \bm{\Lambda} \mathbf{F}^{-1}$ can be computed by means of singular value decomposition (SVD). The value of $G$ is obtained by the square of the largest singular value, whereas its optimal conditions, i.e.,~optimal perturbations, can be computed via the right singular eigenvector~\cite{Schmid1}. Following the notation in Ref.~\cite{Hanifi1}, the maximum value of $G$ over all $x$ is denoted as $G_{max}$, and its maximum over all $\beta$ and $\omega$, is referred to as $G_{opt}$, with the optimal spanwise wavenumber $\beta_{opt}$, optimal frequency $\omega_{opt}$, and optimal location $x_{opt}$, respectively. For the purpose of code validation, transient-growth results in Ref.~\cite{Tumin1} are reproduced and reported in Appendix \ref{sec:app1}. Additionally, for the non-ideal gas regime, a grid-independence study is performed for a highly non-ideal reference case.

\section{\label{sec:3}Flow cases}
\begin{figure}[!b]
\centering
\includegraphics[angle=-0,trim=0 0 0 0, clip,width=0.6\textwidth]{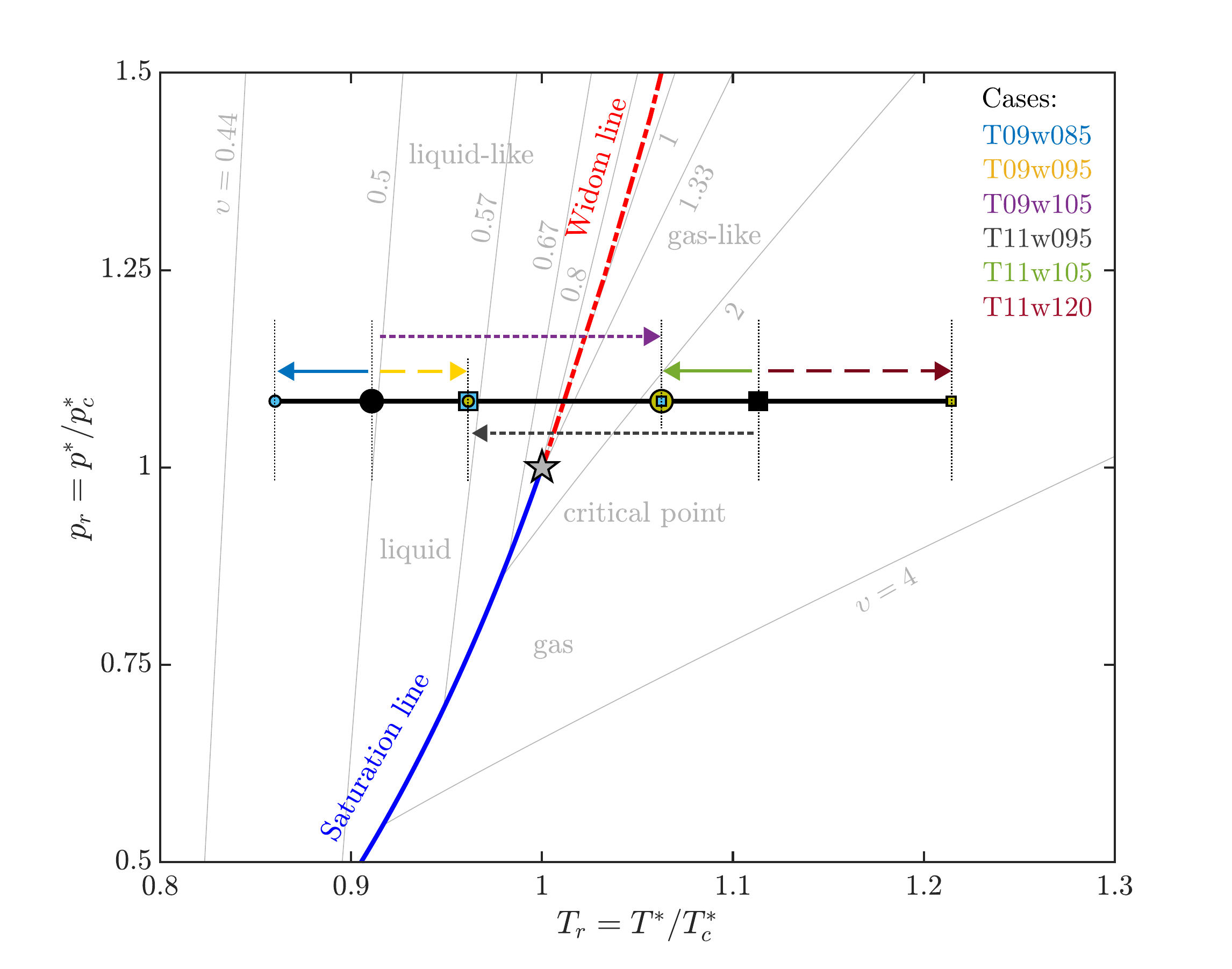}
\caption{\label{fig:fig1} Reduced temperature-pressure $(T_r$--$p_r)$ diagram with isolines of reduced specific volume ($\upsilon_r$): isobar at $p^*=\SI{80}{bar}$ for $\text{CO}_2$ with the corresponding cases (lines) of Table~\ref{tab:tableBF}. Note that the lines are offset for a better representation. }
\end{figure}

Before conducting the non-modal stability analysis, base-flow profiles must be known first. A constant pressure of $p^*=\SI{80}{bar}$, which is for $\text{CO}_2$ $1.084$ times larger than the critical pressure $p^*_c$, is selected. Note that the closer the ratio $p^*/p^*_c$ to unity, the stronger the thermodynamic non-ideality. To better assess the thermodynamic states considered in this study, a $T_r$--$p_r$ diagram is shown in Fig.~\ref{fig:fig1} together with isolines of reduced specific volume  ($\upsilon^*_r=1/\rho^*_r=\upsilon^*/\upsilon^*_c$). Along the isobar (thick black line), two different configurations are considered, with four having a subcritical temperature with respect to free-stream temperature of $T^*_\infty/T^*_{pc}=0.90$ (marked by a black circle ($\CIRCLE$) in Fig.~\ref{fig:fig1}), and four with a supercritical free-stream temperature of $T^*_\infty/T^*_{pc}=1.10$ (marked by a black square ($\blacksquare$) in Fig.~\ref{fig:fig1}), where $T^*_{pc}$ is the temperature at the Widom line ($T^*_{pc}/T^*_{c}=1.012$ at $p^*=\SI{80}{bar}$). Note that while the pressure is always supercritical, cases that are subcritical in temperature are referred to as subcritical hereafter, and the same applies to supercritical. By varying the wall temperature per case, one can control the non-ideal gas effects, which are maximal in the proximity of the Widom line. Therefore, for both free-stream configurations, one case is considered in which the temperature within the boundary layer crosses the Widom line (defined as transcritical hereafter). All cases are listed in Table~\ref{tab:tableBF} and represented in Fig.~\ref{fig:fig1}.

\begin{table}[!b]
\caption{\label{tab:tableBF} Flow parameters at $p^*=\SI{80}{bar}$ for $\text{CO}_2$ and $M=10^{-3}$. $T^*_w$ is the wall temperature, and $\theta$ is the non-dimensional compressible momentum thickness, as defined in \cite{White1}. Cases can be found in Fig.~\ref{fig:fig1}.}
\begin{ruledtabular}
\begin{tabular}{w{c}{0.15\textwidth}w{c}{0.08\textwidth}|w{c}{0.08\textwidth}w{c}{0.08\textwidth}w{c}{0.08\textwidth}w{c}{0.08\textwidth}w{c}{0.08\textwidth}w{c}{0.08\textwidth}}
 Regime  & $T^*_\infty/T^*_{pc}$ & $T^*_w/T^*_{pc}$ & $T^*_w/T^*_\infty$ & Wall & $\theta$ & Case & Line style  \\ \hline & \\[-0.6em]
 \multirow{3}{*}{\shortstack[l]{\\[-1ex]Subcrit.~temp.}}  & \multirow{3}{*}{0.90} & 0.85 & 0.944  & cooling & 0.7080 & T09w085 & \textcolor{mycolor1}{\rule[0.8ex]{0.75cm}{1pt}}  \\
 & & 0.90 & 1.0 & isotherm & 0.6641 & T09w090 &\textcolor{mycolor2}{\rule[0.8ex]{0.2cm}{1pt}} \textcolor{mycolor2}{\rule[0.8ex]{0.1cm}{1pt}} \textcolor{mycolor2}{\rule[0.8ex]{0.2cm}{1pt}}  \\
 & &  0.95 & 1.056 & heating & 0.6178 & T09w095 &  \textcolor{mycolor3}{\rule[0.8ex]{0.17cm}{1pt}} \textcolor{mycolor3}{\rule[0.8ex]{0.17cm}{1pt}} \textcolor{mycolor3}{\rule[0.8ex]{0.17cm}{1pt}}  \\[0.3em] \cline{1-2} & \\[-0.6em]
 \multirow{2}{*}{\shortstack[l]{\\[1ex]Transcrit.~temp.}}  & 0.90 &   1.05 & 1.167 & heating & 0.4608 & T09w105 & \textcolor{mycolor4}{\rule[0.8ex]{0.09cm}{1pt}} \textcolor{mycolor4}{\rule[0.8ex]{0.09cm}{1pt}} \textcolor{mycolor4}{\rule[0.8ex]{0.09cm}{1pt}} \textcolor{mycolor4}{\rule[0.8ex]{0.09cm}{1pt}} \\[0.3em] \cline{2-8} & \\[-0.6em]
  &  1.10  & 0.95 & 0.864 & cooling & 0.9779 & T11w095 & \textcolor{mycolor8}{\rule[0.8ex]{0.09cm}{1pt}} \textcolor{mycolor8}{\rule[0.8ex]{0.09cm}{1pt}} \textcolor{mycolor8}{\rule[0.8ex]{0.09cm}{1pt}} \textcolor{mycolor8}{\rule[0.8ex]{0.09cm}{1pt}} \\[0.3em] \cline{1-2} & \\[-0.6em]
 \multirow{3}{*}{\shortstack[l]{\\[0.1ex]Supercrit.~temp.}}  & \multirow{3}{*}{1.10} &  1.05 & 0.955 & cooling & 0.6931 & T11w105 & \textcolor{mycolor5}{\rule[0.8ex]{0.75cm}{1pt}}   \\
  &   & 1.10 & 1.0 & isotherm & 0.6641 & T11w110 & \textcolor{mycolor6}{\rule[0.8ex]{0.2cm}{1pt}} \textcolor{mycolor6}{\rule[0.8ex]{0.1cm}{1pt}} \textcolor{mycolor6}{\rule[0.8ex]{0.2cm}{1pt}}  \\
  & &  1.20 & 1.091 & heating & 0.6290 & T11w120   &  \textcolor{mycolor7}{\rule[0.8ex]{0.17cm}{1pt}} \textcolor{mycolor7}{\rule[0.8ex]{0.17cm}{1pt}} \textcolor{mycolor7}{\rule[0.8ex]{0.17cm}{1pt}} \\[0.3em]
\end{tabular}
\end{ruledtabular}
\end{table}
In all configurations, a Mach number of $10^{-3}$ is chosen to reduce the complexity of the physical problem, ruling out strong acoustic effects on the flow stability. On the upper third of Table~\ref{tab:tableBF}, cases with $T^*_\infty/T^*_{pc}=0.90$ (liquid-like free stream) are presented. The temperature profiles for these cases remain subcritical, except for case T09w105, which becomes transcritical, i.e., $T^*_\infty/T^*_{pc}<1$ and $T^*_w/T^*_{pc}>1$. Furthermore, it holds: $Ec_\infty=5.27 \times 10^{-7}$, $Pr_\infty=2.11$, and a unit Reynolds number of $Re_u=\SI{5.317}{\times 10^6 \, m^{-1}}$. Cases with $T^*_\infty/T^*_{pc}=1.10$ (gas-like free stream) are shown in the lower third of Table \ref{tab:tableBF}. By mirroring case T09w105 on the Widom line, while maintaining the same $\Delta T$, case T11w095 results as the transcritical case for a gas-like free stream. Furthermore, it is: $Ec_\infty=9.92 \times 10^{-8}$, $Pr_\infty=1.22$, and a unit Reynolds number of $Re_u=\SI{2.209}{\times 10^6 \, m^{-1}}$.

\begin{figure}[!tb]
\vspace{-0.6cm}
\centering
\includegraphics[angle=-0,trim=0 0 0 0, clip,width=0.80\textwidth]{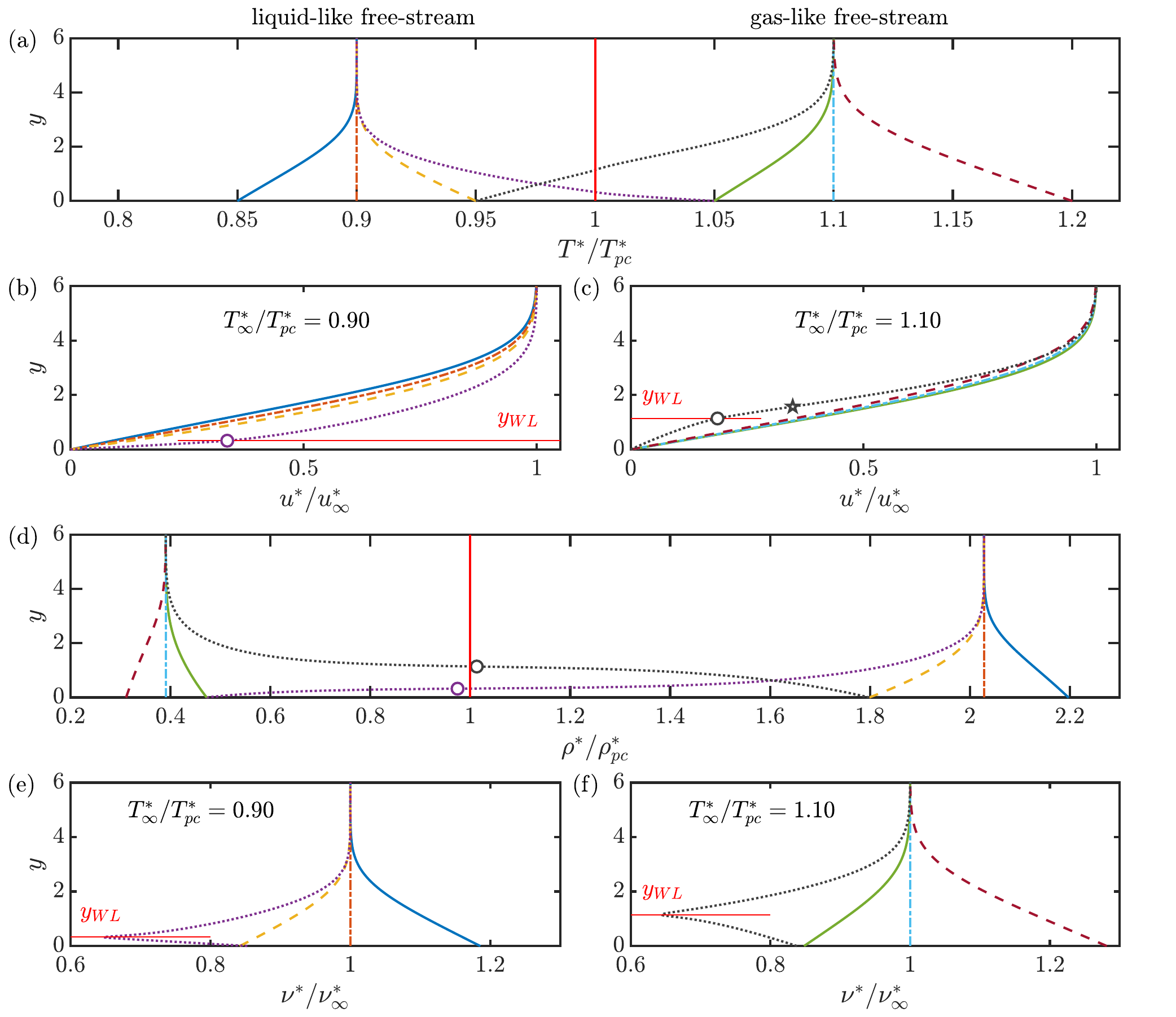}
\caption{\label{fig:fig2} Base-flow profiles for the considered cases: (a) temperature, (b,c) streamwise velocity, (d) density and (e,f) kinematic viscosity over the dimensionless wall-normal coordinate $y$. See Tab.~\ref{tab:tableBF} for line legend. The red solid line (\textcolor{red}{\rule[0.5ex]{0.2cm}{1pt}}) in (a,d) indicates the Widom line. The location of the generalized inflection point is marked by a black circle (\textcolor{black}{$\circ$}) symbol in (b--d). In (c), the location of the inflection point is marked by a black star (\textcolor{black}{$\smallstar$}) symbol. The location of the Widom line is indicated with $y_{WL}$ in (b,c) and (e,f).}
\end{figure}

The base-flow profiles for all cases in Table~\ref{tab:tableBF} are shown in Fig.~\ref{fig:fig2}. Dimensional quantities are made dimensionless by the corresponding Widom-line quantities, except for the streamwise velocity $u$ and kinematic viscosity $\nu$ (see Eq.~\eqref{eq:variables}). Temperature, streamwise velocity, density and kinematic viscosity are plotted over the wall-normal coordinate $y$, which is made dimensionless by the local Blasius length scale $\delta^*$ (see Eq.~\eqref{eq:blasius}). As the temperature profile crosses the Widom line (see Fig.~\ref{fig:fig2}(a)), regardless of the free-stream temperature, large variations in thermodynamic and transport properties are noticeable (e.g., density in Fig.~\ref{fig:fig2}(d)). As a result, both transcritical cases T09w105 and T11w095 exhibit inflectional base-flow profiles according to the definition of the generalized inflection point, i.e., $d(\bar{\rho} \ d\bar{u}/dy)/dy=0$. This GIP is located in the proximity of the Widom line, as indicated in Figs.~\ref{fig:fig2}(b--d) with a black circle (\textcolor{black}{$\circ$}) symbol. This leads to a minimum near the Widom line of kinematic viscosity $\nu$ (see Figs.~\ref{fig:fig2}(e,f)), as already observed in Ref.~\cite{Bugeat1}. Moreover, for case T11w095, an inflectional $\bar{u}(y)$-profile, i.e., $d^2\bar{u}/dy^2=0$, is found above the Widom line (marked by a black star (\textcolor{black}{$\smallstar$}) symbol), due to a strong wall cooling similar to Ref.~\cite{Ren3}.
\section{\label{sec:4}Non-modal analysis}
In this section, we present the non-modal and modal stability analyses of the boundary-layer flows. After calculating the maximum and optimal energy amplifications in Sec.~\ref{sec:40}, Section \ref{sec:41} provides profiles of the optimal disturbances for selected cases, with special focus on the transcritical regime. Subsequently, the influence of Reynolds number and wall temperature on transient growth is investigated. Finally, in Sec.~\ref{sec:44}, a comparison between non-modal and modal growth is provided.
\subsection{\label{sec:40}Optimal amplifications}
\begin{figure}[!tb]
\includegraphics[angle=-0,trim=0 0 0 0, clip,width=1.0\textwidth]{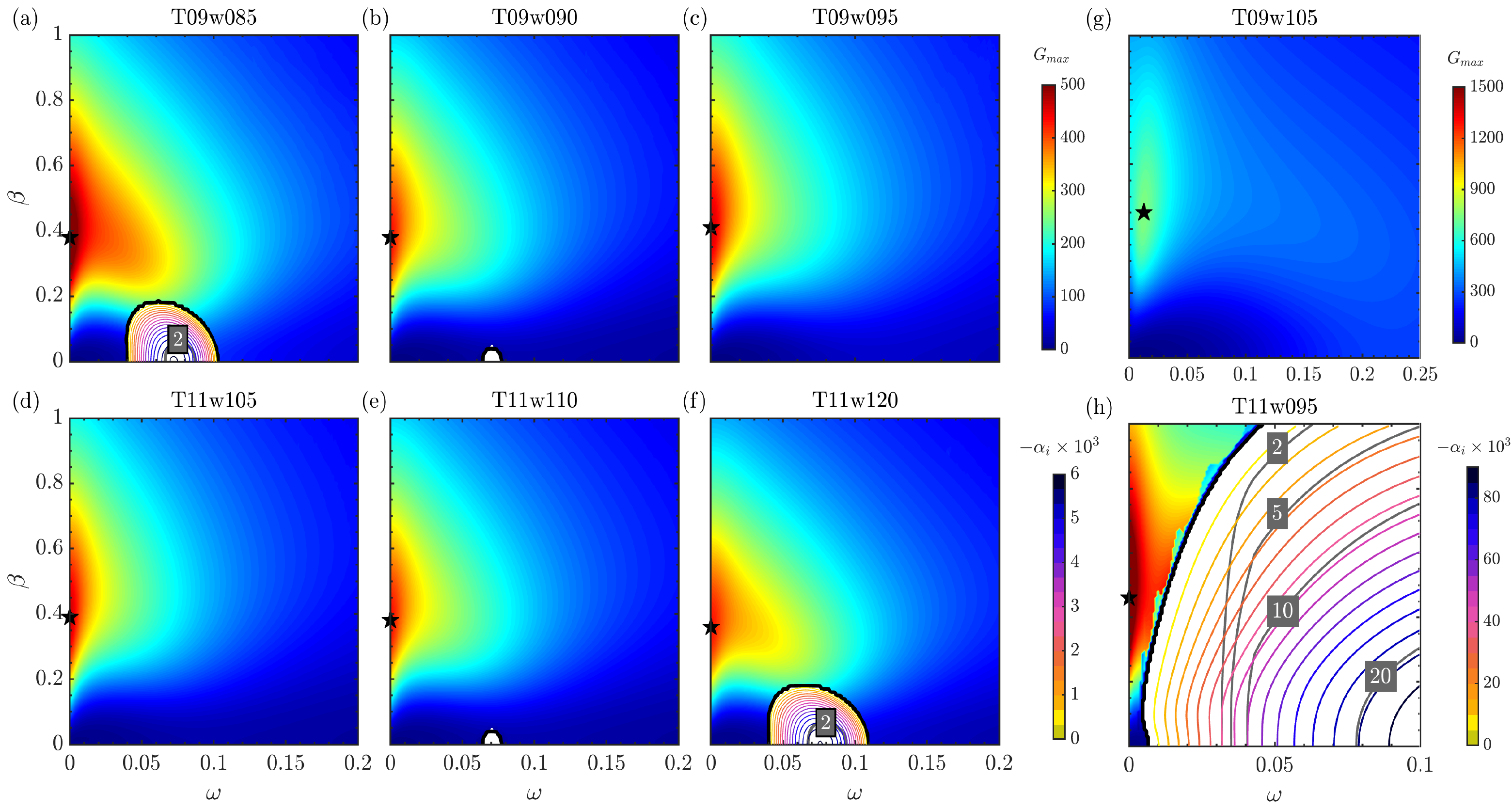}
\caption{\label{fig:fig3}Contour plot of $G_{max}$ and growth rate $-\alpha_i$ (modally unstable regions) against the angular frequency $\omega$ and spanwise wavenumber $\beta$ at $Re_\delta=300$ for the subcritical and supercritical cases: (a) T09w085, (b) T09w090, (c) T09w095, (d) T11w105, (e) T11w110, (f) T11w120, (g) T09w105, (h) T11w095. $G_{opt}$, i.e.,~$\max\{G_{max}\}$, is denoted with a black star (\textcolor{black}{$\filledstar$}) symbol. $N$-factor iso-contours are indicated in gray.}
\end{figure}

Optimal growth has been studied for the subcritical (T09w090, T09w095, T09w085), supercritical (T11w110, T11w105, T11w120), and transcritical (T09w105, T11w095) cases. An initial local Reynolds number of $Re_\delta=300$ is chosen, and the optimization process is conducted up to $x=3000$ until the maximum value of $G(x)$, i.e.,~$G_{max}$, is found. Each contour plot in Fig.~\ref{fig:fig3} has been computed on a grid of $\Delta \omega=0.001$ $\times$ $\Delta \beta=0.01$. The colored contour represents $G_{max}(\omega,\beta)$, and the largest $G_{max}$, or rather $G_{opt}$, is marked with a black star ({$\filledstar$}). Secondary contour regions display an unstable mode of spatial growth rate $-\alpha_i$ with contour lines equally spaced between 0 and $\max\{-\alpha_{i}\}$. %Here, the maximum energy amplification is infinite,
Here, the optimization process is stopped. Furthermore, the integral amplification, e.g., $N$-factor, contours are indicated. For their calculation, the local growth rate has been integrated from $Re_\delta=300$ to the optimal transient location at $Re_{\delta,opt}$, given by $x_{opt}$. A more comprehensive comparison between non-modal and modal $N$-factors is later performed in Sec.~\ref{sec:44}.

In the non-transcritical cases, as the wall temperature increases towards to the Widom line, either through an increase in $T_w$ (subcritical regime) or a reduction in $T_w$ (supercritical regime), modal amplification is completely damped. The unstable regions in Fig.~\ref{fig:fig3}(a) and Fig.~\ref{fig:fig3}(f) correspond to the only modal instability present in the subcritical and supercritical regimes, according to Ref.~\cite{Ren2}. They are of Tollmien-Schlichting-wave type. Overall, wall cooling in the liquid-like region (see T09w085 in Fig.~\ref{fig:fig3}(a)) shows similar transient growth as wall cooling in the gas-like region (see T11w105 in Tab.~\ref{fig:fig3}(d)). Cases T09w090 and T11w110 retain an identical energy amplification, reproducing the isothermal incompressible limit, which is independent of the considered gas law. With respect to the transcritical cases in Figs.~\ref{fig:fig3}(g,h), $G_{opt}$ significantly increases compared to the non-transcritical cases, especially in the case of cooling across the Widom line (T11w095) with $G_{opt}=1750$. Note that, while for case T09w105 modal instabilities are not present at $Re_\delta=300<Re_{cr}$ (with $Re_{cr}$ being the critical Reynolds number), case T11w095 is highly modally unstable (only one single mode) already at $Re_\delta=300$. In contrast to the non-transcritical cases and case T11w095, wall heating across the Widom line (case T09w105) causes a shift of $\omega_{opt}$ from 0 to finite values. For instance, in Fig.~\ref{fig:fig3}(g), it is $\omega_{opt}=0.013$. A similar shift is observed when a temporal analysis is performed in Appendix~\ref{sec:app2}. The physical mechanism responsible for the shift is explained in Sec.~\ref{sec:41}. Differently than in the non-transcritical cases, case T09w105 unveils a weakly decaying behavior of $G_{max}$ for large angular frequency $\omega$ around $\beta_{opt}$. It is interesting to note that at $\beta=0$, there is a distinct increase in $G_{max}$ along the $\omega$-axis with a sub-optimal amplification peak of $G_{max}\approx 282$ at $\omega \approx 0.23$. A further analysis of the sub-optimal energy amplification is performed in Sec.~\ref{subsec:4b5}.

\begin{figure}[!t]
\includegraphics[angle=-0,trim=0 0 0 0, clip,width=1.0\textwidth]{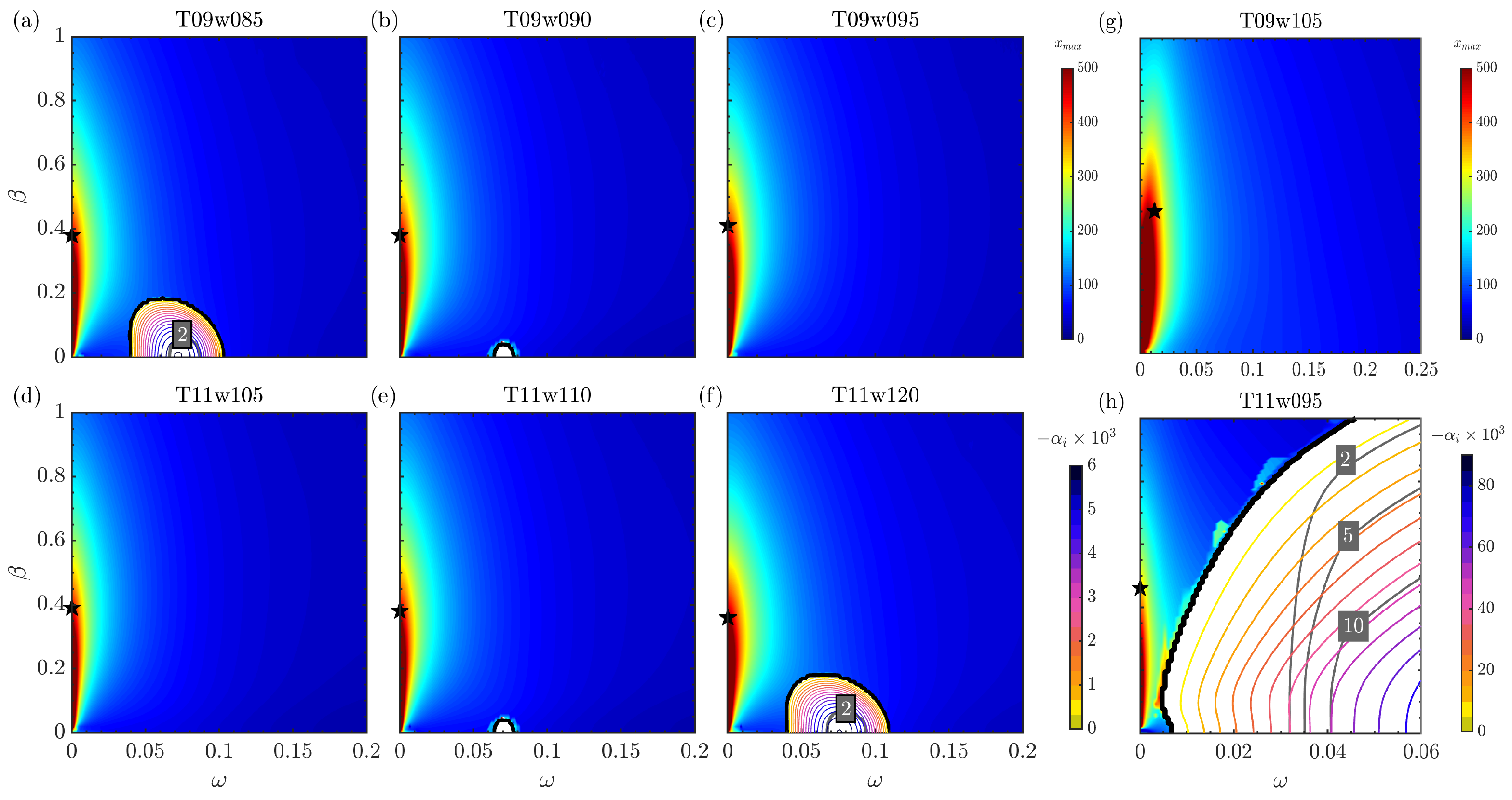}
\caption{\label{fig:fig4}Contour plot of $x_{max}$ and growth rate $-\alpha_i$ (modally unstable regions) against the angular frequency $\omega$ and spanwise wavenumber $\beta$ at $Re_\delta=300$ for the subcritical and supercritical cases: (a) T09w085, (b) T09w090, (c) T09w095, (d) T11w105, (e) T11w110, (f) T11w120, (g) T09w105, (h) T11w095. The black star (\textcolor{black}{$\filledstar$}) symbol refers to $G_{opt}$, i.e.,~$\max\{G_{max}\}$, according to Fig.~\ref{fig:fig3}. $N$-factor iso-contours are indicated in gray.}
\end{figure}

Contours of the maximum location $x_{max}$ are displayed in Fig.~\ref{fig:fig4} for the same $\Delta \omega$ $\times$ $\Delta \beta$ of Fig.~\ref{fig:fig3}. The optimum transient growth is always located at large streamwise locations (long-distance amplification) in agreement with the observations of Refs. \cite{Corbett1} and \cite{Bitter1} for incompressible and compressible boundary layers. Similarly, for small $x$ (short-distance amplification), $G_{opt}$ is always detected at large $\omega$-values before shifting to $\omega=0$ as $x$ increases. This phenomenon is observed for all non-transcritical cases and case T11w095, except for the non-trivial case T09w105 in Fig.~\ref{fig:fig4}(g) where $\omega_{opt}$ is finite.

A summary of transient growth for the subcritical, transcritical, and supercritical regimes is reported in Table~\ref{tab:SD_M0001_T0.9andT1.1}. Note that the value of $G_{kin.opt}$ is calculated using the kinetic part and a frozen internal energy of $\mathbf{M}$ in Eq.~\eqref{eq:energymatrix}, with $\mathbf{M}=\operatorname{diag} \left(1, \bar{\rho}, \bar{\rho}, \bar{\rho},  1 \right)$ (see Appendix \ref{sec:app3}). Furthermore, the effect of the energy norm on the optimal growth is analyzed.
\begin{table}[!tb]
\caption{\label{tab:SD_M0001_T0.9andT1.1}Summary of spatial transient growth characteristics at $Re_\delta=300$: subcritical, transcritical and supercritical cases.}
\begin{ruledtabular}
\begin{tabular}{ccccccccc}
    Case & $G_{opt}$ & $G_{kin.opt}$ & $x_{opt}$ & $Re_{\delta,opt}$ & $\omega_{opt}$ & $\beta_{opt}$ & $\max\{-\alpha_{i}\}$   \\ \hline & \\[-0.6em]
    T09w085 & 540 &  476 & 464 & 478.7 &  \multirow{3}{*}{0} & 0.377 & $0.569 \times 10^{-2}$   \\
    T09w090 & 437 & 437 & 462 & 478.1 & & 0.381 & $0.238 \times 10^{-3}$ \\
    T09w095 & 460 & 397 & 434 & 469.2 & & 0.408 & -  \\[0.3em] \hline & \\[-0.6em]
        T09w105 & 758 & 490 & 422 & 465.4 & 0.013 & 0.45 & -  \\
    T11w095 & 1750 & 1.6704 $\times 10^4$ & 346 & 440.2 & 0 & 0.46 & 0.089  \\[0.3em] \hline & \\[-0.6em]
    T11w105 & 466 & 457 & 447 & 473.4 &  \multirow{3}{*}{0} & 0.391 & -   \\
    T11w110 & 437 & 437 & 462 & 478.1 & & 0.381 & $0.238 \times 10^{-3}$ \\
    T11w120 & 434 & 414 & 490 & 486.8 & & 0.364 & $0.566 \times 10^{-2}$ \\[0.3em]
\end{tabular}
\end{ruledtabular}
\end{table} 
From Table~\ref{tab:SD_M0001_T0.9andT1.1}, it is evident that, given the same temperature ratio $T^*_w/T^*_\infty$, the optimal energy amplification increases more when the wall is cooled compared to the incompressible reference case, as density increases for both regimes. In fact, the largest $G_{opt}$ is found for case T09w085, with the largest density ratio, i.e.,~$\rho^*_w/\rho^*_\infty=1.0834$, among all non-transcritical cases of Tab.~\ref{tab:tableBF}. This confirms the trend encountered in the ideal-gas results of Ref.~\cite{Tumin1} at $M_\infty=0.5$. Regarding the optimal energy amplification with frozen internal energy $G_{kin.opt}$, its greater impact in noticed in the subcritical regime, where $G_{opt}$ drops by more than $10\%$. For case T09w105, $G_{opt}$ significantly increases due to non-ideal gas effects, whereas $G_{kin.opt}$ consistently drops (density ratio at the wall: $\rho^*_w/\rho^*_\infty \approx 0.23$). When the gas-like free stream is cooled down at the wall to a liquid-like regime (case T11w095 with $\rho^*_w/\rho^*_\infty \approx 4.61$), the optimal energy amplification is strongly dampened by the non-ideal gas effects, i.e., $G_{opt}\ll G_{kin.opt}$, while $\omega_{opt}$ is found at $\omega=0$. Concerning the ratio between kinetic $\hat{E}_{kin.}$ and total energy $\hat{E}$, obtained by evaluating Eqs.~\eqref{eq:realE_ekin} and \eqref{eq:realE}, respectively, for $\hat{q}=\hat{q}_{opt}$, a value of $32\%$ is found for case T09w105, whereas of $47\%$ for case T11w095. This demonstrates how optimal thermal fluctuations become relevant whenever the Widom line is crossed, i.e.,~for large property variations. An analysis of the optimal energy amplification and spanwise wavenumber as a function of the wall temperature is performed in Sec.~\ref{sec:43b}.
\subsection{\label{sec:41}Optimal perturbations}
\subsubsection{Subcritical and supercritical regimes}
\begin{figure}[!tb]
\centering
\includegraphics[angle=-0,trim=0 0 0 0, clip,width=1.0\textwidth]{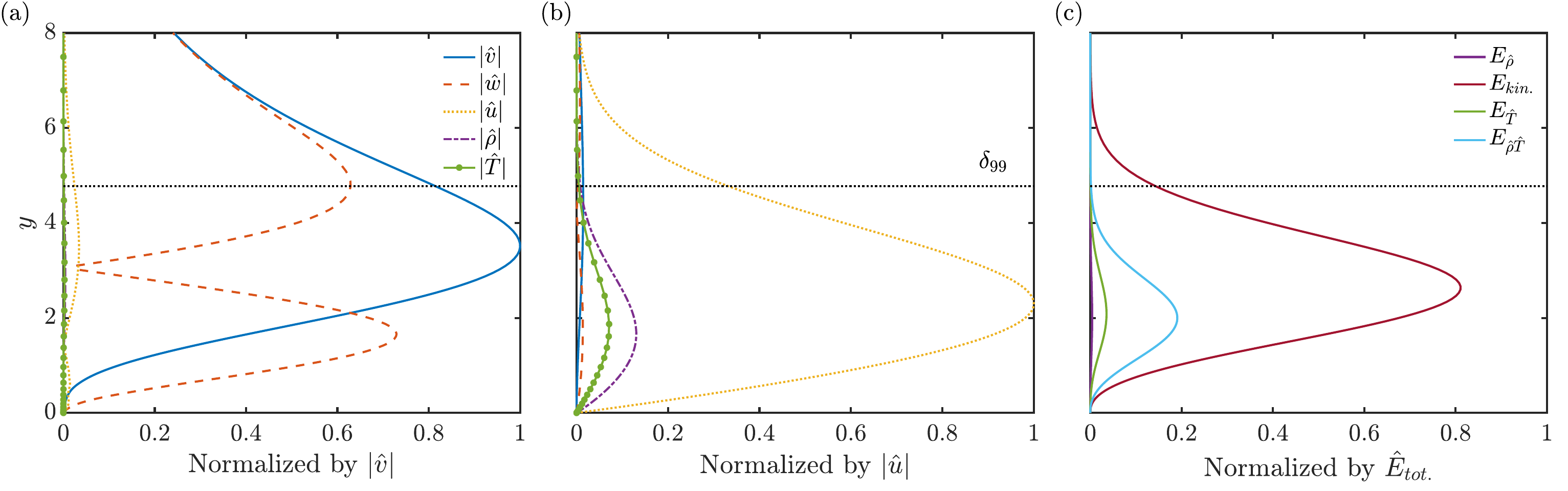}
\caption{\label{fig:fig5}Absolute value of the optimal disturbances at $Re_\delta=300$ over wall-normal distance $y$ for T09w095 of Tab.~\ref{tab:SD_M0001_T0.9andT1.1}: (a) initial, (b) output, (c) output energy amplitude obtained by Eqs.~\eqref{eq:realE_ekin} and \eqref{eq:realE_eint} with respect to the total energy. The boundary-layer thickness is indicated by $\delta_{99}$.}
\end{figure}

For all subcritical and supercritical cases in Fig.~\ref{fig:fig3} and Tab.~\ref{tab:SD_M0001_T0.9andT1.1}, $G_{opt}$ occurs at $\omega=0$, indicating an input streamwise vortex typical of the ideal-gas regime \cite{Hanifi1,Tumin1,Bitter1}, as shown for case T09w095 in Fig.~\ref{fig:fig5}(a). Similar profiles are found for the other non-transcritical cases of Tab.~\ref{tab:SD_M0001_T0.9andT1.1} (not shown for brevity), as the initial perturbation energy, i.e., Eq.~\eqref{eq:realE}, is entirely of kinetic nature.
Figure \ref{fig:fig5}(b) illustrates the flow response to the streamwise vortex at $x=x_{opt}$ for case T09w095. The output optimal disturbance mainly consists of streamwise velocity, taking the form of streamwise (high-low velocity) streaks, with negligible temperature and density disturbances due to small corresponding mean-flow gradients. Similar streamwise velocity streaks are recovered for all subcritical and supercritical cases in Tab.~\ref{tab:SD_M0001_T0.9andT1.1} (only T09w095 displayed for brevity). The formation of velocity streaks has been explained by the lift-up effect \cite{Landahl1}. Note that, for case T09w095, the kinetic energy fraction $\hat{E}_{kin.}/\hat{E}$ decreases from $99.9\%$ to $86\%$ after amplification, whereas for the subcrtical cases, this drop is only up to $5\%$. Fig.~\ref{fig:fig3}(c) exhibits that the total energy amplitude is predominantly kinetic (large $\hat{E}_{kin.}/\hat{E}$), while the non-diagonal terms $\hat{E}_{\rho'T'}$ in matrix $\mathbf{M}$ of Eq.~\eqref{eq:energymatrix} contribute less than $20\%$ due to small base-flow thermodynamic derivatives. This indicates a weak influence non-ideality on the transient growth in the subcritical and supercritical regimes, with optimal perturbations resembling those in the incompressible ideal-gas regime \cite{Andersson1,Tumin1}.

\subsubsection{Transcritical regime}

\begin{figure}[!tb]
\centering
\includegraphics[angle=-0,trim=0 0 0 0, clip,width=1.0\textwidth]{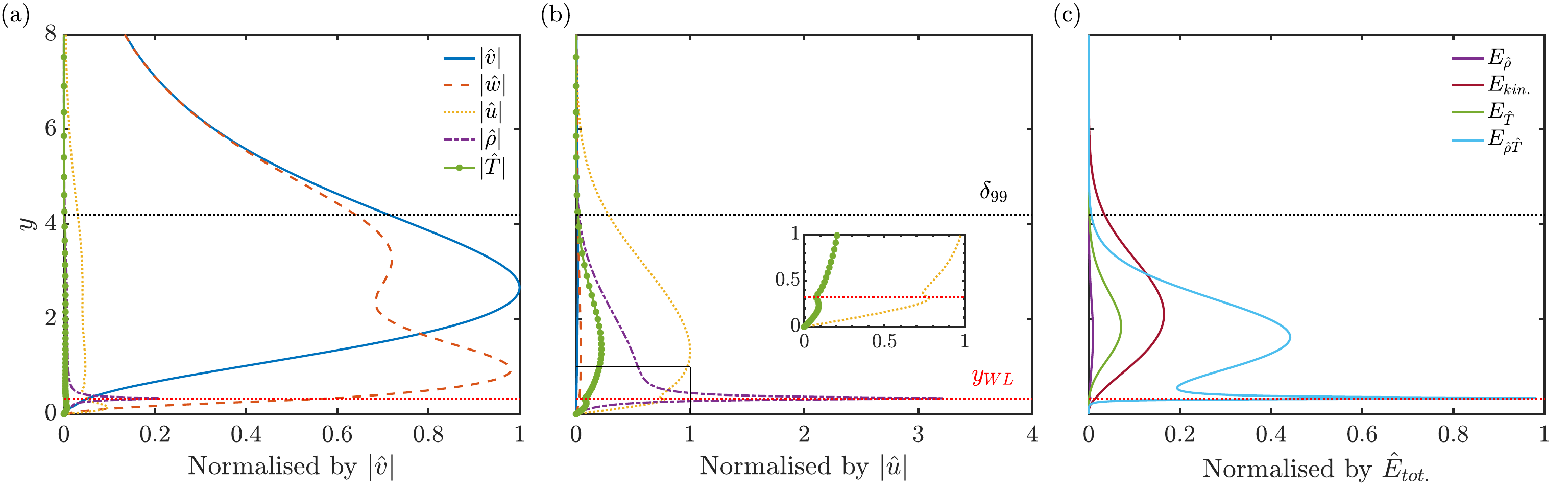}
\caption{\label{fig:fig6}Absolute value of the optimal disturbances at $Re_\delta=300$ over wall-normal distance $y$ for T09w105 of Tab.~\ref{tab:SD_M0001_T0.9andT1.1}: (a) initial, (b) output, (c) output energy amplitude obtained by Eqs.~\eqref{eq:realE_ekin} and \eqref{eq:realE_eint} with respect to the total energy amplitude. The boundary-layer thickness and the location of the Widom line are indicated by $\delta_{99}$ and $y_{WL}$, respectively.}
\end{figure}

Figs.~\ref{fig:fig6}(a,b) present the optimal initial and output perturbations for the transcritical case T09w105, respectively. The initial disturbance at $x=0$ takes the form of counter-rotating vortices, similar to the other two regimes. The resulting velocity streaks at maximum amplification have strong thermal $\hat{\rho}$ and $\hat{T}$ components. A notable strong density disturbance, absent in the non-transcritical cases (see e.g., Fig.~\ref{fig:fig5}(b)), peaks at the Widom line due to the abrupt mean density gradient observed in Fig.~\ref{fig:fig2}(d). In Fig.~\ref{fig:fig5}(c) the energy amplitude of the disturbance components normalized by the total energy indicates that $\hat{E}_{\hat{\rho}\hat{T}}$ reaches almost unity at the Widom line (largest non-ideal gas effects), with negligible kinetic energy contribution. On the contrary, the kinetic energy peak occurs in the liquid-like regime (high-viscosity region) around $y \approx 0.35$, alongside the maximum temperature disturbance and a secondary peak of $\hat{E}_{\hat{\rho}\hat{T}}$. Similar to the non-transcritical cases in Fig.~\ref{fig:fig5}, the resulting optimal streaks are regions of excess and defect of streamwise momentum in the streamwise direction.

\begin{figure}[!tb]
\centering
\includegraphics[angle=-0,trim=0 0 0 0, clip,width=1.0\textwidth]{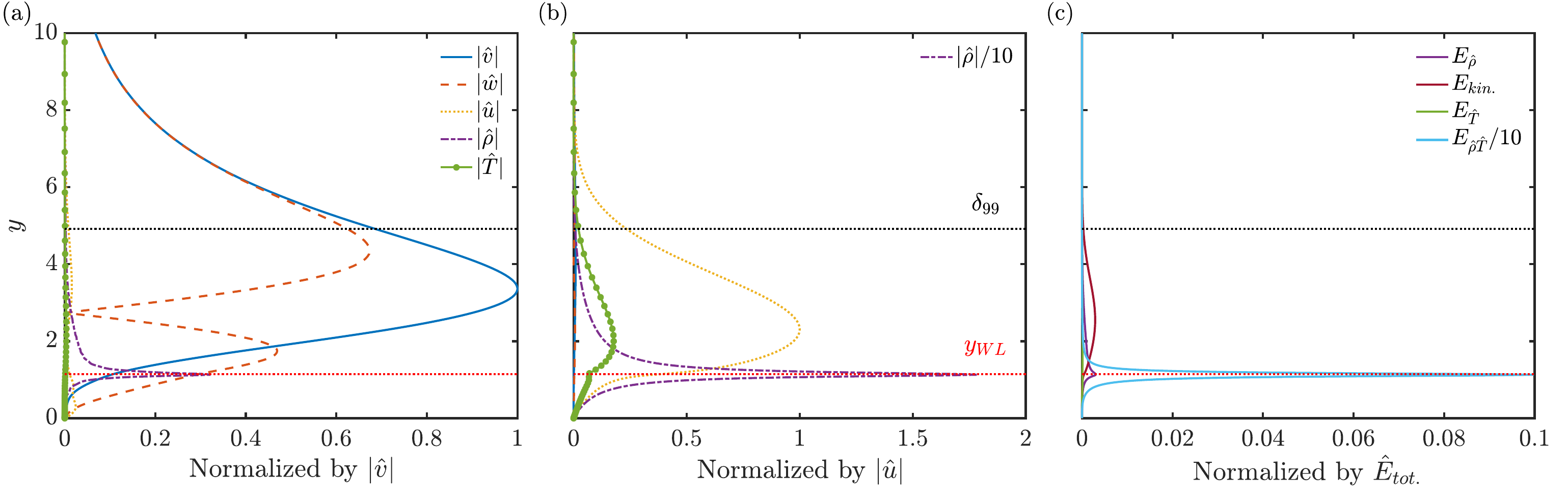}
\caption{\label{fig:fig7}Absolute value of the optimal disturbances at $Re_\delta=300$ over wall-normal distance $y$ for T11w095 of Tab.~\ref{tab:SD_M0001_T0.9andT1.1}: (a) initial, (b) output, (c) output energy amplitude obtained by Eqs.~\eqref{eq:realE_ekin} and \eqref{eq:realE_eint} with respect to the total energy amplitude. The boundary-layer thickness and the location of the Widom line are indicated by $\delta_{99}$ and $y_{WL}$, respectively.}
\end{figure}

The optimal perturbations for the transcritical case T11w095, where the temperature profile is cooled over the Widom line, are examined hereafter. Fig.~\ref{fig:fig3}(h) reveals that streamwise-independent streaks ($\omega_{opt}=0$, $\beta_{opt}=0.46$) are the optimal disturbances, differing from case T09w105. This is confirmed by the wall-normal profiles of optimal perturbations in Figs.~\ref{fig:fig7}(a) and \ref{fig:fig7}(b), driven by the lift-up effect. Unlike the other cases, thermal streaks, i.e.,~$\hat{\rho}$ and $\hat{T}$, are even more significant, with the density disturbance reaching a factor 18 at the Widom line due to a larger mean density gradient than in case T09w105. Fig.~\ref{fig:fig7}(c) displays the disturbance energy amplitude, showing the maximum total energy amount at the Widom line, attributed to the abrupt gradients of the thermo-physical properties (see Fig.~\ref{fig:fig2}). Here, non-ideality is greatest, while the kinetic energy amount is almost negligible throughout the boundary layer. 

\subsubsection{Transcritical regime: vortex tilting}
The mechanism of streak generation is best explained by the lift-up effect or vortex tilting \cite{Butler1,Brandt1}. For a more accurate physical interpretation of the Widom line's effect on the optimal perturbation profiles, we consider the inviscid compressible vorticity, denoted as $\omega_i$, equation as
\begin{equation}
		\dfrac{\partial \omega_i}{\partial t } 
		+ u_j \dfrac{\partial \omega_i}{\partial x_j}= - \omega_i  \dfrac{\partial u_j}{\partial x_j} 
		 + \dfrac{\epsilon_{ijk}}{\rho^2} \dfrac{\partial \rho}{\partial x_j}\dfrac{\partial p}{\partial x_k},
   \label{eq:vort_general}
\end{equation}
where $\epsilon_{ijk}$ is the Levi-Civita symbol, and the second term on the RHS is the baroclinic term. After superposition and linearization of Eq.~\eqref{eq:vort_general} (see Appendix \ref{sec:app5}), the wall-normal vorticity perturbation $\omega'_y$ component becomes 
\begin{equation}
    	\dfrac{\partial \omega'_y}{\partial t } + \bar{u} \dfrac{\partial \omega'_y}{\partial x} =  - \dfrac{\partial v'}{\partial z} \dfrac{\partial \bar{u}}{\partial y}, 
     \label{eq:vorty}
\end{equation}
where the baroclinic term is zero, $\partial \bar{u}/\partial y$ is the spanwise mean-flow vorticity, and $\partial v'/\partial z$ is the perturbation strain rate of the initial vortex \cite{Farrell1}. The latter tilts the mean-flow vorticity into the wall-normal $y$ direction, increasing $\omega'_y$. Thus, the vortex-tilting mechanism depends on the initial disturbance strain rate $\partial v'_{in}/\partial z$, which can be assessed via the initial streamwise vorticity $\omega'_{x,in}=\partial w'_{in}/\partial y - \partial v'_{in} /\partial z$.

\begin{figure}[tb]
\centering
\includegraphics[angle=-0,trim=0 0 0 0, clip,width=1.0\textwidth]{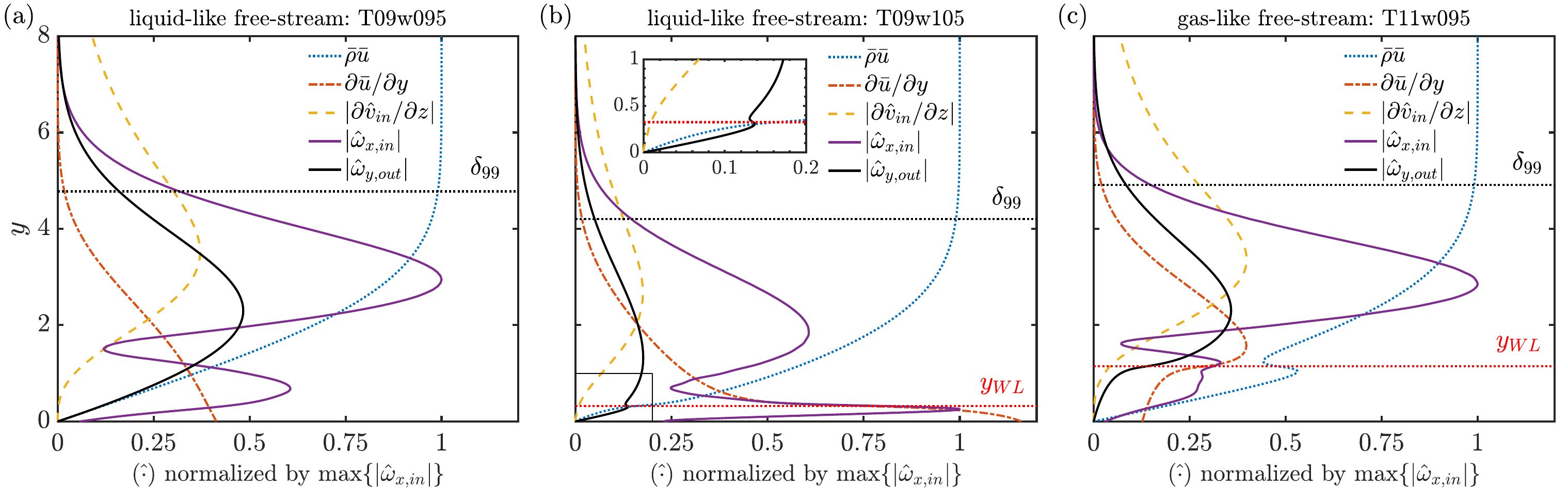}
\caption{\label{fig:fig8}Vortex tilting: (a) case T09w095, (b) case T09w105, (c) case T11w095. All disturbance terms $(\hat{\cdot})$ are normalized by $\max\{|\hat{\omega}_x|\}$. The boundary-layer thickness and the location of the Widom line are indicated by $\delta_{99}$ and $y_{WL}$, respectively.}
\end{figure}

Figs.~\ref{fig:fig8}(a,b) illustrate the initial streamwise vorticity, perturbation strain rate, and base-flow streamwise momentum for subcritical case T09w095 and transcritical case T09w105, respectively. The streamwise vorticity $|\hat{\omega}_{x,in}|$ exhibits two peaks: a lower one due to wall boundary conditions ($\hat{v}$ and $\hat{w}$ are zero at the wall) not corresponding to a streamwise vortex, and an upper peak, far from the wall, corresponding to the vortex structure responsible for the largest amount of wall-normal redistribution ($\partial \hat{v}_{in}/\partial z$ is largest). Here, $\hat{\omega}_{y,out}$ (black line) peaks, consistent across all the non-transcritical cases of Tab.~\ref{tab:tableBF}.

In the transcritical case T09w105, as shown in Fig.~\ref{fig:fig8}(b), the Widom line affects the streamwise vorticity $\hat{\omega}_{x,in}$, peaking due to the large $y$-gradient of spanwise perturbation velocity $\hat{w}_{in}$ (not displayed). However, in this region, the low wall-normal displacement (see $\hat{v}_{in}$ in Fig.~\ref{fig:fig6}(a)) of a larger streamwise momentum (see blue dotted line at $y=y_{WL}$ in comparison to case T09w095) reduces the contribution of the $\hat{w}_{in}$-peak to the lift-up effect. Additionally, the perturbation strain rate resembles the subcritical case shown in Fig.~\ref{fig:fig8}(a), but the Widom line significantly enhances the base-flow shear near the wall (see red dash-dotted line in Fig.~\ref{fig:fig8}(b)), considerably affecting, in turn, the vortex-tilting term ($|\partial \bar{u}/\partial y \cdot \partial \hat{v}_{in}/\partial z|$). As a result, $\hat{\omega}_{y,out}$ is substantially altered at the Widom line, exhibiting a secondary peak (inset of Fig.~\ref{fig:fig8}(b)), which corresponds to the $\hat{u}_{out}$-peak in Fig.~\ref{fig:fig6}(b). Thus, the streamwise-velocity streaks have two distinct peaks, the larger near the maximum perturbation strain rate and a smaller at the Widom line. On the other hand, it is important to note that the $\hat{u}_{out}$-peak at the Widom line in Fig.~\ref{fig:fig6}(b) minimally affects the energy amplification, as internal energy dominates the $G$-increase in this transcritical case, and not the kinetic energy (see $G_{opt}$ vs.~$G_{kin.opt}$ in Tab.~\ref{tab:SD_M0001_T0.9andT1.1}).

Similar to the transcritical wall-heating case depicted in Fig.~\ref{fig:fig8}(b), the vortex tilting of the transcritical wall-cooling case is examined in Fig.~\ref{fig:fig8}(c). As with case T09w105, the perturbation strain rate $\partial \hat{v}_{in}/\partial z$ (Fig.~\ref{fig:fig8})(c) remains unaffected by the Widom line. Instead, its influence is evident in $\hat{\omega}_{x,in}$, causing a substantial reduction in the lower peak near the wall. Below the Widom line, in the liquid-like regime (see the large $\bar{\rho}_w$ in Fig.~\ref{fig:fig2}(d)), we observe a considerable increase in the horizontal momentum near $y=y_{WL}$ (blue dotted line, Fig.~\ref{fig:fig8}(c)). This suggests a stronger lift up in this region. Indeed, the resulting peak in $\hat{\omega}_{y,out}$ occurs just above the Widom line, where the vortex-tilting term $|\partial \bar{u}/\partial y \cdot \partial \hat{v}_{in}/\partial z|$ is largest. Unlike Fig.~\ref{fig:fig8}(b), there is no secondary $\hat{u}_{out}$-peak at the Widom line in this case, as the mean-flow vorticity $\partial \bar{u}/\partial y$ near the wall is small. These observations suggest that the resulting streaks have higher amplitude than those in case T09w105, attributed to a stronger vortex-tilting term $|\partial \bar{u}/\partial y \cdot \partial \hat{v}_{in}/\partial z|$. This behavior is supported by the greater optimal energy amplification of case T11w095 compared to case T09w105: from Tab.~\ref{tab:SD_M0001_T0.9andT1.1}, $G_{opt,\mathrm{T11w095}}>G_{opt,\mathrm{T09w105}}$. 

\subsubsection{Transcritical regime: lift-up effect and Orr mechanism}
\label{sec:414}

\begin{figure}[tb]
\centering
\includegraphics[angle=-0,trim=0 0 0 0, clip,width=0.85\textwidth]{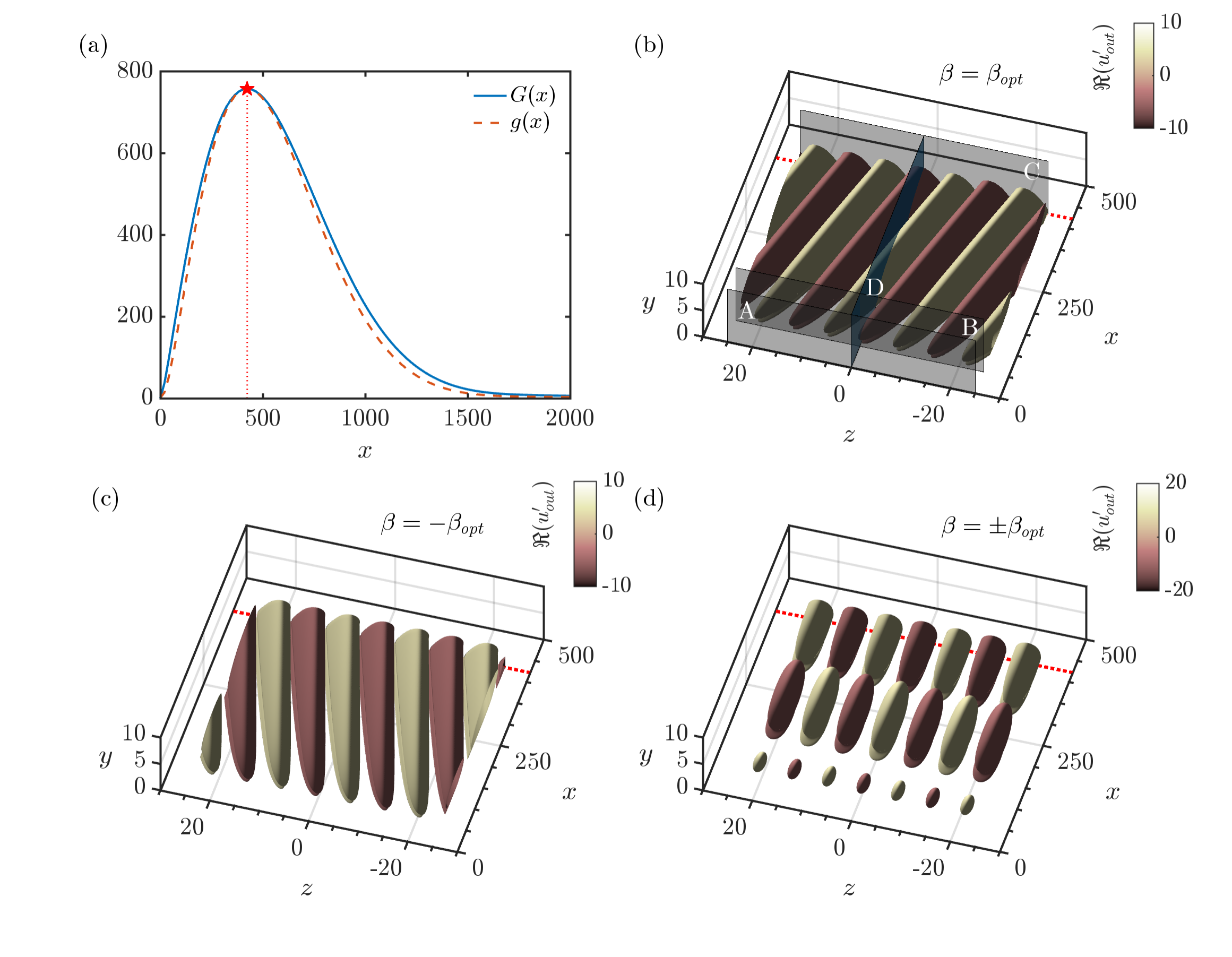}
\caption{\label{fig:fig9}Case T09w105 ($Re_\delta=300$, $\omega_{opt}=0.013$, $\beta_{opt}=0.45$): (a) envelope of the optimal transient growth $G(x)$ (\textcolor{mycolor1}{\rule[0.5ex]{0.3cm}{1pt}}) and transient growth $g(x)$ of the optimal perturbation (\textcolor{mycolor2}{\rule[0.5ex]{0.11cm}{1pt}} \textcolor{mycolor2}{\rule[0.5ex]{0.11cm}{1pt}}) over the streamwise direction $x$, the red dotted line indicates the location of the optimal perturbation ($g(x_{opt})=G(x_{opt})$); (b) iso-contours of the streamwise velocity perturbations ($\Re(u'_{out})=\pm4.0$) for $\beta=\beta_{opt}$ corresponding to their transient growth $g$; (c) iso-contours of the streamwise velocity perturbations ($\Re(u'_{out})=\pm4.0$) for $\beta=-\beta_{opt}$; (d) superposition of iso-contours ($\Re(u'_{out})=\pm8.0$) with $\pm\beta_{opt}$. The red dashed line indicates the optimal growth location ($x=x_{opt}$). Planes: $\text{A}$, $\text{B}$, $\text{C}$ and $\text{D}$ of Fig.~\ref{fig:fig10}.}
\end{figure}

\begin{figure}[!tb]
\centering
\includegraphics[angle=-0,trim=0 0 0 0, clip,width=0.95\textwidth]{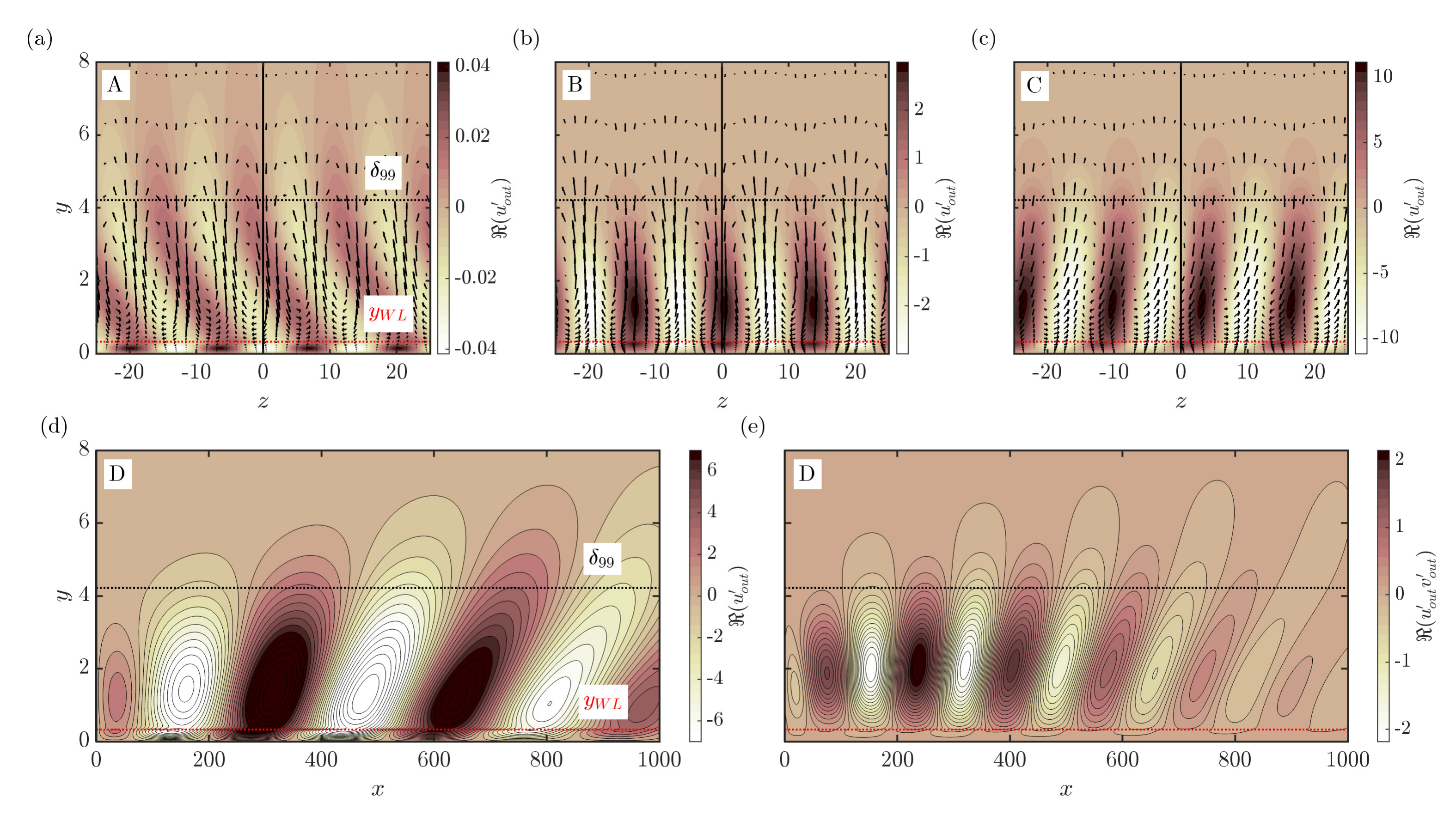}
\caption{\label{fig:fig10}Case T09w105. Contours of the optimal streamwise velocity perturbation $u'_{out}$ and velocity vectors of the resulting cross-stream perturbation velocity $|\vec{V'_{out}}|=\sqrt{v'^2_{out}+w'^2_{out}}$ on an $z$-$y$ plane: (a) $x=0$, (b) $x=50$, (c) $x=x_{opt}=425$. Contours of the optimal perturbation on an $x$-$y$ plane at $z=0$: (d) streamwise velocity $u'_{out}$, (e) Reynolds stress $u'_{out}v'_{out}$. Note that $\delta_{99}$ and $y_{WL}$ are constant due to the parallel mean-flow assumption used in the transient-growth analysis. Capital A, B, C, D correspond to planes in Fig.~\ref{fig:fig9}.}
\end{figure}

For both transcritical cases, a detailed analysis of spatial evolution of the optimal amplification is conducted to comprehend its structure. Beginning with case T09w105, Fig.~\ref{fig:fig9}(a) presents the transient growth $G(x)$ at $\omega_{opt}$ and $\beta_{opt}$, obtained from Eq.~\eqref{eq:trans_opt}. This represents the optimal gain envelope over all possible initial conditions. To calculate the disturbance profiles over the streamwise distance $x$, we assume the same initial condition at $x=0$ for the optimal perturbation and perform space marching up to $x=x_{opt}$, following $g(x)$ (red dashed line). At this location, the energy amplification $g(x)$ is equal to the maximum value of the $G$-envelope (solid blue line), while remaining lower for $x \neq x_{opt}$. For the calculation of $g(x)$, Eq.~\eqref{eq:trans_opt} utilizes $\mathbf{F}$ associated with the right singular eigenvector at $x=x_{opt}$. Fig.~\ref{fig:fig9}(b) provides a three-dimensional (3-D) snapshot of the streamwise velocity perturbations, revealing oblique elongated structures. These results are computed based on the optimal perturbation chosen from $g(x)$ of Fig.~\ref{fig:fig9}(a).
The physical mechanism of the lift-up effect is evident in the cross-stream slices ($y$-$z$ plane) of Figs.~\ref{fig:fig10}(a--c), where contours of the output streamwise velocity perturbation $u'_{out}$ are superimposed over the velocity vectors of the cross-stream perturbation velocity $|\vec{V'_{out}}|=\sqrt{v'^2_{out}+w'^2_{out}}$ at three different streamwise locations. Resulting high-velocity streaks ($\Re(u'_{out})>0$) form where counter-rotating vortices pull high momentum fluid downward, and vice versa for low-velocity streaks ($\Re(u'_{out})<0$). This behavior extends to the density and temperature streaks, where streamwise vortices act with the mean-density and against the mean-temperature profile, respectively. At $x=x_{opt}=425$, the lift-up mechanism maximizes streak amplification. The crossing of the Widom line introduces secondary $u'_{out}$-peaks, notably in the near-wall region, especially at early $x$, as demonstrated by the vortex-tilting mechanism in Fig.~\ref{fig:fig8}(b).
Despite the 2-D parallel flow assumption, vortices in Figs.~\ref{fig:fig10}(a--c) are not symmetrical with respect to the wall-normal axis but inclined in the spanwise direction. Within a very short distance ($x \approx 50$), they are first erected before being tilted in the other direction farther downstream. This observation suggests the presence of another mechanism, contributing to additional energy amplification in the cross-plane spanwise component \cite{Hack1}. Notably, the shift of the optimal amplification $G_{opt}$ to finite frequencies, visible in Fig.~\ref{fig:fig3}(d), implies that the Orr mechanism \cite{Orr1} enhances the optimal amplification.

Examining a wall-normal slice ($x$-$y$ plane) in Figs.~\ref{fig:fig10}(d,e) reveals the presence of this additional mechanism. Initially misaligned with the mean shear, the output streamwise velocity perturbations in Fig.~\ref{fig:fig10}(d) gradually incline in the direction of shear as they move downstream. During this process, the disturbance extracts energy from the mean-shear energy through the action of the Reynolds stress $u'_{out}v'_{out}$ (Fig.~\ref{fig:fig10}(e)), which increases due to conservation of circulation in an $x$-$y$ plane \cite{Parente1}. Subsequently, disturbance energy is returned to the mean flow, with viscosity effects becoming relevant at large $x$. The largest absolute value of $u'_{out}v'_{out}$ occurs at $x \approx 240$, farther upstream than $x_{opt}=425$. The Orr mechanism enhances cross-stream velocity, leading to an additional rise of the streamwise velocity perturbations \cite{Farrell1}, which peak at $x=x_{opt}$. This positive interplay between Orr mechanism, originally located at $\beta=0$ and $\omega>0$, and lift-up effect, originally located at $\beta>0$ and $\omega=0$, results in optimal transient growth at finite frequencies and spanwise wavenumbers, e.g., $\omega_{opt}>0, \beta_{opt}>0$. Such interaction was previously discovered in constant shear flows \cite{Butler1,Farrell1} and more recently in strongly stratified flows \cite{Saikia1} under the Boussinesq assumption \cite{Parente1}.

When observing the oblique perturbations in Fig.~\ref{fig:fig9}(b), one may question why the optimal streaks exhibit such an orientation with a propagation angle $\Psi$, with $\tan(\Psi)=\beta/\alpha$. In the context of the two-dimensional spanwise-periodic underlying base flow, achieving the same optimal growth is possible with a spanwise wavenumber $\beta$ as well as with $-\beta$ (see \cite{Butler1}). This implies that identical 3-D structures are obtained with opposite inclination to the wall-normal axis (Fig.~\ref{fig:fig9}(c)). In addition, the Reynolds stress mechanism observed in Figs.~\ref{fig:fig10}(d--e) is exactly replicated. Since we are considering eigenvalues and eigenvectors of a linear amplification problem \cite{Luchini1}, a superposition of the two oblique waves can be calculated. As depicted in Fig.~\ref{fig:fig9}(d), this results in a checkered wave pattern, arising from multiple standing waves in the spanwise direction with a resultant zero propagation angle. These streamwise and spanwise alternating structures resemble the initial stage of an oblique transition (see, for instance, Ref.~\cite{Berlin1}). In fact, the (non-linear) generation of a streamwise vortex by two least-damped Orr-Sommerfeld oblique waves with $(\omega,\pm \beta)$ is followed by the transient growth of vortex-generated streaks. An analysis of the superposed cross-stream slices in Figs.~\ref{fig:fig10}(a--c) unveils symmetrical structures in the spanwise direction, featuring zero cross-stream velocity.\par
\begin{figure}[!tb]
\centering
\includegraphics[angle=-0,trim=0 0 0 0, clip,width=0.85\textwidth]{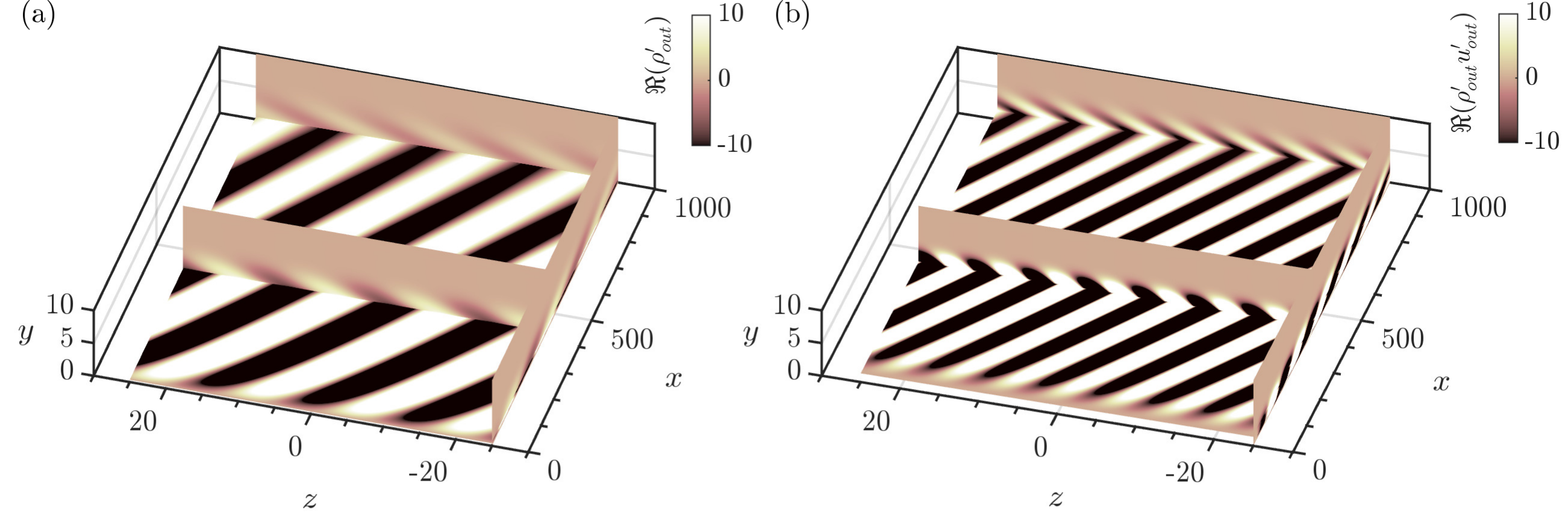}
\caption{\label{fig:11}Case T09w105 ($Re_\delta=300$, $\omega_{opt}=0.013$, $\beta_{opt}=0.45$). Slices of 3-D contours of optimal disturbances: (a) density, (b) $x$-momentum. }
\end{figure}

Oblique elongated structures are evident not only for the dynamic streaks but also for the thermal streaks, as previously observed in Fig.~\ref{fig:fig6}. Thus, slices of 3-D density perturbations are shown in Fig.~\ref{fig:11}(a). Note that a phase difference of $\pi$ exists between high-low-density and -temperature streaks, akin to observations in three-dimensional compressible ideal-gas boundary layers \cite{Tempelmann1}. Similar perturbation shapes are observed for the $x$-momentum in Fig.~\ref{fig:11}(b), where regions of high and low density correspond to regions of low and high streamwise velocity, respectively. 

In the transcritical wall-cooling case T11w095, optimal perturbations correspond to velocity streaks, alongside strong density and temperature streaks, which are streamwise-independent as indicated by $G_{opt}$ at $\omega_{opt}=0$ and $\beta_{opt}=0$ in Fig.~\ref{fig:fig3}(h). These structures resemble those in the subcritical and supercritical regimes, albeit with significantly larger thermal streaks amplitudes.

\subsubsection{Sub-optimal growth}
\label{subsec:4b5}
As previously pointed out, optimal transient growth is found at $x_{opt} \gg 1$ due to a combined Orr and lift-up mechanism. However, in case T09w105, as shown in Fig.~\ref{fig:fig3}(g), strong sub-optimal growth is evident at short distances from the initial location. This non-trivial phenomenon is further investigated hereafter. 

\begin{figure}[!t]
\includegraphics[angle=-0,trim=0 0 0 0, clip,width=1.0\textwidth]{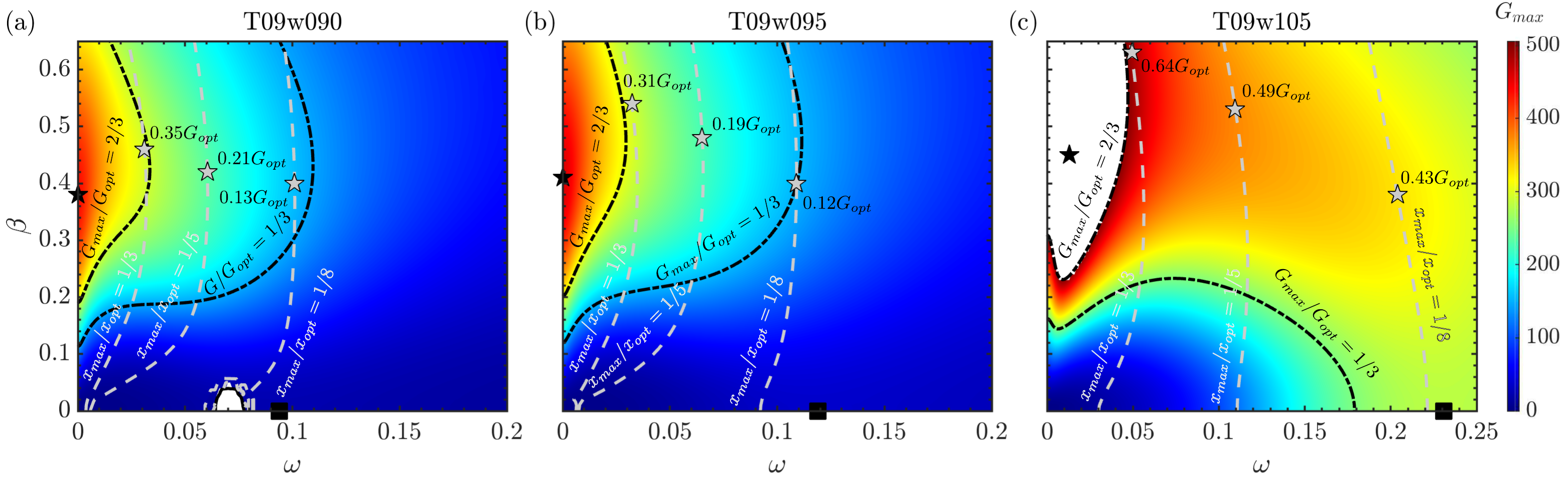}
\caption{\label{fig:12}Contour plot of $G_{max}$ in the $\omega$--$\beta$ space at $Re_\delta=300$ for cases: (a) T09w090, (b) T09w095, and (c) T09w105. $G_{opt}$, i.e.,~$\max\{G_{max}\}$, is denoted with a black star (\textcolor{black}{$\filledstar$}) symbol. The black dash-dotted lines represent $G_{max}/G_{opt}=2/3$ and $G_{max}/G_{opt}=1/3$, whereas the grey dashed lines represent $x_{max}/x_{opt}=1/3$, $x_{max}/x_{opt}=1/5$, and $x_{max}/x_{opt}=1/8$. The black square ($\blacksquare$) symbols denote $\max\{G_{max}(\omega,\beta=0)\}$, wheres the grey star (\textcolor{gray}{$\filledstar$}) symbols denote $\max\{G_{max}(x_{max}/x_{opt}=1/3)\}$, $\max\{G_{max}(x_{max}/x_{opt}=1/5)\}$, and $\max\{G_{max}(x_{max}/x_{opt}=1/8)\}$. }
\end{figure}

In Fig.~\ref{fig:12}, we reconsider the maximum energy amplification $G_{max}$ in the $\omega$--$\beta$ space for the wall-heating cases T09w090 (isothermal incompressible limit), T09w095, and T09w105, with black dash-dotted isolines denoting $G_{max}/G_{opt}=2/3$ and $G_{max}/G_{opt}=1/3$. Notably, the region between $G_{max}/G_{opt}=1/3$ and $2/3$ is significantly larger when the base-flow temperature crosses the Widom line, as seen in case T09w105. Energy amplifications exceeding $G_{opt}/3$ are observed along the $\omega$-axis for spanwise-uniform disturbances, demonstrating the Orr mechanism's substantial contribution to the energy amplification for the transcritical wall-heating case. In fact, $\max\{G_{max}(\omega,\beta=0)\}$ (black square ($\blacksquare$) symbol) in case T09w105 exceeds those of cases T09w095 and T09w090  by a factor of $7.8$ and $12.3$, respectively. Simultaneously, in Fig.~\ref{fig:12}(c), the non-negligible sub-optimal growth peaks much earlier in space than the global optimal (black star ($\filledstar$) symbol) at large $\omega$ and low $\beta$. Grey dashed isolines indicate a ratio of $x_{max}$ to $x_{opt}$ of about $1/8$, with $x_{opt}=422$ according to Tab.~\ref{tab:SD_M0001_T0.9andT1.1}. This confirms the
%(temporal)
observations of Refs.~\cite{Corbett1} and \cite{Butler1} in the temporal framework, where sub-optimal disturbances, associated with short-time scales, are no longer streamwise uniform but reach their maximum amplitude more rapidly. Moreover, tracking the streamwise evolution of $G_{max}$ (grey stars) in Fig.~\ref{fig:12}(c) reveals highly oblique disturbances at small $x$ given the large $\omega_{max}$, with the initial disturbance no longer resembling a streamwise vortex. At large $x$, similar disturbance shapes as in Fig.~\ref{fig:fig6} are recovered. This behavior aligns with previous findings for ideal gas at $M_\infty=5.0$ \cite{Bitter1}. 

\begin{figure}[!tb]
\centering
\includegraphics[angle=-0,trim=0 0 0 0, clip,width=0.85\textwidth]{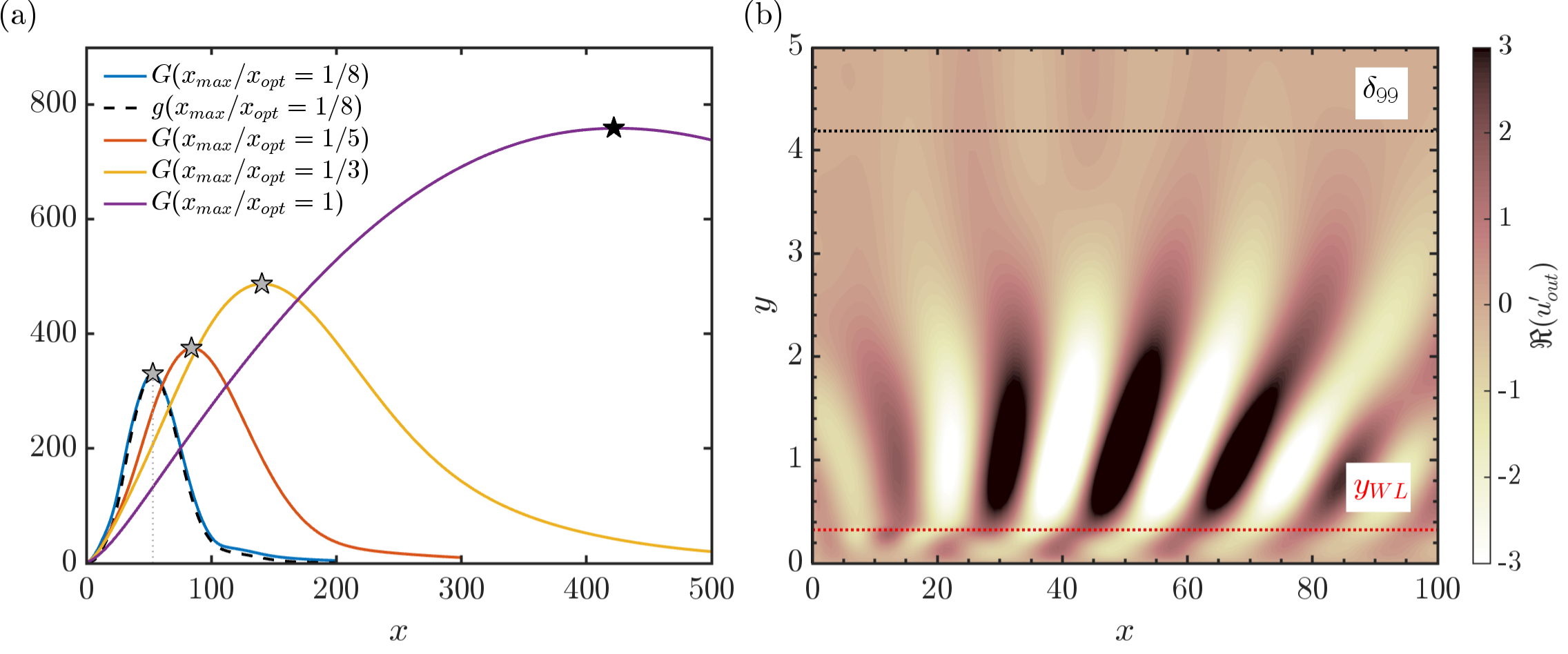}
\caption{\label{fig:13}Case T09w105. (a) Over the streamwise direction $x$: envelope of the largest sub-optimal transient growth at $x_{max}/x_{opt}=1/8$ (\textcolor{mycolor1}{\rule[0.5ex]{0.3cm}{1pt}}), $x_{max}/x_{opt}=1/5$ (\textcolor{mycolor2}{\rule[0.5ex]{0.3cm}{1pt}}), $x_{max}/x_{opt}=1/3$ (\textcolor{mycolor3}{\rule[0.5ex]{0.3cm}{1pt}}), sub-optimal transient growth $g(x_{max}/x_{opt}=1/8)$ of the optimal perturbation (\textcolor{black}{\rule[0.5ex]{0.11cm}{1pt}} \textcolor{black}{\rule[0.5ex]{0.11cm}{1pt}}), and envelope of the (global) optimal transient growth at $\omega_{opt}=0.013$ and $\beta=0.45$ (\textcolor{mycolor4}{\rule[0.5ex]{0.3cm}{1pt}}, black star indicates the global maximum as in Fig.~\ref{fig:12}(c)); the grey dotted line indicates the location of the maximum energy growth (grey star as in Fig.~\ref{fig:12}(c)). (b) Contours of the output streamwise velocity perturbations corresponding to the black dotted line $g(x_{max}/x_{opt}=1/8)$ ($x$-$y$ plane at $z=0$).}
\end{figure}

In Fig.~\ref{fig:13}(a), a comparison of optimal and sub-optimal energy amplification over the streamwise distance is presented. The global optimal transient growth, marked by a black star symbol in Fig.~\ref{fig:12}(c), is considered alongside the largest sub-optimal transient growths, peaking at $x=[x_{opt}/3,x_{opt}/5,x_{opt}/8]$ (grey star symbols in Fig.~\ref{fig:12}(c)). Notably, at $x \approx 52$ from the initial location, the local maximum energy amplification $G_{max}(x_{max}/x_{opt}=1/8)$ exceeds the optimal energy amplification at $x_{opt}$ by a factor. This signifies a near-optimal amplification of $G_{max}/G_{opt}\approx 0.43$ at a distance eight times shorter than $x_{opt}$, highlighting the relevance of sub-optimal energy growth. Recalling the analysis of the oblique perturbations in Sec.~\ref{sec:414}, an analysis of the initial condition yielding maximum sub-optimal gain at $x_{max}/x_{opt}=1/8$, denoted as $g(x_{max}/x_{opt}=1/8)$ in Fig.~\ref{fig:13}(a), is performed. At this location, the sub-optimal spanwise wavenumber and frequency are $\beta=0.38$ and $\omega=0.204$, respectively. In Fig.~\ref{fig:13}(b), a strong tilting of the flow structures characteristic of the Orr mechanism is noticeable, resulting in larger spanwise velocity components compared to those in Fig.~\ref{fig:fig10}. This leads to oblique structures with a propagation angle $\Psi$ of approximately $42^{\circ}$, in contrast to the nearly streamwise-independent global optimal structures in Fig.~\ref{fig:fig9}(b,c) with $\Psi \approx 2^{\circ}$. This result confirms the observation of Ref.~\cite{Corbett1}, where larger local optima can be achieved at $x\ll x_{opt}$ via efficient energy extraction from the mean flow through the Orr mechanism. As $x$ increases and $\omega_{opt} \rightarrow 0$, the lift-up mechanism becomes dominant, and (nearly) streamwise-independent structures are recovered.

\subsection{Effect of initial Reynolds number \label{sec:43a}}

\begin{figure}[!tb]
\centering
\includegraphics[angle=-0,trim=0 0 0 0, clip,width=0.85\textwidth]{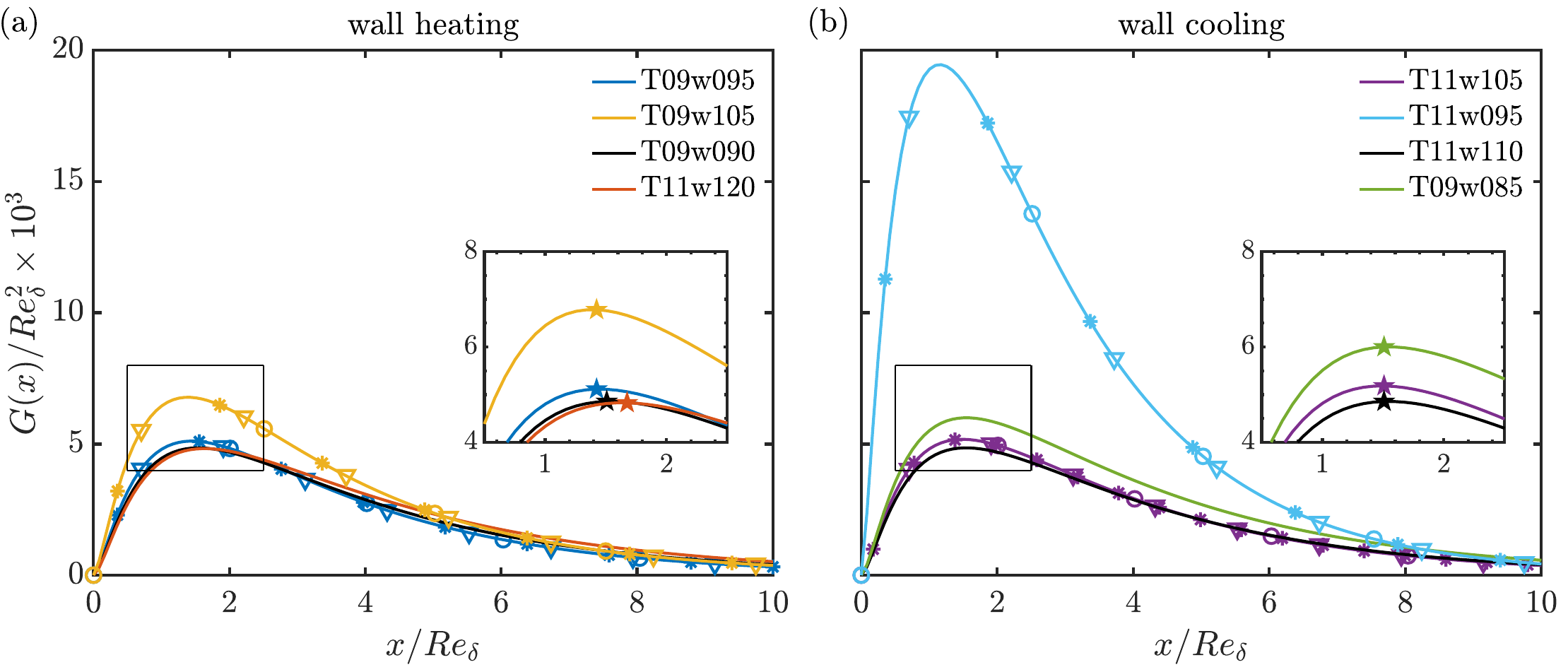}
\caption{\label{fig:14}Streamwise-independent disturbances. Energy amplification $G(x)/Re^2_\delta$ over the streamwise distance $x/Re_\delta$ at $\beta=\beta_{opt}(Re_\delta=300)$ and $\omega=0$. Lines in (a): T09w095, T09w105, T09w090, T11w120. Lines in (b): T11w105, T11w095, T11w110, T09w085. Symbols: circle (\textcolor{black}{$\circ$}) ($Re_\delta=300$), asterisk (\textcolor{black}{$\ast$}) ($Re_\delta=1000$), triangle (\textcolor{black}{$\triangledown$}) ($Re_\delta=3000$). Cases T09w090, T11w120, T11w110, T09w085 are displayed without symbols for better representation, but they also obey the scaling. In the insets, $G_{max}/Re^2_\delta$ is marked by a colored star (\textcolor{black}{$\filledstar$}) symbol. }
\end{figure}

In Sec.~\ref{sec:41}, all investigations were performed at a constant initial Reynolds number of $Re_\delta=300$. For both incompressible and compressible boundary-layer flows under the ideal-gas assumption, streamwise-independent modes scale according to Refs.~\cite{Gustavsson1,Hanifi2,Tumin1}: for the spatial theory, $G$ varies quadratically with the local Reynolds number, and $x$ scales linearly with $Re_\delta$. Thus, a similar analysis is performed for the non-ideal gas cases in Tab.~\ref{tab:tableBF}. The optimal energy amplification is calculated for initial Reynolds numbers $Re_\delta=300, 1000,$ and $3000$, using the same $\beta_{opt}$ at $Re_\delta=300$ and $\omega=0$. Figs.~\ref{fig:14}(a) and \ref{fig:14}(b) illustrate the scaling relations for cases with wall heating from a liquid-like free stream and wall cooling from a gas-like free stream, respectively. Curves at different $Re_\delta$ collapse for the same temperature difference, confirming the scaling law's validity not only at subcritical and supercritical temperatures, but also with the presence of strong non-ideality across the Widom line.

\subsubsection{Transcritical wall heating: scaling laws}
\begin{figure}[!tb]
\centering
\includegraphics[angle=-0,trim=0 0 0 0, clip,width=1.0\textwidth]{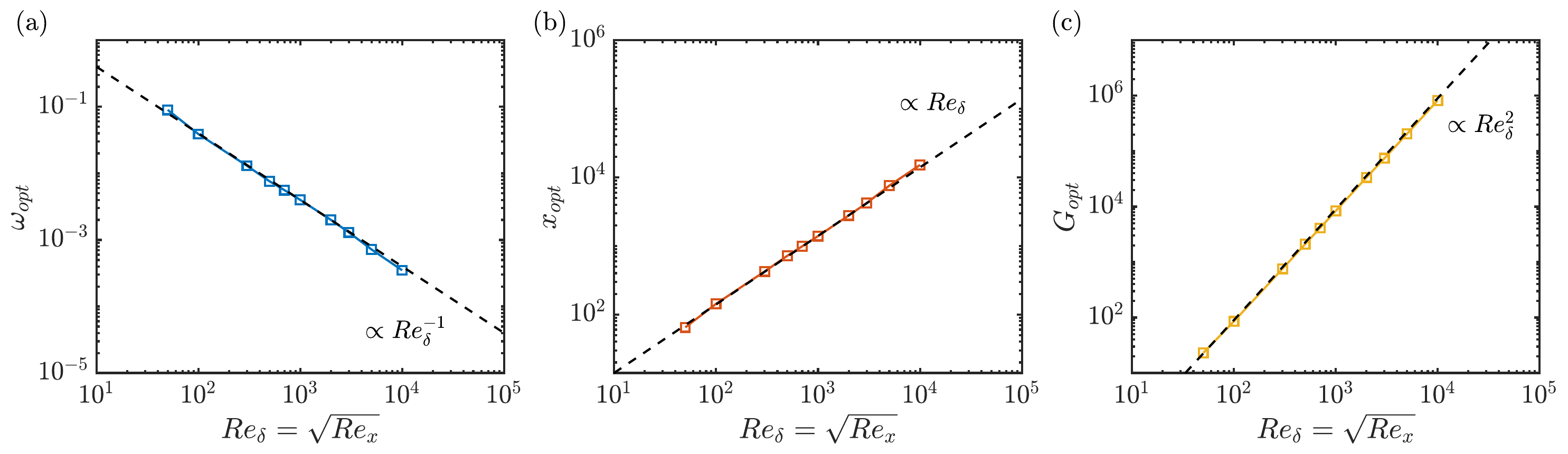}
\caption{\label{fig:15}Case T09w105: (a) optimal frequency, (b) optimal location, (c) optimal energy amplification. The dashed black lines indicate the power-law approximations with corresponding exponent $n$.}
\end{figure}

\begin{figure}[!tb]
\centering
\includegraphics[angle=-0,trim=0 0 0 0, clip,width=0.95\textwidth]{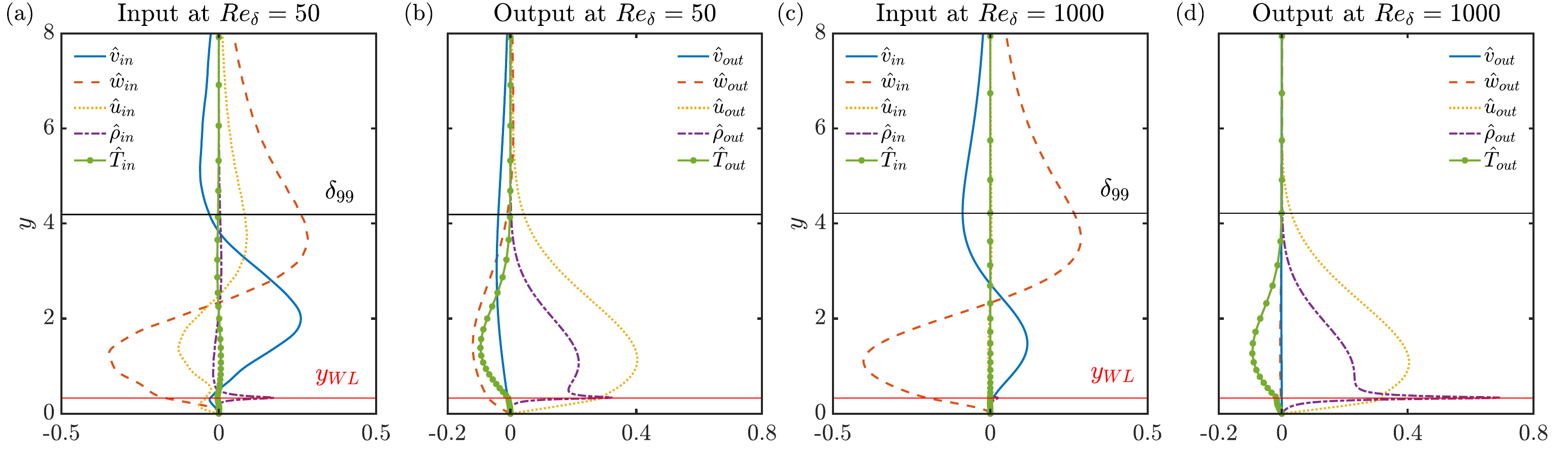}
\caption{\label{fig:16}Case T09w105. Wall-normal profiles of optimal input (a,c) and output (b,d) disturbances (real part) at two different input Reynolds numbers: (a,b) $Re_\delta=50$, (c,d) $Re_\delta=1000$. The boundary-layer thickness and the location of the Widom line are indicated by $\delta_{99}$ and $y_{WL}$, respectively. }
\end{figure}

The transcritical case T09w105, previously identified with streamwise-modulated streaks at $Re_\delta=300$, is further examined. In Figs.~\ref{fig:15}(a) and \ref{fig:15}(b), the optimal frequency $\omega_{opt}$ and optimal location $t_{opt}$ are plotted against the initial Reynolds number for the optimal amplification $G_{opt}$ shown in Fig.~\ref{fig:15}(c). As $Re_\delta$ increases, the optimal energy amplification shifts to lower frequencies, while the spanwise wavenumber remains nearly unaltered (not displayed here for the sake of brevity). This shift can be interpreted as a reduction in the Orr mechanism, with the mere lift-up effect becoming dominant as $\omega \rightarrow 0$. The decrease in $\omega_{opt}$ can be well approximated by $\omega_{opt} \propto Re^{n}_\delta$, with $n=-1$, similar to the relation $\omega_{opt} \propto Re^{-0.8}_\delta$ obtained by Ref.~\cite{Hack1} when including non-parallel effects. This trend suggests that optimal perturbations have zero frequency in the inviscid limit of $Re\rightarrow\infty$ \cite{Hack1}. Moreover, power laws for $G_{opt}$ and $x_{opt}$, akin to those for streamwise-independent perturbations (see Fig.~\ref{fig:14}), hold even when the Orr mechanism is at play. Hence, the relations $x_{opt} \propto Re_\delta$ and $G_{opt} \propto Re^{2}_\delta$ are valid regardless of the underlying physical mechanism.
At $Re_\delta=50$ and $1000$, the optimal perturbations are presented in Figs.~\ref{fig:16}(a,b) and \ref{fig:16}(c,d), respectively. As expected, the spanwise and wall-normal velocity components are most pronounced initially. However, a notable input streamwise velocity amplitude at $Re_\delta=50$, especially near the Widom line, highlights the influence of the Orr mechanism at low Reynolds numbers. In contrast, at $Re_\delta=1000$, the lift-up effect predominates, with $\hat{u}_{in} \approx 0$. Additionally, examining the output spanwise velocity perturbation $\hat{w}_{out}$ reveals that at $Re_\delta=50$ with $\omega_{opt} \gg 0$, the Orr mechanism significantly amplifies the spanwise velocity, consistent with Ref.~\cite{Jiao1}. 

\begin{figure}[tb]
\centering
\includegraphics[angle=-0,trim=0 0 0 0, clip,width=0.9\textwidth]{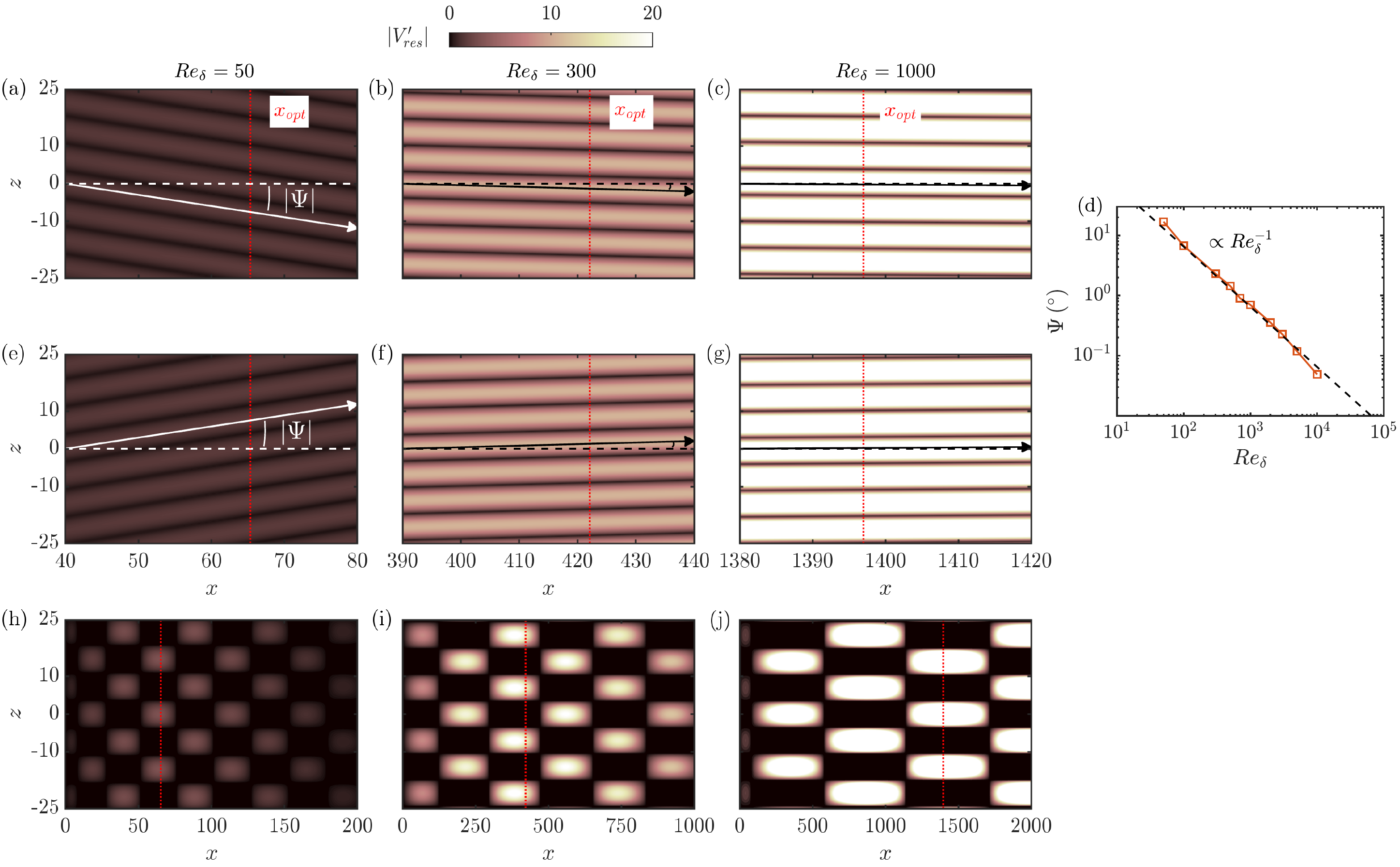}
\caption{\label{fig:17}Case T09w105. In (a--c) and (e--j), contour plots in the $x$-$z$ plane ($y=0.77$) of the resulting streaks' tangential velocity $|\vec{V}'_{res}|$: (a,e,h) $Re_\delta=50$, (b,f,i) $Re_\delta=300$, and (c,g,j) $Re_\delta=1000$; (a--c) $\beta=\beta_{opt}$, (e--g) $\beta=-\beta_{opt}$, and (h--j) $\beta=\pm\beta_{opt}$. The arrow represents the velocity vector with propagation angle $\Psi$. In (d), development of the streaks' propagation angle $\Psi$ as a function of the initial Reynolds number. The dashed black line indicates the power-law approximation with exponent $n=-1$.}
\end{figure}

Recalling the streaks' shape in Fig.~\ref{fig:11} and the evolution of the optimal frequency in Fig.~\ref{fig:15}, the propagation angle of the oblique streaks likely depends on the initial Reynolds number. To explore this, snapshots of the resulting optimal streak tangential velocity $|\vec{V}'_{res}|=\sqrt{u'^2+w'^2}$ on an $x$-$z$ plane are illustrated in Figs.~\ref{fig:17}(a,e,h), Figs.~\ref{fig:17}(b,f,i), and Figs.~\ref{fig:17}(c,g,j) at $Re_\delta=50$, $300$, and $1000$, respectively. Optimal perturbations are obtained with $\beta=\beta_{opt}$ in Figs.~\ref{fig:17}(a--c), with $\beta=-\beta_{opt}$ in Figs.~\ref{fig:17}(e--g), and with $\beta=\pm\beta_{opt}$ in Figs.~\ref{fig:17}(h--j). To calculate the streaks' propagation angle $\tan(\Psi)=\beta_{opt}/\alpha$, the optimal spanwise wavenumber is derived from $G_{opt}(\beta_{opt},\omega_{opt})$. The streamwise wavenumber $\alpha$ is extracted as follows: the spatial evolution of the wall-normal maximum of the streamwise perturbation velocity is evaluated at at $z=0$ (see plane D in Fig.~\ref{fig:fig9}(b)), determining the streamwise wavelength $\lambda_x$ from the wave period of the oscillating streamwise velocity, and thus $\alpha=2\pi/\lambda_x$. This method achieves a good agreement between the calculated $\Psi$ and the actual streaks' orientation. Repeating the procedure at different $Re_\delta$, the dependence of $\Psi$ on the Reynolds number is displayed on a logarithmic axis in Fig.~\ref{fig:17}(d). As the optimization proceeds downstream, streaks progressively evolve into 2-D structures due to the frequency shift of their optimal transient growth. Since $\beta_{opt}$ remains nearly unchanged while $\omega_{opt}$ scales as $Re^{-1}_\delta$, the propagation angle $\Psi$ follows the same power law, approaching zero as $Re_\delta \rightarrow \infty$. Note that this analysis uses a locally-parallel spatial optimization procedure, so generalizing the actual streak structure to the entire flat-plate distance requires advanced non-local methods, such as a PSE-based formulation \cite{Tempelmann1,Paredes1}. As highlighted in Fig.~\ref{fig:fig9}(c), oblique streaks can have both positive and negative propagation angles, owning the same patterns, with their superposition in Figs.~\ref{fig:17}(i--k) depending on the initial Reynolds number. When moving downstream, the checkerboard wave pattern stretches due to a larger streamwise wavenumber.

\subsubsection{Transcritical wall heating: role of Mode II}
\begin{figure}[!t]
\centering
\includegraphics[angle=-0,trim=0 0 0 0, clip,width=0.90\textwidth]{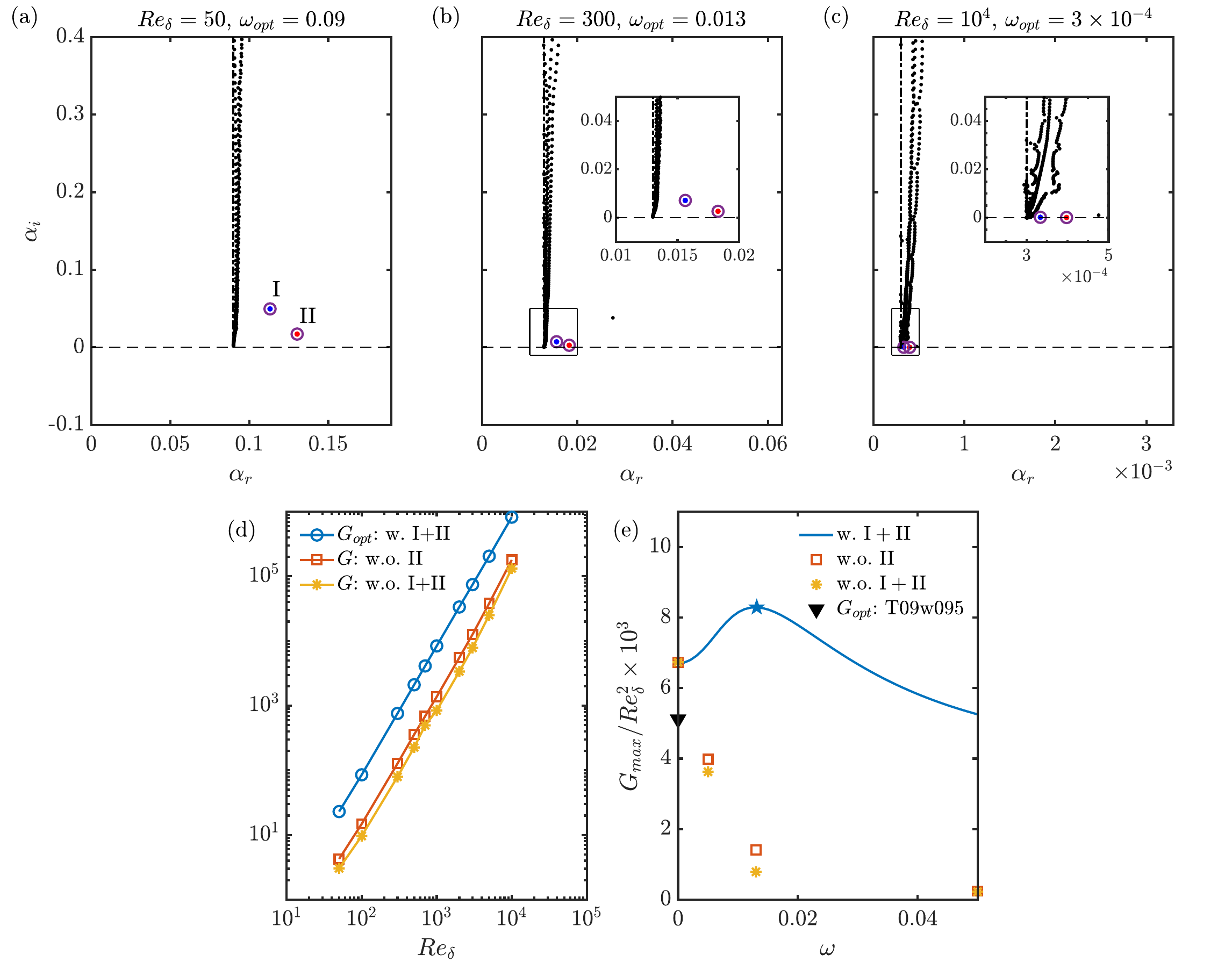}
\caption{\label{fig:18}Case T09w105. Eigenspectrum at (a) $Re_\delta=50$, (b) $Re_\delta=300$, and (c) $Re_\delta=10^4$ (higher mesh resolution needed), $\omega=\omega_{opt}$. Blue and red dots represent Mode I and II, respectively. The 2 modes with the largest contribution to the optimal growth are circled in purple. The black dash-dotted line refers to the continuous spectra according to Ref.~\cite{Tumin4}. In (d), the optimal energy amplification is plotted over the local Reynolds number for the full spectrum (blue line); note that $G$ for the full spectrum without Mode II (red line) and full spectrum without Mode I and II (yellow line) is not anymore $G_{opt}$ for these cases. In (e), the maximum energy amplification is plotted over the frequency for the full spectrum (blue line with $G_{opt}$ marked by a blue ($\filledstar$) symbol), full spectrum w.o.~Mode II (red line), and full spectrum w.o.~Mode I and II (yellow line). $G_{opt}$ of case T09w095 (marked by a black ($\blacktriangledown$) symbol.}
\end{figure}

To understand the frequency shift in the optimal energy amplification for the transcritical wall-heating case T09w105, we analyze the underlying eigenspectrum, forming the eigenvector basis for matrix $\mathbf{F}$ (see Sec.~\ref{subsec:2c4}), used in calculating the energy growth $G$ in Eq.~\eqref{eq:trans_opt}. Figs.~\ref{fig:18}(a--c) show the eigenspectra at $Re_\delta=50$, $300$, and $10^4$ with the corresponding $\omega=\omega_{opt}$, obtained by solving the eigenvalue problem in Eq.~\eqref{eq:EVP}. As identified by Ref. \cite{Ren2}, two discrete modes (Mode I and II), here damped, appear when the base-flow temperature crosses the Widom line from a liquid-like free-stream to a gas-like wall. Mode I, resembling the Tollmien-Schlichting mode, always exists, regardless of the thermodynamic regime. In Figs.~\ref{fig:18}(a--c), the only continuous branches at $M_\infty=10^{-3}$ belong to the vorticity and entropy modes (indicated by the vertical dash-dotted line with $\alpha=\omega$). For 3-D disturbances, i.e., $\beta=\beta_{opt}>0$, two vorticity modes and one entropy mode are found in agreement with Ref.~\cite{Tumin4}. When moving downstream, the continuous branches become more distinct, and the two discrete modes approach the vorticity/entropy modes. At $Re_\delta=10^4$, in fact, both Mode I and II nearly synchronize with the continuous spectrum at a phase velocity of $c_r \approx 1$. As synchronization progresses from $Re_\delta=50$ to $10^4$, the two eigenmodes contributing to the largest optimal amplification are tracked (see purple circles in Figs.~\ref{fig:18}(a--c)), which are computed via the right singular eigenvector associated to the largest singular value of $\mathbf{F} \bm{\Lambda} \mathbf{F}^{-1}$ in Eq.~\eqref{eq:trans_opt}. The modes with the highest transient-growth contribution are consistently Mode I and II, indicating that, regardless of the initial Reynolds number, these two slowly-decaying modes are the main contributors to energy amplification, while the other modes in the continuous spectrum decay rapidly and play a minor role. For ideal gas at $M_\infty=5.0$, a similar process for discrete modes was witnessed by Ref.~\cite{Bitter1}, where optimal disturbances were oblique for $x<x_{opt}$, unlike the continuous-spectrum interaction in \cite{Schmid2}. However, in Ref.~\cite{Bitter1}, the least-damped discrete mode corresponds to Mack's second mode, which differs from the current Mode II in supercritical fluids \cite{Ren2}.

Since transient growth is mathematically linked to the non-orthogonality of the eigenvectors in the initial value problem, analyzing the underlying eigenspectrum is essential for our study \cite{Corbett1}. Ref.~\cite{Gustavsson1} notes that the individual and least-damped Orr-Sommerfeld mode can interfere with the continuous branches, leading to the largest energy amplification. Therefore, we investigate how the two discrete least-damped modes contribute to transient growth and energy amplification in a spatial framework. First, $G_{opt}$ at finite $\beta_{opt}$ and $\omega_{opt}$, based on the full spectrum of Figs.~\ref{fig:18}(a--c), is shown by a blue line in Fig.~\ref{fig:18}(d), coinciding with Fig.~\ref{fig:15}(c). Next, Mode II is excluded from the optimization, producing a modified eigenspectrum similar to the non-transcritical cases. With the same optimal frequency and spanwise wavenumber, the modified optimal amplification $G_{\mathrm{w.o.II}}$ (red line in Fig.~\ref{fig:18}(d)) is calculated. Excluding Mode II results in a significant drop in $G$, with $G_{opt}/G_{\mathrm{w.o.II}} \approx 5$. Discarding both Mode I and II yields the energy amplification $G_{\mathrm{w.o.I+II}}$ (yellow line in Fig.~\ref{fig:18}(d)), which experiences a slight decrease compared to $G_{\mathrm{w.o.II}}$. Hence, the substantial drop in $G$ is primarily due to the increase in non-ideality when crossing the Widom line, and consequently, the presence of Mode II. Fig.~\ref{fig:18}(e) displays the maximum energy amplification over the frequency at a constant spanwise wavenumber of $\beta=\beta_{opt}$. Omitting Mode II from the initial value problem shifts the peak optimal growth (blue star (\textcolor{mycolor1}{$\filledstar$}) symbol) from finite frequency values (blue line) to $\omega=0$ (red line), resembling non-transcritical cases in Fig.~\ref{fig:fig3}, where streamwise-independent disturbances were found. Comparing $G_{opt,\mathrm{w.o.II}}$ at $\omega=0$ with $G_{opt}$ of case T09w095 (black ($\blacktriangledown$) symbol) reveals similar values, reflecting the spectra similarity. 

Physically, considering the results from Sec.~\ref{sec:40} and \ref{sec:41} in relation to Mode II, one might infer that the Orr mechanism disappears once Mode II is omitted from the eigenspectrum. Nevertheless, even in the non-transcritical cases where $G_{opt}$ is at $\omega=0$, the Orr mechanism is present at $\omega>0$ \cite{Farrell1}, though it results in sub-optimal energy growth (see Fig.~\ref{fig:12}). This is typical of wall-bounded shear flows, where the dominant energy growth is associated with 2-D streaks \cite{Butler1}. In unbounded infinite-shear flows, the combination of Orr and lift-up mechanism has been revealed to be optimal, as transient growth is inherently three-dimensional relying on the interplay of these two mechanisms \cite{Kaminski1}. In terms of modal analysis, with the appearance of Mode II and its high contribution to transient growth, an additional degree of non-orthogonality is introduced \cite{Schmid2}, enhancing the interaction, already for sub-optimal energy growth (see Fig.~\ref{fig:12}), between the lift-up effect and Orr mechanism. The latter significantly amplifies the spanwise velocity \cite{Jiao1}, as observed in Figs.~\ref{fig:fig10}(a--c) with inclined structures in the spanwise direction and in Fig.~\ref{fig:16}(b,d) with larger $\hat{w}_{out}$ at lower Reynolds numbers. This behavior is supported by the cases in Tab.~\ref{tab:SD_M0001_T0.9andT1.1}, where only one discrete least-damped mode is present, and the interplay of lift-up and Orr mechanism only produces sub-optimal energy growth (see Fig.~\ref{fig:12}).

\subsection{Effect of wall temperature \label{sec:43b}}

\begin{figure}[!tb]
\centering
\includegraphics[angle=-0,trim=0 0 0 0, clip,width=0.80\textwidth]{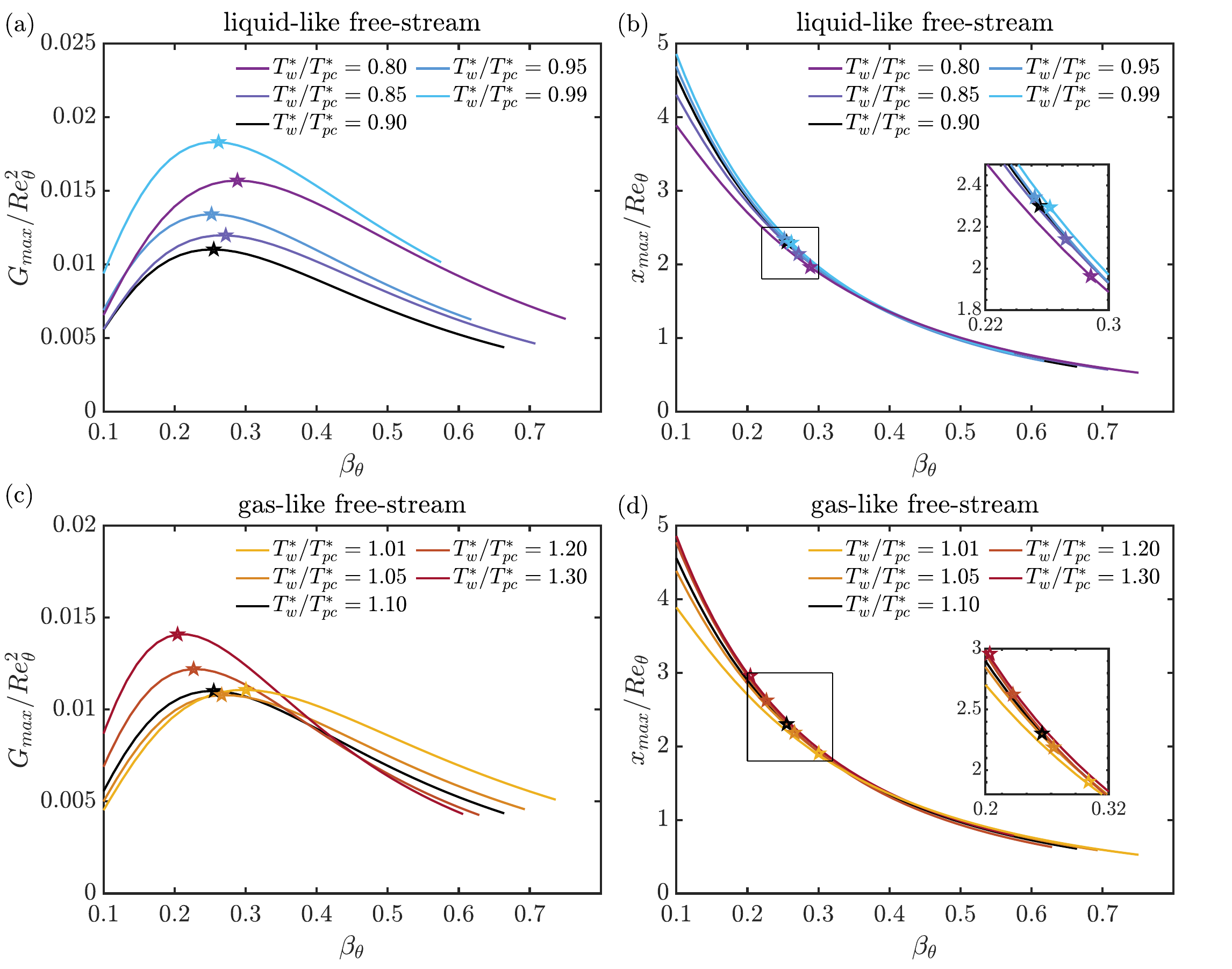}
\caption{\label{fig:19}Subcritical and supercritical regime. Maximum energy amplification $G_{max}/Re^2_\theta$ and its distance $x_{max}/Re_\theta$ at different $T^*_w/T^*_{pc}$: (a,b) $T^*_\infty/T^*_{pc}=0.90$, (c,d) $T^*_\infty/T^*_{pc}=1.10$. The optimal amplification $G_{opt}$ is indicated with a colored star ($\smallstar$) symbol.}
\end{figure}

The influence of wall temperature on transient growth is examined. For the non-transcritical cases in Tab.~\ref{tab:SD_M0001_T0.9andT1.1}, a preliminary trend was evident in Fig.~\ref{fig:fig3}. In the subcritical and supercritical regimes, wall cooling exhibits higher amplification rates than wall heating. Here, consistent with Corbett \& Bottaro \cite{Corbett1}, the influence of the non-dimensional compressible momentum thickness $\theta=\int_0^\infty \rho u (1-u)dy$ (see Ref.~\cite{White1}) is considered, with its value given in Tab.~\ref{tab:tableBF}. Figs.~\ref{fig:19}(a,c) and Figs.~\ref{fig:19}(b,d) plot the maximum energy amplification $G_{max}/Re^2_\theta$ and its distance $x_{max}/Re_\theta$ over the rescaled spanwise wavenumber $\beta_\theta=\beta\theta$, respectively. Streamwise-independent modes are considered, as they are the most non-modally amplified perturbations in Fig.~\ref{fig:fig3}. The wall temperature is varied for subcritical free-stream conditions with $T^*_\infty/T^*_{pc}=0.90$, as shown in Fig.~\ref{fig:fig3}(a), and for supercritical free-stream conditions with $T^*_\infty/T^*_{pc}=1.10$, as shown in Fig.~\ref{fig:fig3}(b).
The impact of a diabatic wall is more pronounced in the subcritical regime (Fig.~\ref{fig:19}(a)), where both wall cooling and wall heating cause a greater increase in $G$ than in the supercritical regime of Fig.~\ref{fig:19}(c). As $\theta$ decreases with increasing $T^*_w/T^*_\infty$ in both regimes, the highest $G_{max}/Re^2_\theta$ occurs with wall heating. Notably, in the supercritical regime, the lowest $G_{opt}/Re^2_\theta$ is found not with $T^*_w/T^*_\infty=1.0$ but with a slightly negative wall-temperature gradient. Ref.~\cite{Corbett1} noted the universal behavior of the momentum thickness scaling for a Falkner-Skan boundary layer. In fact, with a $\theta$-scaling, the optimal spanwise wavenumber $\beta_{\theta,opt}$ was independent of the mean-flow pressure gradient, a behavior nearly replicated in Fig.~\ref{fig:19}(a) for a liquid-like free-stream, where $\beta_{\theta,opt}$ ranges from $0.25$ to $0.29$. Even more noteworthy, $t_{max}/Re_\theta$ exhibits nearly exact scaling with $\beta_\theta$ in both regimes (Figs.~\ref{fig:19}(b,d)). \par
\begin{figure}[!tb]
\centering
\includegraphics[angle=-0,trim=0 0 0 0, clip,width=0.80\textwidth]{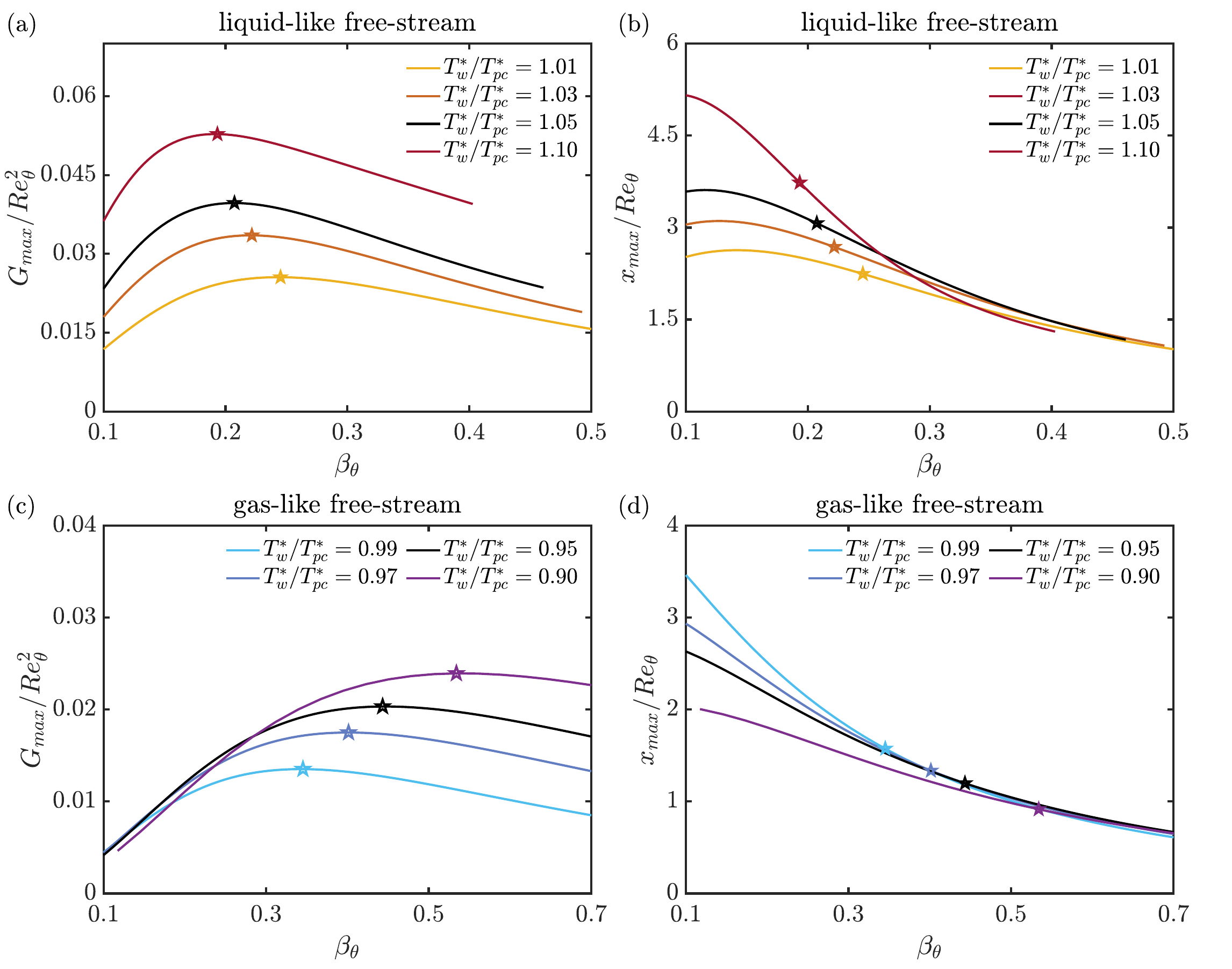}
\caption{\label{fig:20}Transcritical regime. Maximum energy amplification $G_{max}/Re^2_\theta$ and its distance $x_{max}/Re_\theta$ at different $T^*_w/T^*_{pc}$: (a,b) $T^*_\infty/T^*_{pc}=0.90$, (c,d) $T^*_\infty/T^*_{pc}=1.10$. The optimal amplification $G_{opt}$ is indicated with a colored star ($\smallstar$) symbol.}
\end{figure}
A similar investigation is conducted for the transcritical wall-heating case T09w105, previously shown in Fig.~\ref{fig:fig3}(d). Here, the free-stream temperature is constant at $T^*_\infty/T^*_{pc}=0.90$, while the wall temperature increases from $T^*_w/T^*_{pc}=1.01$ to $T^*_w/T^*_{pc}=1.10$ in the gas-like regime. Consequently, the momentum thickness $\theta$ decreases from $0.533$ to $0.403$, and the height of the Widom-line crossing relative to the boundary-layer thickness, i.e.,~$y_{WL}/\delta_{99}$, rises. Figs.~\ref{fig:20}(a,b) present results at four different wall temperatures. Note that optimal growth shifts to finite frequencies just after the Widom is crossed, with streamwise-modulated streaks observed up to $T^*_w/T^*_{pc}=1.10$. Beyond this ratio, energy amplification at very low frequencies dominates before $G_{opt}$ is found at $\omega=0$. Hence, optimal streamwise-independent streaks with $\Psi=\ang{0}$ occur for a wall-heating factor $T^*_w/T^*_\infty$ greater than $1.22$. In Figs.~\ref{fig:20}(a,b), a wall-temperature increase above $T^*_{pc}$ no longer follows the $\theta$ scaling as in Figs.~\ref{fig:19}(a,b). A even larger shift of $\beta_{\theta,opt}$ is observed in Fig.~\ref{fig:20}(c,d), despite good overlap in $x_{max}/Re_\theta$ for large $\beta_\theta$. Here, the free-stream temperature is constant at $T^*_\infty/T^*_{pc}=1.10$, while the wall temperature varies from $T^*_w/T^*_{pc}=0.99$ to $T^*_w/T^*_{pc}=0.90$ in the liquid-like regime. The momentum thickness $\theta$ increases from $0.803$ to $1.173$, with the height of the Widom-line crossing increasing due to wall cooling. Note also that the boundary-layer thickness grows with stronger transcritical wall cooling, achieving a similar growth of the Widom-line crossing with the same $\Delta T=|T^*_\infty-T^*_w|$ for both transcritical wall heating and cooling. However, wall-normal gradients of base-flow properties at the Widom line, e.g.,~$|\partial \bar{\rho}/\partial y|_{WL}$, are larger for the transcritical wall-cooling case for the same $\Delta T$, leading to larger $G_{opt}$, scaled with $Re^2_{\delta}$. Overall, it is evident that with strong non-ideality the momentum thickness is no longer the universal scaling parameter in transient growth.

\begin{figure}[!tb]
\centering
\includegraphics[angle=-0,trim=0 0 0 0, clip,width=0.95\textwidth]{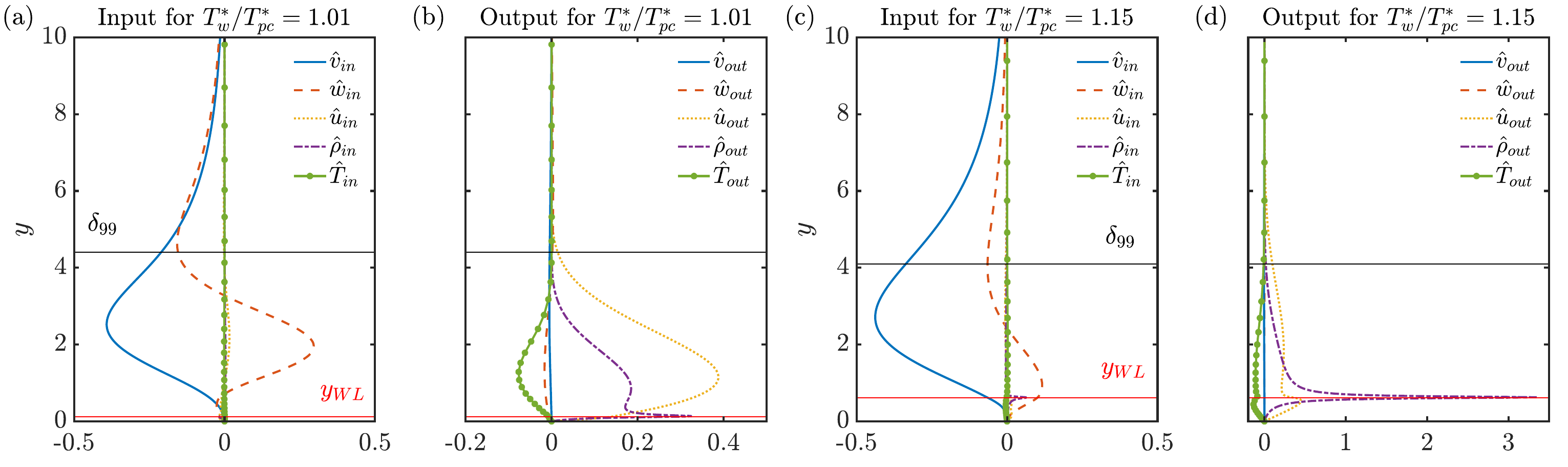}
\caption{\label{fig:21}Transcritical wall heating at $Re_\delta=300$. Wall-normal profiles of optimal input (a,c) and output (b,d) disturbance at two different wall temperatures: (a,b) $T^*_w/T^*_{pc}=1.01$, (c,d) $T^*_w/T^*_{pc}=1.15$. The boundary-layer thickness and the location of the Widom line are indicated by $\delta_{99}$ and $y_{WL}$, respectively. }
\end{figure}

To highlight the influence of transcritical wall heating on the optimal perturbations, the wall-normal profiles of optimal input and output disturbances at $T^*_w/T^*_{pc}=1.01$ and $T^*_w/T^*_{pc}=1.15$ are displayed in Figs.~\ref{fig:21}(a,b) and (c,d), respectively. The Reynolds number is $Re_\delta=300$. At $T^*_w/T^*_{pc}=1.01$, with $G_{opt}$ at $\omega_{opt} \approx 0.01$, the Orr mechanism results in small but non-zero streamwise velocity and larger spanwise velocity for the input perturbations, causing cross-stream inclination of the perturbation structures (Figs.~\ref{fig:fig10}(a--c)). At $T^*_w/T^*_{pc}=1.15$, optimal amplifications are found at zero frequency with $\hat{u}_{in}=0$ and a smaller $\hat{w}_{in}$ than in Fig.~\ref{fig:21}(a). Despite the Widom line moving away from the wall, the input disturbances' shape remains nearly identical, as seen in Ref.~\cite{Tumin2} for compressible ideal-gas boundary layers. On the other hand, the output profiles are greatly affected by different transcritical wall heating. The amplitude of the density perturbation (purple dash-dotted line) at the Widom line increases with its local base-flow gradient, with $\partial \bar{\rho}/\partial y|_{WL}$ at $T^*_w/T^*_{pc}=1.15$ being $1.5$ times larger than at $T^*_w/T^*_{pc}=1.01$, resulting in stronger density streaks. Regarding the output streamwise velocity perturbation (yellow dotted line), its peak shifts from the middle of the boundary layer in Fig.~\ref{fig:21}(b) to the Widom line in Fig.~\ref{fig:21}(d), due to a stronger vortex-tilting term ($|\partial \bar{u}/\partial y \cdot \partial \hat{v}_{in}/\partial z|$) around the Widom line, driven by larger perturbation strain rate away from the wall. Thus, the highest increase in $\hat{\omega}_y$ occurs around the Widom line, with the secondary peak of $\hat{u}_{out}$ in the boundary layer's center being less amplified compared to case $T^*_w/T^*_{pc}=1.01$. 
For the transcritical wall-cooling cases (Figs.~\ref{fig:20}(c,d)), similar features to the transcritical wall-heating cases are found (not shown here for the sake of brevity). However, due to a larger $\partial \bar{\rho}/\partial y|_{WL}$, the output density perturbations exhibit significantly larger amplitudes than the output velocity perturbations. The vortex-tilting term $|\partial \bar{u}/\partial y \cdot \partial \hat{v}_{in}/\partial z|$ is greatly enhanced by wall cooling, increasing both mean-flow vorticity and perturbation strain rate. This term peaks above $y_{WL}$, which rises as $T^*_w/T^*_{pc}$ decreases. As a result, the wall-normal vorticity perturbation $\hat{\omega}_{y,out}$ shows a similar profile to Fig.~\ref{fig:fig8}(c), but more pronounced around the Widom line.

\begin{figure}[!tb]
\centering
\includegraphics[angle=-0,trim=0 0 0 0, clip,width=0.95\textwidth]{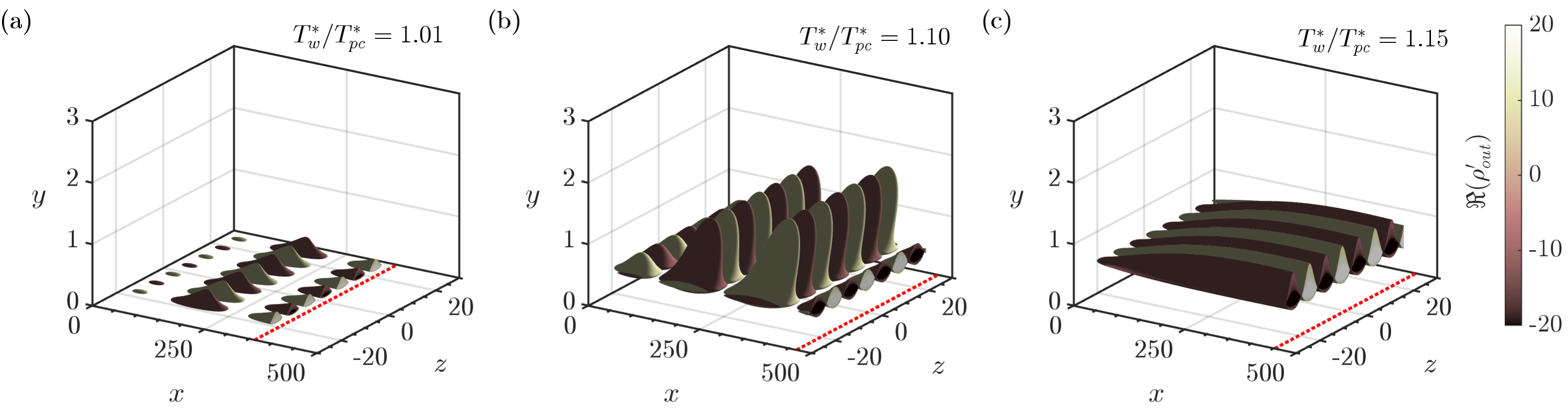}
\caption{\label{fig:22}Transcritical wall heating at $Re_\delta=300$. Iso-contours of the optimal density perturbations ($\Re(\rho'_{out})=\pm8.0$): (a) $T^*_w/T^*_{pc}=1.01$ (superposition), (b) $T^*_w/T^*_{pc}=1.10$ (superposition), (c) $T^*_w/T^*_{pc}=1.15$. The red dashed line indicates the optimal growth location ($x=x_{opt}$).}
\end{figure}

Fig.~\ref{fig:22} visualizes the 3-D optimal density perturbation structures for $T^*_w/T^*_{pc}=1.01, 1.10, 1.15$. For the first two cases, perturbations with $\beta=\pm \beta_{opt}$ are superimposed, while for $T^*_w/T^*_{pc}=1.15$ optimal disturbances are streamwise-independent. As $T^*_w/T^*_{pc}$ increases to $1.10$, $\omega_{opt}$ also increases, causing oblique structures to have a larger propagation angle and a shorter streamwise wavelength, resulting in a strong alternate checkered wave pattern. Additionally, at $T^*_w/T^*_{pc}=1.01$, the largest density structures are confined to the Widom-line region. As the wall temperature increases to $T^*_w/T^*_{pc}=1.10$, these structures extend into the mid-section of the boundary layer, exhibiting significant amplitudes. At $T^*_w/T^*_{pc}=1.15$ (Fig.~\ref{fig:22}(c)), even stronger density perturbations are present, yet they remain predominantly located around the Widom-line region.

\section{\label{sec:44}Comparison between modal and non-modal growth}

In this study, energy amplification has been maximized at $x_{opt}$, yet these locations may be too large to justify the locally parallel assumption, as $x$ is non-dimensionalized by the Blasius length scale $\delta$. For every case in Tab.~\ref{tab:tableBF} $x_{opt}$ is proportional to $Re_{\delta,0}$, the initial local Reynolds number. This suggests that the optimal disturbance requires $O(Re_{\delta,0}\cdot\delta)$ to grow from the initial location $x_0$. As $\delta \propto \sqrt{x} \propto Re_\delta$, the growth of the boundary-layer thickness, which is proportional to the Blasius length scale \cite{White1}, corresponds to a factor $\sqrt{x_{opt}}/\sqrt{x_0}$ or $Re_{\delta,opt}/Re_{\delta,0}$ from $x_0$ to $x_{opt}$. Consequently, a shorter $x$ near the leading edge is required for the same increase in $\delta$, where the locally parallel and boundary-layer assumption are more likely to fail. Yet, previous studies on the non-parallel assumption \cite{Andersson1,Tumin2} have revealed similar optimal distances to those of this study, to Ref.~\cite{Butler1} for incompressible flows, and to Ref.~\cite{Bitter1} for hypersonic flows. For this reason, further investigations are not undertaken hereafter.

With large $x_{opt}$, transient growth might be irrelevant for transition to turbulence due to a earlier exponential growth of modal instabilities. Thus, it is necessary to assess which mechanism is more likely to lead to non-linear breakdown. To properly compare non-modal with modal calculations, the integral amplification $N$ is evaluated, as $G$ in Eq.~\eqref{eq:trans_opt} is optimized over a prescribed space interval \cite{Schmid2}. Hence, the integral spatial amplification rate, namely the $N$-factor, is used. The $N$-factor for modal instabilities is calculated as 
\begin{subequations}
\begin{gather}
    N=\int_{x_0}^{x} -\alpha_i(\omega/Re_\delta,\beta) dx, \quad N_{modal}(x)=\max_{\omega/Re_\delta,\beta} \{ N(x,\omega/Re_\delta,\beta) \}, \tag{34a,b}
    \label{eq:Nfac_modal}
\end{gather}
\end{subequations}
where $x_0$ is the streamwise location for neutral instability (branch I), and $N_{modal}(x)$ represents the envelope of all possible modal $N$-factors over all frequencies and spanwise wavenumbers. An analysis of the modal growth of 3-D disturbances for cases in Tab.~\ref{tab:tableBF} is reported in Sec.~\ref{sec:app4}. In summary, for all cases, the maximum growth rate at a constant frequency, i.e., $\max_{\omega=\text{const.}} \{-\alpha_i\}$, is dependent on $\beta$, and thus 2-D modes are not always the most unstable. Yet, the maximum integral amplification $N$ is always found for 2-D modes, i.e., $N_{modal}(x)=\max_{\forall \omega} \{N(\beta=0)\}$.

Contrary to modal growth, the $N$-factor for non-modal (optimal) growth aligns with Refs.~\cite{Bitter1,Tempelmann1,Levin1} and is expressed as
\begin{eqnarray}
&&N_{opt}(x,Re_{\delta,0})=0.5\ln\left(G\right)
\label{eq:Nfac_opt}
\end{eqnarray}
where $N_{opt}$ depends on the initial Reynolds number $Re_{\delta,0}$ and $G$ is chosen as the optimal amplification $G_{opt}$. Note that, while short-distance energy amplification can be larger than $G_{opt}(Re_{\delta,0})$ at $x \ll x_{opt}$, $G_{opt}$ is the global maximum for each varying $Re_{\delta,0}$. In both non-transcritical and transcritical wall-cooling cases, $\omega_{opt}$ is zero. However, in the transcritical wall-heating case, $\omega_{opt}$ varies with $Re_{\delta,0}$ due to a non-constant streaks' propagation angle (see Fig.~\ref{fig:15}(a)), so $G$ must be adapted in the streamwise direction. Moreover, $\beta_{opt}$ remains almost constant with respect to the chosen $Re_{\delta,0}$ and is initially selected as the optimal one at $Re_{\delta,0}=300$ (see Tab.~\ref{tab:SD_M0001_T0.9andT1.1}) before optimization at different $Re_{\delta,0,N}$ (initial optimal location). For a strict comparison with integral amplifications, the calculation of $N_{opt}$ excludes modal growth, ensuring that only finite $G_{opt}$ from stable eigenmodes are considered. The envelope of $N_{opt}$, $N_{opt,env.}$, represents the maximum transient growth at each streamwise location. Despite the boundary-layer growth, non-modal calculations are performed under the locally parallel assumption. Non-parallel results of Ref.~\cite{Tumin2} indicate a difference in $N_{opt}$ of only $0.4$--$0.5$ \cite{Bitter1}, which is deemed acceptable for comparing to modal growth.

Tab.~\ref{tab:Nfac_M0001_T09andT11} summarizes the comparison between non-modal and modal growth for the non-transcritical cases.
\begin{table}[!tb]
\caption{\label{tab:Nfac_M0001_T09andT11}Summary of the $N$-factor comparison between modal and non-modal growth for subcritical and supercritical cases. $Re_{\delta,0,N}$, $Re_{\delta,N}$, and $\beta_N$ indicate the initial optimal Reynolds number, the local Reynolds number at which $N_{modal}$ exceeds $ N_{opt}$, and the optimal spanwise wavenumber at $Re_{\delta,N}$, respectively. $N(Re_{\delta,N})$ is the $N$-factor at this position, with its corresponding energy amplification $G_{opt}(Re_{\delta,N})=\exp\{N(Re_{\delta,N})\}^2$. Cases T09w090 and T11w110 are identical.}
\begin{ruledtabular}
\begin{tabular}{cccccc}
  Case & $Re_{\delta,0,N}$ & $Re_{\delta,N}$ & $\beta_{N}$ & $N(Re_{\delta,N})$ & $G_{opt}(Re_{\delta,N})$   \\ \hline & \\[-0.6em]
    T09w085 & 440 & 605 & 0.45 & 3.4 & 947 \\
    T09w090 & 760 & 1040 & 0.46 & 3.9 & 2314 \\
    T09w095 & 1470 & 2010 & 0.49 & 4.6 & 9725  \\[0.3em] \hline & \\[-0.6em]
    T11w105 & 1240 & 1705 & 0.47 & 4.4 & 6966 \\
    T11w120 & 425 & 595 & 0.44 & 3.3 & 739  \\[0.3em]
\end{tabular}
\end{ruledtabular}
\end{table} 
Transient growth, present in the subcritical flow region below the critical Reynolds number, achieves a maximum $N$-factor of 4.6 up to $Re_\delta=2000$, corresponding to energy amplifications of about $10^4$, ($G=\exp(N)^2$). This occurs only when the wall is heated in the liquid-like regime or cooled in the gas-like regime towards the Widom line. Conversely, modal growth dominates the transition to turbulence once the wall is cooled in the subcritical regime or heated in the supercritical regime. Accurate transition predictions require knowledge of the initial disturbance energy and the underlying receptivity process. $N$ values for non-linear breakdown or transition to turbulence depend on various factors, such as disturbance sources \cite{Bitter1}. For instance, transient growth can be significant in noisy environments with high free-stream turbulence levels $\mathrm{Tu}$ \cite{Levin1} or discrete wall roughness elements \cite{Reshotko2}. While the disturbance structures here are similar to those in ideal-gas studies, accurate analogies are limited due to a lack of transition experiments and unknown receptivity mechanisms in boundary layers at supercritical pressure. Hence, it remains uncertain whether transient growth is critical for the cases in this study. 
\begin{figure}[!b]
\centering
\includegraphics[angle=-0,trim=0 0 0 0, clip,width=0.85\textwidth]{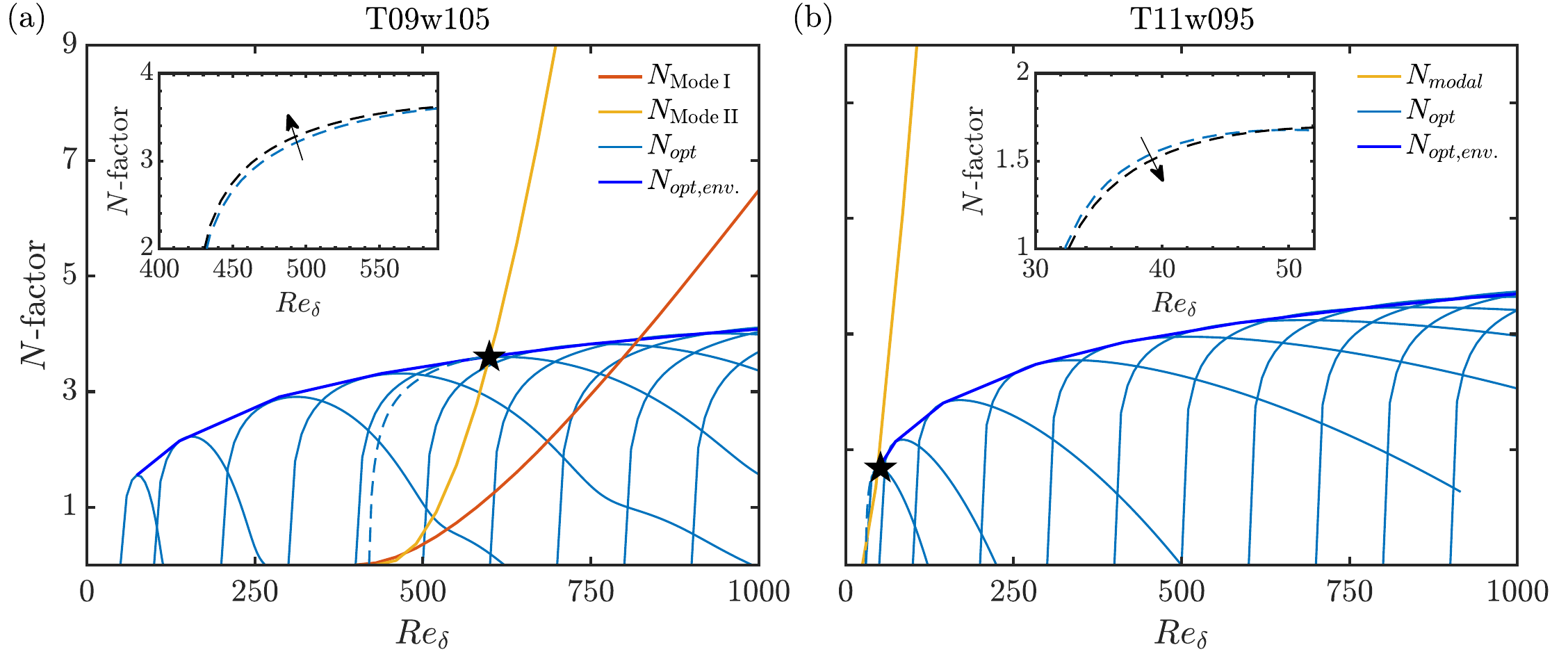}
\caption{\label{fig:23}$N$-factor calculation: $N_{modal}$ (\textcolor{mycolor2}{\rule[0.5ex]{0.3cm}{1pt}}) and (\textcolor{mycolor3}{\rule[0.5ex]{0.3cm}{1pt}}) are the $N$-factors obtained from Mode I and II, respectively, according to the modal stability analysis; $N_{opt}$ (\textcolor{mycolor1}{\rule[0.5ex]{0.3cm}{1pt}}) is the $N$-factor corresponding to transient growth, and $N_{opt,env.}$ (\textcolor{mycolor9}{\rule[0.5ex]{0.3cm}{1pt}}) is the envelope of $N_{opt}$. (a) T09w105, (b) T11w095. The optimal growth originating from the optimal initial position $x_{0,N}$ is indicated with a dashed blue line. The location $x_N$ at which $N_{modal}=N_{opt}$ is denoted with a black star (\textcolor{black}{$\filledstar$}). In the inset, the optimal growth is varied over the spanwise wavenumber (black arrow); black dashed line illustrates the maximum optimal growth at $\beta_N$.}
\end{figure}

The analysis in Tab.~\ref{tab:Nfac_M0001_T09andT11} is also applied to the transcritical cases in Fig.~\ref{fig:23}. For case T09w105, $N_{opt}$-calculations involve the modification of $\omega_{opt}$ as a function of $Re_{\delta,0}$, as illustrated in Fig.~\ref{fig:15}(a). Overall, the transcritical cases exhibit significantly stronger modal growth than the non-transcritical cases. With transcritical wall heating (case T09w105), the highly unstable Mode II, resulting from inviscid instability (see GIP in Fig.~\ref{fig:fig2}(b), despite the fuller velocity profile), is present (yellow line). Mode I (red line) is also more unstable than its counterpart in the liquid-like regime, given the same wall-to-free-stream temperature ratio $T^*_w/T^*_\infty=1.167$. For case T09w105, larger growth rates of Mode II compared to Mode I and the low non-modal energy amplification let the modal instability dominate further downstream around $Re_\delta \approx 590$. This suggests that at these thermodynamic transcritical conditions, growth occurs below the critical Reynolds number (also called subcritical growth in hydrodynamic stability theory), but with modest amplification rates. As noted for the non-transcritical cases, transient growth might be critical only if a trivial analogy is drawn with the other ideal-gas transient-growth studies.\par
When the wall temperature is cooled over the Widom line (case T11w095), an even stronger modal instability, also of inviscid nature, is found. In this transcritical wall-cooling case, only one mode is unstable regardless of frequency and Reynolds number. The GIP moves away from the wall, due the immense viscosity gradient at the wall \cite{Ren3}, while the $\bar{u}(y)$-profile becomes inflectional above the Widom line, leading to a strong inviscid instability. As shown in Fig.~\ref{fig:23}(b), the exponential growth (yellow line) surpasses the algebraic growth near the flat-plate leading edge due to this inviscid instability. Thus, subcritical disturbance growth is irrelevant here. Note that the boundary-layer assumption may be inadequate here due to leading-edge effects. Regarding the optimal spanwise wavenumber in the insets of Fig.~\ref{fig:23}, one can recognize that the optimal $N$-factor is not affected by a $\beta$-change. A summary of the $N$-factor analysis is provided in Tab.~\ref{tab:Nfac_M0001_trans}.
\begin{table}[!tb]
\caption{\label{tab:Nfac_M0001_trans}Summary of the $N$-factor comparison between modal and non-modal growth for transcritical cases. $Re_{\delta,0,N}$, $Re_{\delta,N}$, and $\beta_N$ indicate the initial optimal Reynolds number, the local Reynolds number at which $N_{modal}$ exceeds $N_{opt}$, and the optimal spanwise wavenumber at $Re_{\delta,N}$, respectively. $N(Re_{\delta,N})$ is the $N$-factor at this position, with its corresponding energy amplification $G_{opt}(Re_{\delta,N})=\exp\{N(Re_{\delta,N})\}^2$.}
\begin{ruledtabular}
\begin{tabular}{cccccc}
Case & $Re_{\delta,0,N}$ & $Re_{\delta,N}$ & $\beta_{N}$ & $N(Re_{\delta,N})$ & $G_{opt}(Re_{\delta,N})$   \\ \hline & \\[-0.6em]
    T09w105 & 420 & 590 & 0.53 & 3.5 & 1097 \\[0.3em] \hline & \\[-0.6em]
    T11w095 & 30 & 52 & 0.40 & 1.7 & 30 \\[0.3em]
\end{tabular}
\end{ruledtabular}
\end{table}

\begin{figure}[!t]
\centering
\includegraphics[angle=-0,trim=0 0 0 0, clip,width=0.85\textwidth]{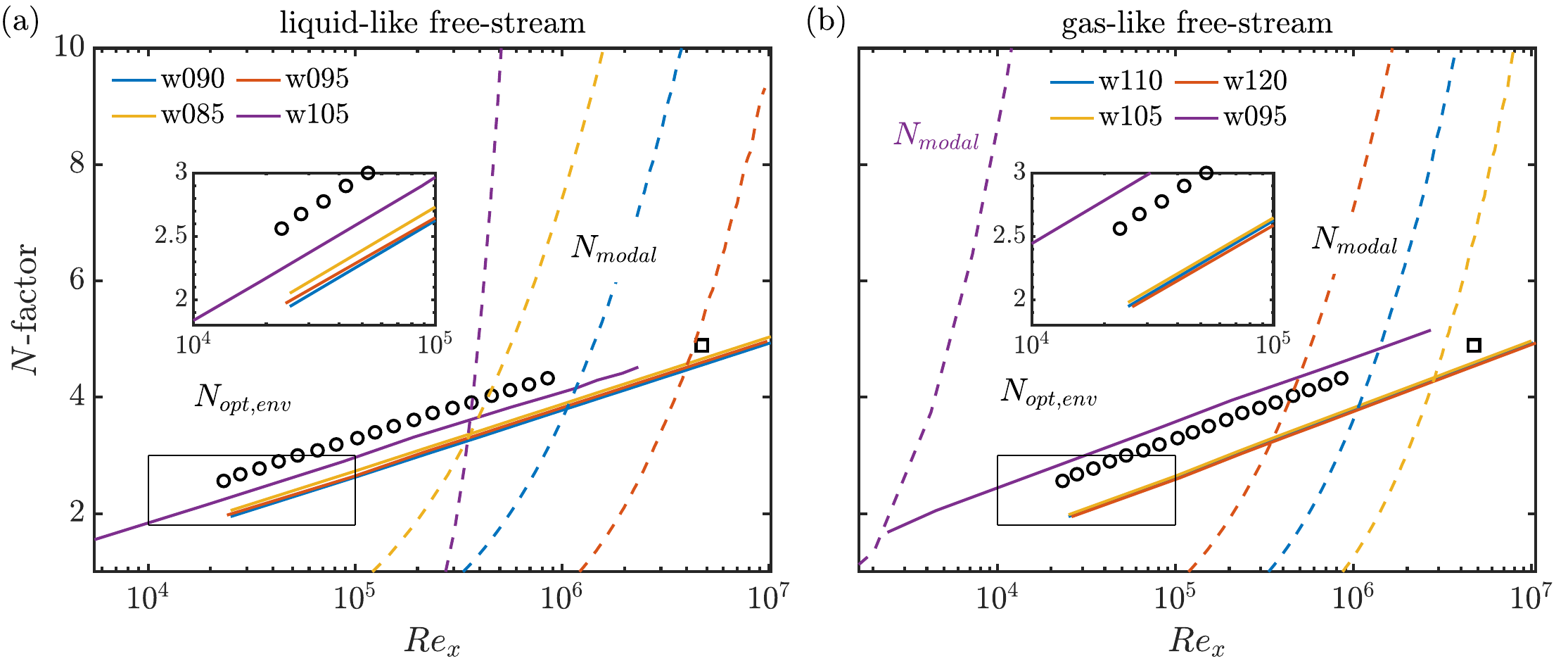}
\caption{\label{fig:24}Envelope curves of optimal non-modal and modal growth: (a) liquid-like free-stream cases with $T^*_{\infty}/T^*_{pc}=0.90$, (b) gas-like free-stream cases with $T^*_{\infty}/T^*_{pc}=1.10$. $N_{opt,env.}$: continuous line, $N_{modal}$: dashed line. For case T09w105, one common modal $N$-factor is calculated for Mode I and II. The black circle (\textcolor{black}{$\circ$}) and square (\textcolor{black}{$\square$}) symbols refer to the adverse and zero pressure gradient cases of Ref.~\cite{Levin1}, respectively.}
\end{figure}

Fig.~\ref{fig:24} presents a final comparison of all supercritical cases, with Fig.~\ref{fig:24}(a) for a liquid-like free stream (cases T09w090, T09w085, T09w095, T09w105) and Fig.~\ref{fig:24}(b) for a gas-like free stream (cases T11w110, T11w105, T11w120, T11w095). Non-modal growth is represented as straight line on a logarithmic scale. Non-transcritical cases exhibit $N_{opt,env.}$-factors that are unaffected by minor wall-to-free-stream temperature variation near the Widom line. For the transcritical cases, cooling over the Widom line in Fig.~\ref{fig:24}(b) significantly increases $N_{opt,env.}$, with $N_{opt,env.,\mathrm{T11w095}}/N_{opt,env.,\mathrm{T11w110}}\approx 1.34$. Nevertheless, this rise is even more pronounced for modal growth (see dashed lines). When examining the $N$-factors in Fig.~\ref{fig:24} for a Falkner-Skan boundary layer with adverse (APG, Hartree parameter $\beta_H=-0.1$) and zero pressure (ZPG) gradient as in Ref.~\cite{Levin1}, three main conclusions arise. First, an increase in the APG has a similar effect on transient growth as heating or cooling over the Widom line in a ZPG boundary layer at supercritical pressure, with transcritical wall cooling revealing a larger amplification. In case T11w095, exponential growth with a strong inviscid instability dominates from the leading edge. Secondly, Ref.~\cite{Levin1} predicts that with increased APG, the required level of free-stream turbulence for bypass transition decreases. In fact, when $\mathrm{Tu}>3$, $Re_{x,tr}$ was seen to shift to values below $10^5$. If confirmed for a non-ideal boundary-layer flow at supercritical pressure, transient growth could be critical for transition in case T09w095, as non-modal growth in Fig.~\ref{fig:24}(a) dominates over modal growth below $Re_{x} \approx 3 \times 10^5$. Lastly, comparing the ZPG case from Ref.~\cite{Levin1} with case T09w090 reveals a difference of $N\approx 0.4$--0.5, demonstrating the robustness of the locally parallel assumption used in this study for non-modal optimization against an adjoint-based optimization algorithm with parabolized stability equations.

\section{\label{sec:5}Conclusion}

Transient growth of diabatic flat-plate boundary layers is examined at supercritical pressure. Supercritical carbon dioxide ($\text{CO}_2$) at a constant pressure of $\SI{80}{bar}$ (with a critical pressure of $\SI{73.9}{bar}$) is selected as a representative non-ideal fluid in the proximity of the Widom line. Optimal energy amplifications and disturbances are calculated in the spatial framework by means of a singular value decomposition of the locally-parallel, linearized compressible Navier-Stokes equations. To account for non-ideality and strong thermo-physical property variations around the Widom line, a novel non-ideal energy norm is derived in Sec.~\ref{sec:2} such that pressure work is eliminated. The new norm is equal to Chu's \cite{Chu1} and Mack's \cite{Hanifi1} norms under the ideal-gas assumption.

This study explores the subcritical, supercritical, and transcritical regimes (relative to the pseudo-boiling temperature of $\text{CO}_2$, $T^*_{pc}=\SI{307.7}{K}$), determined by prescribing free-stream and wall temperatures at a fixed Mach number of $10^{-3}$ (see Sec.~\ref{sec:3}). In the subcritical (liquid-like: $T^*_\infty/T^*_{pc}=0.90$ with $T^*_w<T^*_{pc}$) and supercritical (gas-like: $T^*_\infty/T^*_{pc}=1.10$ with $T^*_w>T^*_{pc}$) regimes, wall cooling enhances transient growth. For the subcritical cases, the optimal streamwise distance $x_{opt}$ decreases, while the optimal spanwise wavenumber $\beta_{opt}$ increases as the wall temperature approaches the Widom line. Conversely, in the supercritical regime, the opposite trend is observed. In these two weakly non-ideal regimes, the maximal energy amplification and its distance scale best with the compressible momentum thickness, in agreement with the Falkner-Skan boundary layer of Ref.~\cite{Corbett1} under the ideal-gas assumption. The optimal energy amplification is detected at finite spanwise wavenumbers, yet the optimal frequency is zero, indicating that the optimal disturbances correspond to counter-rotating vortices, which evolve into streamwise-independent streaks. This corresponds to the well-known lift-up mechanism, in agreement with ideal-gas results. Moreover, scaling laws, e.g.,~$G \propto Re_\delta^2$ and $x_{max} \propto Re_\delta$, consistently hold.

When the wall temperature crosses the Widom line (transcritical wall-heating: $T^*_\infty/T^*_{pc}=0.90$ with $T^*_w>T^*_{pc}$; transcritical wall-cooling: $T^*_\infty/T^*_{pc}=1.10$ with $T^*_w<T^*_{pc}$), optimal structures are no longer exclusively streamwise independent. In the transcritical wall-heating regime, a finite optimal frequency suggests the involvement of the Orr mechanism. In fact, analysis of the Reynolds stress reveals that optimal disturbance structures are initially tilted against the mean shear before being re-oriented to it. A strong contribution of the Orr mechanism is observed for sub-optimal transient growth, where larger local energy-amplification peaks than the global optimal one can be achieved for $x\ll x_{opt}$. This leads to highly-oblique sub-optimal disturbance structures, which do not originate from a streamwise vortex. The Orr mechanism exists also for perturbations with a negative spanwise wavenumber. Thus, the resulting superposition forms a checkerboard pattern, analogous to the oblique-transition mechanism in Ref.~\cite{Berlin1}. Moreover, streamwise-modulated streaks in the transcritical wall-heating regime exhibit significant streamwise velocity disturbances and strong thermal components ($\hat{\rho}$ and $\hat{T}$) around the Widom line. While similar large density streaks are also observed in the transcritical wall-cooling regime, they are not oblique. Despite comparable $|T^*_w-T^*_\infty|$,  transcritical wall-cooling yields higher optimal energy amplification rates $G_{opt}/Re^2_\delta$. This increase can be explained by the vortex-tilting mechanism, which highlights the alignment of the output wall-normal vorticity with the highest wall-normal displacement above the Widom line. Conversely, a misalignment between the two in the transcritical wall-heating regime results in a smaller $G$, albeit with a secondary peak of the optimal streamwise velocity at the Widom line.

When considering the effect of initial Reynolds number, scaling laws hold for both optimal streamwise-independent and -modulated disturbances. Particularly, in the case of optimal oblique streaks in transcritical wall-heating, we observe a dependency of the optimal frequency $\omega_{opt}$ on $Re^{-1}_\delta$, while the optimal energy $G_{opt}$ scales with $Re^2_\delta$. This implies that as $Re_\delta \rightarrow \infty$, the Orr mechanism is not active, leading to the recovery of streamwise-independent structures. Consequently, the streaks' propagation angle $\Psi$ becomes proportional to $Re^{-1}_\delta$. Investigating the underlying eigenspectrum in the transcritical wall-heating regime reveals the influence of the non-orthogonal eigenfunctions on optimal growth. Specifically, the stable transcritical Mode II is found to actively participate in the interplay between lift-up and Orr mechanism. If this mode is intentionally excluded from the optimization procedure, both mechanisms yield a significantly reduced sub-optimal growth only, resulting in modified optimal transient growth solely associated with streamwise-independent disturbances, growing via the lift-up mechanism.

In relation to the effect of wall temperature on transient growth, we note that the momentum-thickness scaling is ineffective in the transcritical regime, in contrast to the non-transcritical regimes. In the transcritical wall-heating regime, increasing the wall temperature initially shifts optimal frequencies to higher values. However, once the Widom line moves away from the near-wall layer, optimal energy growth is found at $\omega=0$. Stronger wall heating leads to a rise in the density perturbation amplitudes around the Widom line due to greater mean-flow density gradients. In contrast, in the transcritical wall-cooling regime, $G_{opt}$ remains at $\omega_{opt}=0$, while $\beta_{opt}$ slightly increases. Notably, a 5\% increase only in wall temperature can result in a 30\% to 70\% gain in $G_{opt}$. Furthermore, similar optimal disturbance structures are detected when, in the transcritical regime, the optimal perturbations are streamwise independent.

A comparison between modal and non-modal growth across all regimes is presented in Sec.~\ref{sec:44}. Initially, the modal instability of oblique perturbations is studied for both non-transcritical and transcritical cases. At the low Mach number of $10^{-3}$ employed in this study, 2-D modes exhibit the highest amplifications, both locally (across most of the frequency spectrum) and integrally. This suggests that, as shown in the oblique modal stability analysis at finite Mach numbers in Ref.~\cite{Ren2}, decreasing the Mach number (compressibility effects) shifts the largest possible local and integral amplification from a 3-D to a 2-D mode, regardless of the considered thermodynamic regime at supercritical pressure. Notably, no critical $N$-factors of transition have been experimentally measured for flows at supercritical pressure so far, hence transition-prediction analogies can only be drawn in relation to ideal-gas transient-growth results. In the non-transcritical regimes, transient growth is marginally affected by the non-ideal thermodynamic regime. Within their best-case-scenarios (subcritical wall-heating and supercritical wall-cooling), modal growth prevails over transient growth at amplification levels around $N \approx 4$. On the other hand, in the transcritical regime, characterized by a highly inflectional base flow (inviscid instability), modal instability dominates, especially for transcritical wall-cooling, where no subcritical transition below the critical Reynolds number is likely to occur. In this scenario, the increase in the non-modal $N$-factor, i.e.,~$N_{opt}$, is significant compared to the non-transcritical wall-cooling cases, similar to the effect of an adverse pressure gradient in ideal-gas transient-growth calculations. A similar analogy applies to transcritical wall-heating, albeit with lower transient-growth amplifications, reaching up to $N \approx 3.5$ before a strong modal amplification emerges. Nevertheless, a critical Reynolds number of approximately $3\times 10^5$ could be sufficiently large to favor transition via transient growth, particularly when the level of free-stream turbulence is high.

Discussions on the influence of an increasing reduced pressure on the N-factor comparison indicate that, for transcritical wall-heating, transient growth is likely to be the prevailing route to turbulence. Conversely, for transcritical wall-cooling, modal growth consistently dominates.

\begin{acknowledgments}
The European Research Council (grant no.~ERC-2019-CoG-864660, Critical) has supported the investigations presented in this paper. The authors acknowledge the use of computational resources of DelftBlue supercomputer, provided by Delft High Performance Computing Centre (\url{https://www.tudelft.nl/dhpc}).
\end{acknowledgments}

\appendix

\section{Supercritical $\text{CO}_2$}
\begin{figure}[!b]
\centering
\includegraphics[angle=-0,trim=0 0 0 0, clip,width=0.4\textwidth]{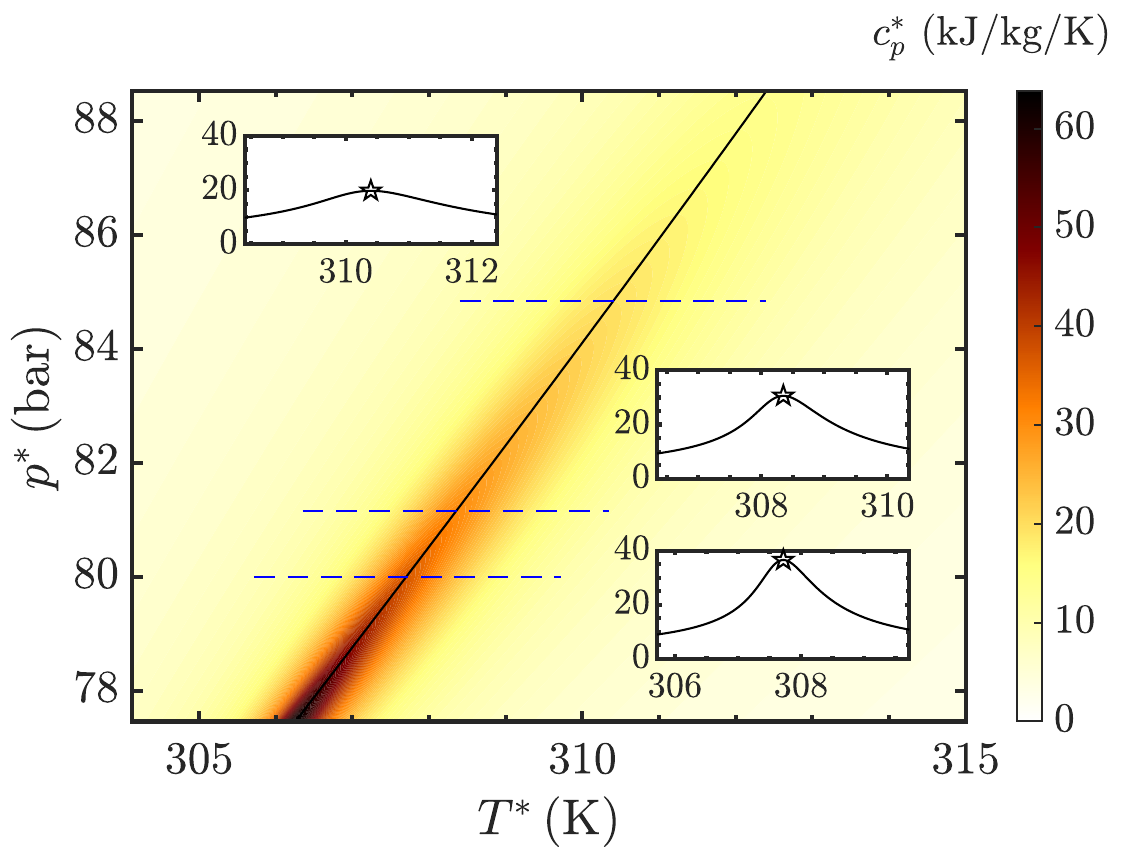}
\caption{\label{fig:a1}Evolution of the pseudo-boiling temperature $T^*_{pc}$ (\textcolor{mycolor10}{\rule[0.5ex]{0.3cm}{1pt}}) over $p^*$ with contours of $c^*_p$. Three isobars (\textcolor{mycolor9}{\rule[0.5ex]{0.2cm}{1pt}} \textcolor{mycolor9}{\rule[0.5ex]{0.2cm}{1pt}}) are drawn: (a) $T^*_{pc}(p^*_c=\SI{80}{bar})=\SI{307.7}{K}$, (b) $T^*_{pc}(p^*_c=\SI{81.15}{bar})=\SI{308.3}{K}$, and (c) $T^*_{pc}(p^*_c=\SI{84.8}{bar})=\SI{310.4}{K}$.  }
\end{figure}

The material dependent parameters of $\text{CO}_2$ are provided in Table~\ref{tab:tableCO2}.
\begin{table}[!b]
\caption{\label{tab:tableCO2}Thermodynamic properties of $\text{CO}_2$: gas constant ($R^*_g$), critical pressure ($p^*_c$), and critical temperature ($T^*_c$).}
\begin{ruledtabular}
\begin{tabular}{ccc}
$R^*_g \ (\SI{}{J/kg/K})$ & $p^*_c \ (\SI{}{bar})$ & $T^*_c \ (\SI{}{K})$   \\ \hline & \\[-0.6em]
 $188.9$ & $73.9$ & $304.1$ \\[0.3em]
\end{tabular}
\end{ruledtabular}
\end{table}
The evolution of the pseudo-boiling temperature as a function of pressure and temperature is plotted in Fig.~\ref{fig:a1} along with contours of $c^*_p$.

\section{\label{sec:app1b}Modal stability: base-flow matrices}
The base-flow matrices of Ref.~\cite{Ren1} are adjusted for $\bar{\chi}$ as a function of $\bar{p}$ and $\bar{T}$. Only the modified terms in the stability equations (see Eq.~\eqref{eq:LSTmatrices}) are reported hereafter. For simplicity, the derivative of a thermodynamic quantity with respect to $\bar{p}$ at constant $\bar{T}$, and vice versa, is denoted as $\partial/\partial \bar{p}$ instead of $\partial/\partial \bar{p}|_{\bar{T}}$ and as $\partial/\partial \bar{T}$ instead of $\partial/\partial \bar{T}|_{\bar{p}}$, respectively. Furthermore, any occurrence of $\partial/\partial \bar{\rho}$ in the matrices of Ref.~\cite{Ren1} changes to $\partial/\partial \bar{p}$ and will not be reported hereafter. The modified elements are:

\begin{equation}
  \left.
      \begin{array}{cc}
\mathbf{L_{t}}(1,1)=\dfrac{\partial \bar{\rho}}{\partial \bar{p}}, \quad \mathbf{L_{t}}(1,5)=\dfrac{\partial \bar{T}}{\partial \bar{p}},
	    \end{array}
	  \right\}
\end{equation}

\begin{equation}
  \left.
      \begin{array}{cc}
\mathbf{L_{x}}(1,1)=\bar{u}\dfrac{\partial \bar{\rho}}{\partial \bar{p}}, \quad \mathbf{L_{x}}(1,5)=\bar{u}\dfrac{\partial \bar{T}}{\partial \bar{p}}, \\[2ex]
\mathbf{L_{x}}(2,1)=1, \quad \mathbf{L_{x}}(2,5)=0, \\[2ex]
\mathbf{L_{x}}(5,2)=Ec_\infty \bar{p}, \quad \mathbf{L_{x}}(5,3)=-\dfrac{2 Ec_\infty \bar{\mu}}{Re_\delta} \dfrac{\partial \bar{u}}{\partial y}, \\
	    \end{array}
	  \right\}
\end{equation}

\begin{equation}
  \left.
      \begin{array}{cc}
\mathbf{L_{y}}(3,1)=1, \quad \mathbf{L_{y}}(3,5)=0, \\[2ex]
\mathbf{L_{y}}(5,1)=-\dfrac{1}{Re_\delta Pr_\infty} \dfrac{\partial \bar{\kappa}}{\partial \bar{p}} \dfrac{\partial \bar{T}}{\partial y}, \quad \mathbf{L_{y}}(5,2)=-\dfrac{2 Ec_\infty  \bar{\mu}}{Re_\delta} \dfrac{\partial \bar{u}}{\partial y}, \\[2ex]
\quad \mathbf{L_{y}}(5,3)=Ec_\infty  \bar{p}, \quad \mathbf{L_{y}}(5,5)=-\dfrac{1}{Re_\delta Pr_\infty} \left( \dfrac{\partial \bar{\kappa}}{\partial y} + \dfrac{\partial \bar{\kappa}}{\partial \bar{T}} \dfrac{\partial \bar{T}}{\partial y} \right), \\
	    \end{array}
	  \right\}
\end{equation}

\begin{equation}
  \left.
      \begin{array}{cc}
\mathbf{L_{z}}(4,1)=1, \quad \mathbf{L_{z}}(4,5)=0, \\[2ex]
\mathbf{L_{z}}(5,4)=Ec_\infty \bar{p}, \\
	    \end{array}
	  \right\}
\end{equation}

\begin{equation}
  \left.
      \begin{array}{cc}
\mathbf{L_{q'}}(2,1)=-\dfrac{1}{Re_\delta} \left( \dfrac{\partial \bar{\mu}}{\partial \bar{p}} \dfrac{\partial^2 \bar{u}}{\partial y^2} +  \dfrac{\partial^2 \bar{\mu}}{\partial \bar{p} \, \partial \bar{T}} \dfrac{\partial \bar{T}}{\partial y} \dfrac{\partial \bar{u}}{\partial y} \right), \quad \mathbf{L_{q'}}(2,5)=-\dfrac{1}{Re_\delta} \left( \dfrac{\partial \bar{\mu}}{\partial \bar{T}} \dfrac{\partial^2 \bar{u}}{\partial y^2} +  \dfrac{\partial^2 \bar{\mu}}{\partial \bar{T}^2} \dfrac{\partial \bar{T}}{\partial y} \dfrac{\partial \bar{u}}{\partial y} \right), \\[3ex]
\mathbf{L_{q'}}(3,1)=0, \quad \mathbf{L_{q'}}(3,5)=0, \\[2ex]
\mathbf{L_{q'}}(5,1)=-\dfrac{Ec_\infty}{Re_\delta} \left[ \dfrac{\partial \bar{\mu}}{\partial \bar{p}} \left(\dfrac{\partial \bar{u}}{\partial y}\right)^2 + \dfrac{1}{Ec_\infty Pr_\infty} \dfrac{\partial^2 \bar{\kappa}}{\partial \bar{p} \, \partial \bar{T}} \left(\dfrac{\partial \bar{T}}{\partial y}\right)^2 + \dfrac{1}{Ec_\infty Pr_\infty} \dfrac{\partial \bar{\kappa}}{\partial \bar{p}} \dfrac{\partial^2 \bar{T}}{\partial y^2} \right], \\[3ex]
\mathbf{L_{q'}}(5,5)=-\dfrac{Ec_\infty}{Re_\delta} \left[ \dfrac{\partial \bar{\mu}}{\partial \bar{T}} \left(\dfrac{\partial \bar{u}}{\partial y}\right)^2 + \dfrac{1}{Ec_\infty Pr_\infty} \dfrac{\partial^2 \bar{\kappa}}{\partial \bar{T}^2} \left(\dfrac{\partial \bar{T}}{\partial y}\right)^2 + \dfrac{1}{Ec_\infty Pr_\infty} \dfrac{\partial \bar{\kappa}}{\partial \bar{T}} \dfrac{\partial^2 \bar{T}}{\partial y^2} \right], \\[2ex]
	    \end{array}
	  \right\}
\end{equation}

\begin{equation}
  \left.
      \begin{array}{cc}
        \mathbf{V_{xx}}(5,5)=\mathbf{V_{yy}}(5,5)=\mathbf{V_{zz}}(5,5)=-\dfrac{\bar{\kappa}}{Re_\delta Pr_\infty}. \\[2ex]
      	    \end{array}
	  \right\}
\end{equation}

\section{\label{sec:app1}Validation: transient growth}
\begin{figure}[!t]
\centering
\includegraphics[angle=-0,trim=0 0 0 0, clip,width=1.0\textwidth]{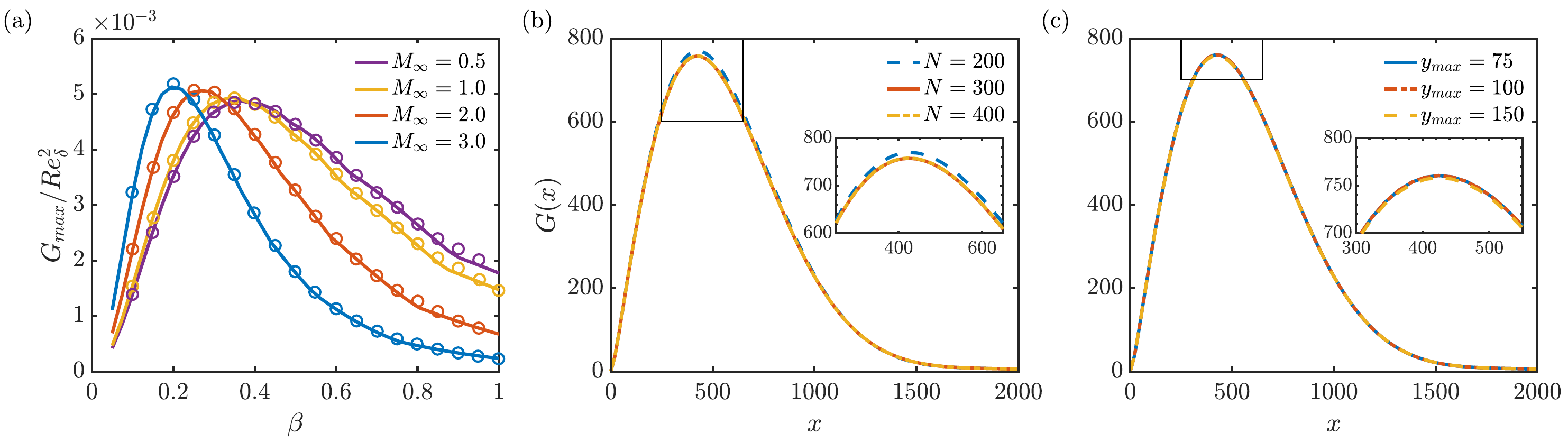}
\caption{\label{fig:a2}(a) ideal gas, comparison of maximum energy amplification: $M_\infty=0.5$ (\textcolor{mycolor4}{\rule[0.5ex]{0.4cm}{1pt}}), $M_\infty=1.0$ (\textcolor{mycolor3}{\rule[0.5ex]{0.4cm}{1pt}}), $M_\infty=2.0$ (\textcolor{mycolor2}{\rule[0.5ex]{0.4cm}{1pt}}), $M_\infty=3.0$ (\textcolor{mycolor1}{\rule[0.5ex]{0.4cm}{1pt}}), Ref.~\cite{Tumin1} marked by a colored circle (\textcolor{black}{$\circ$});~(b) non-ideal gas, influence of mesh size $N$ on the energy amplification $G(x)$: $N=200$ (\textcolor{mycolor1}{\rule[0.5ex]{0.2cm}{1pt}} \textcolor{mycolor1}{\rule[0.5ex]{0.2cm}{1pt}}), $N=300$ (\textcolor{mycolor2}{\rule[0.5ex]{0.4cm}{1pt}}), $N=400$ (\textcolor{mycolor3}{\rule[0.5ex]{0.25cm}{1pt}} \textcolor{mycolor3}{\rule[0.5ex]{0.05cm}{1pt}});~(c) non-ideal gas, influence of mesh height $y_{max}$ on the energy amplification $G(x)$: $y_{max}=75$ (\textcolor{mycolor1}{\rule[0.5ex]{0.4cm}{1pt}}), $y_{max}=100$ (\textcolor{mycolor2}{\rule[0.5ex]{0.2cm}{1pt}} \textcolor{mycolor2}{\rule[0.5ex]{0.05cm}{1pt}}), $y_{max}=150$ (\textcolor{mycolor3}{\rule[0.5ex]{0.2cm}{1pt}} \textcolor{mycolor3}{\rule[0.5ex]{0.2cm}{1pt}}).  }
\end{figure}

The transient-growth analysis presented in this work is validated by reproducing the spatial results of Ref.~\cite{Tumin1} for an ideal gas (adiabatic wall, $T^*_{tot}=\SI{333}{K}$, $Pr_\infty=0.70$, $c^*_p/c^*_v=1.40$, $\omega=0$, $Re_\delta=300$). A good agreement is observed in Fig.~\ref{fig:a2}(a) at different Mach numbers. To obtain a grid-independent solution when using a non-ideal gas, the influence of mesh size $N$ and mesh height $y_{max}$ is investigated. A transcritical wall-heating reference case at constant supercritical pressure is chosen: $p^*_\infty=\SI{80}{bar}$, $T^*_\infty/T^*_{pc}=0.90$, isothermal wall with $T^*_w/T^*_{pc}=1.05$, $M_\infty=10^{-3}$, $Pr_\infty=2.11$, $\omega=0.013$, $\beta=0.45$, and $Re_\delta=300$. The results, shown in Fig.~\ref{fig:a2}(b) for $N$ and Fig.~\ref{fig:a2}(c) for $y_{max}$, respectively, reveal that a value of $N=300$ and $y_{max}=75$ is necessary for mesh independence. Hence, this numerical setup is retained throughout the non-modal calculations in this study.

\section{\label{sec:app2}Temporal vs.~spatial transient growth}
\begin{figure}[t!]
\centering
\includegraphics[angle=-0,trim=0 0 0 0, clip,width=0.85\textwidth]{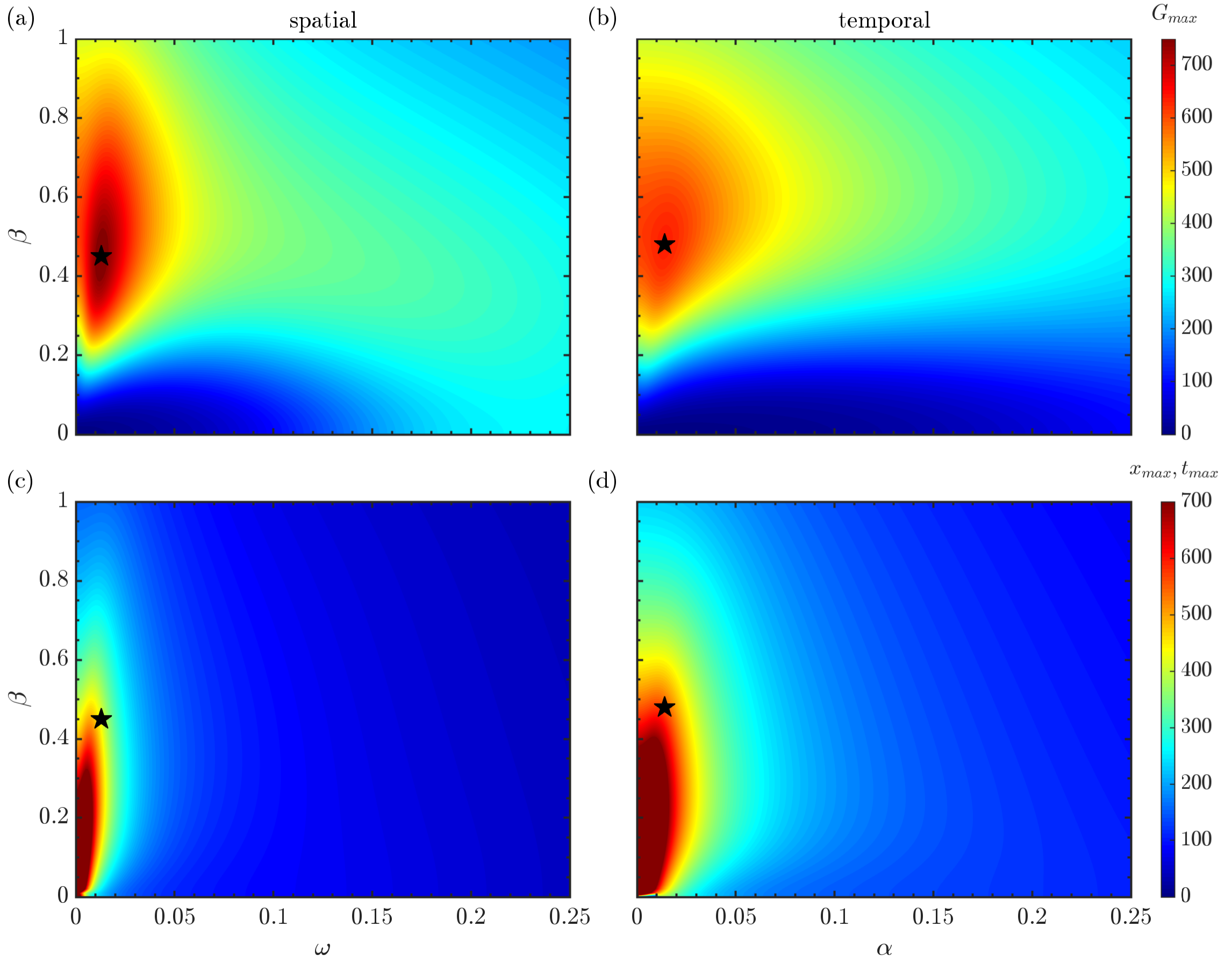}
\caption{\label{fig:a3}Transcritical case T09w105: (a) contour plots of $G_{max}(\omega,\beta)$, (b) contour plots of $G_{max}(\alpha,\beta)$, (c) contour plots of $x_{max}(\omega,\beta)$, and (d) contour plots of $t_{max}(\alpha,\beta)$. $G_{opt}$ is denoted with a black star (\textcolor{black}{$\filledstar$}) symbol.}
\end{figure}

So far, spatial optimal growth has been analyzed for its simpler interpretation in experimental facilities \cite{Tumin1}. However, it is noteworthy to investigate whether similar non-modal behavior is obtained regardless of the considered optimization procedure. For the temporal analysis, initial disturbances are optimized at $t=0$ in order to achieve the maximum temporal energy growth at a later time $t_{max}$ over $\beta$ and streamwise wavenumber $\alpha$ \cite{Hanifi1}. The case selected for the comparison is the transcritical case T09w105, where oblique disturbances are optimal. Figs.~\ref{fig:a3}(a,b) and \ref{fig:a3}(c,d) display the maximum energy amplification $G_{max}$ and the maximum location $x_{max}$ and time $t_{max}$, respectively, for both the spatial and temporal analysis of case T09w105.
The qualitative behavior of $G_{max}$ is nearly identical for low values of $\omega$ and $\alpha$ for both Fig.~\ref{fig:a3}(a) and \ref{fig:a3}(b). At large $\omega$ and $\alpha$, the spatial framework reveals a larger sub-optimal growth near the $\omega$-axis, which is negligible in the temporal framework. In the latter, the location of $G_{opt}$, defined as $\max\{G_{max}(\alpha,\beta)\}$, is found at $\beta_{opt}=0.48$ and $\alpha_{opt}=0.014$. This confirms the observation of Ref.~\cite{Bitter1}, in which most of the modes contributing to the optimal growth belong to the vorticity and entropy branches of the eigenspectrum. In fact, the phase speed $c_{ph,opt}$, obtained as ratio between $\omega_{opt}$ and $\alpha_{opt}$, is nearly 1. With respect to the value of $G_{opt}$ for both frameworks, a difference of about $15\%$ is found. In conclusion, the same physical mechanisms can be observed for both temporal and spatial analysis.

\section{\label{sec:app3}The choice of energy norm}
\begin{figure}[!tb]
\centering
\includegraphics[angle=-0,trim=0 0 0 0, clip,width=0.85\textwidth]{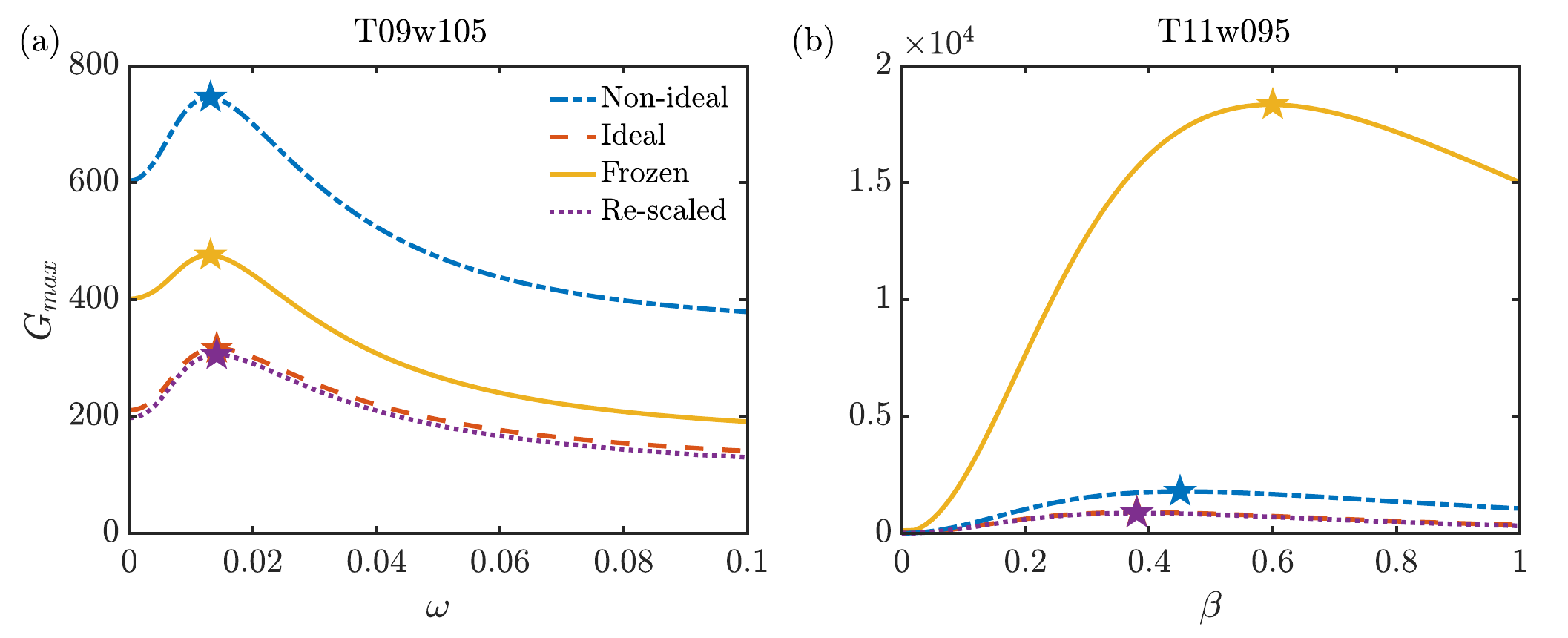}
\caption{\label{fig:a4}Effect of the energy norm on the energy amplification: non-ideal energy norm, ideal-gas energy norm (2), frozen internal energy norm (1), re-scaled 
non-ideal energy norm by an kinetic energy norm (3). (a) T09w105, as a function of $\omega$ for $\beta=\beta_{opt}$; (b) T11w095, as a function of $\beta$ for $\omega=0$. The maximum energy amplification is indicated with a colored star (\textcolor{black}{$\filledstar$}) symbol.  }
\end{figure}

All non-modal optimizations presented in Sec.~\ref{sec:4} are performed with the new energy norm introduced in Sec.~\ref{subsec:2c3}. This norm considers both the kinetic and internal energy of a perturbation in a non-ideal gas flow. One can now question whether the energy amplification is dependent on the choice of the norm. To answer this question and analyze the sensitivity of the results in Sec.~\ref{sec:4}, the energy norm in Eq.~\eqref{eq:realE} is modified. When choosing only the kinetic energy, i.e.,~$\mathbf{M}=\operatorname{diag} \left(0, \bar{\rho}, \bar{\rho}, \bar{\rho},  0 \right)$, an infinite, and thus unphysical, energy amplification is achieved. Instead, the internal energy content is intentionally altered by considering: (1) frozen internal energy, i.e., $\mathbf{M}=\operatorname{diag} \left(1, \bar{\rho}, \bar{\rho}, \bar{\rho},  1 \right)$, (2) internal energy according to ideal gas, e.g.,~$\mathbf{M}=\operatorname{diag} \left( R_g \bar{T}/\bar{\rho}, \bar{\rho}, \bar{\rho}, \bar{\rho},  \bar{\rho} \, c_v/ (Ec_\infty \bar{T}) \right)$. Another alternative, hereafter (3), is to take the non-ideal optimal perturbations obtained in Sec.~\ref{subsec:2c4} and re-scale them by the kinetic energy norm. Thus, following Eq.~\eqref{eq:trans_opt}, the new optimal amplification for (3) is
\begin{equation}
G_{(3)}=\max \dfrac{\| \mathbf{F}_{(3)} \bm{\Lambda}\boldsymbol{\kappa}(0) \|^2_2 }{\|\mathbf{F} \boldsymbol{\kappa}(0) \|^2_2}, \quad \boldsymbol{\kappa}(0)=\mathbf{F}^{-1} \mathbf{r},
	\end{equation} 
where $\mathbf{F}$, $\bm{\Lambda}$, $\mathbf{F}^{-1}$, $\mathbf{r}$ (right singular eigenvector) are obtained with $\mathbf{M}$ as in Eq.~\eqref{eq:energymatrix}, whereas $\mathbf{F}_{(3)}$ is the Cholesky decomposition of $\mathbf{A}_{(3)}=\mathbf{F}_{(3)}^H\mathbf{F}_{(3)}$ (with $\mathbf{A}_{(3),kl} = \int_{0}^{\infty} \hat{\mathbf{q}}^H_k \mathbf{M}_{(3)} \hat{\mathbf{q}}_l \ dy$) as a function of the kinetic energy norm matrix $\mathbf{M}_{(3)}=\operatorname{diag} \left(0, \bar{\rho}, \bar{\rho}, \bar{\rho},  0 \right)$. Energy norms based on (1) and (2) were previously used in Ref.~\cite{Ren1}. In order to account for the largest non-ideal effects on the norm, the two transcritical cases of Tab.~\ref{tab:tableBF}, with their optimal growth reported in Tab.~\ref{tab:SD_M0001_T0.9andT1.1}, are examined. In Fig.~\ref{fig:a4}(a), the maximum energy amplification $G_{max}(\omega)$ at constant $\beta=\beta_{opt}$, i.e., $\beta_{opt}=0.45$, is shown for case T09w105. The optimal energy amplification is found, independently of the energy norm adopted, around $\omega=0.013$ (marked by a colored star ($\filledstar$) symbol). The non-ideal energy norm yields the largest transient growth (blue dash-dotted line). The shape of optimal perturbations displayed in Fig.~\ref{fig:fig6} is not affected by the energy-norm modifications (not presented here for the sake of brevity). In Fig.~\ref{fig:a4}(b), the maximum energy amplification $G_{max}(\beta)$ at constant $\omega=\omega_{opt}$, i.e., $\omega_{opt}=0$, is presented for case T11w095. Here, we notice a greater dependence on the energy norm, especially when the internal energy is kept constant (case (1), yellow continuous line). In this case, there is an actual increase in the internal energy compared to the other norms. Moreover, a shift of the maximum energy amplification is notable. Similar to the transcritical T09w105 case, the shape of the optimal perturbations remains intact except for the density perturbations, where a smaller $\hat{\rho}$-amplitude in case (1) is witnessed. Overall, energy amplifications with the re-scaled energy norm (case (3), purple dotted line) show an analogous transient growth as the ideal-gas norm cases (case (2), red dashed line).

\section{\label{sec:app5}Inviscid vorticity perturbation equation}
The equation for the wall-normal vorticity perturbation $\omega'_y$ is obtained hereafter. The wall-normal vorticity is split as $\omega_y=\bar{\omega}_y+\omega'_y$ and inserted into Eq.~\eqref{eq:vort_general}. Subsequently, the following assumptions, in agreement with Sec.~\ref{subsec:2c}, are drawn: (1) 2-D base flow and locally parallel flow, i.e.,~$\bar{\omega}_y=\bar{v}=\bar{w}=0, \partial(\bar{\cdot})/\partial x=0, \partial(\bar{\cdot})/\partial z=0$, (2) base-flow wall-normal pressure gradient $\partial \bar{p}/\partial y$ is 0, (3) linearization, i.e.,~$q^2_i=q_i q_j=0$. After subtraction of the base flow, the baroclinic term in Eq.~\eqref{eq:vort_general} becomes:
\begin{equation}
     \dfrac{1}{\rho^2} \left( \dfrac{\partial \rho}{\partial z}\dfrac{\partial p}{\partial x} - \dfrac{\partial \rho}{\partial x}\dfrac{\partial p}{\partial z} \right) \quad  \Longrightarrow \quad  \dfrac{1}{\bar{\rho}^2} \left( \dfrac{\partial \rho'}{\partial z}\dfrac{\partial \bar{p}}{\partial x} + \dfrac{\partial \bar{\rho}}{\partial z}\dfrac{\partial p'}{\partial x} - \dfrac{\partial \bar{\rho}}{\partial x}\dfrac{\partial p'}{\partial z} - \dfrac{\partial \rho'}{\partial x}\dfrac{\partial \bar{p}}{\partial z} \right)=0.
\end{equation}
Thus, the final expression of the wall-normal vorticity perturbation equation is
\begin{equation}
    	\dfrac{\partial \omega'_y}{\partial t } + \bar{u} \dfrac{\partial \omega'_y}{\partial x} =  - \dfrac{\partial v'}{\partial z} \dfrac{\partial \bar{u}}{\partial y},
\end{equation}
with absent baroclinic influence.

\section{\label{sec:app4}Modal stability analysis: oblique perturbations}
\begin{figure}[!t]
\centering
\includegraphics[angle=-0,trim=0 0 0 0, clip,width=0.7\textwidth]{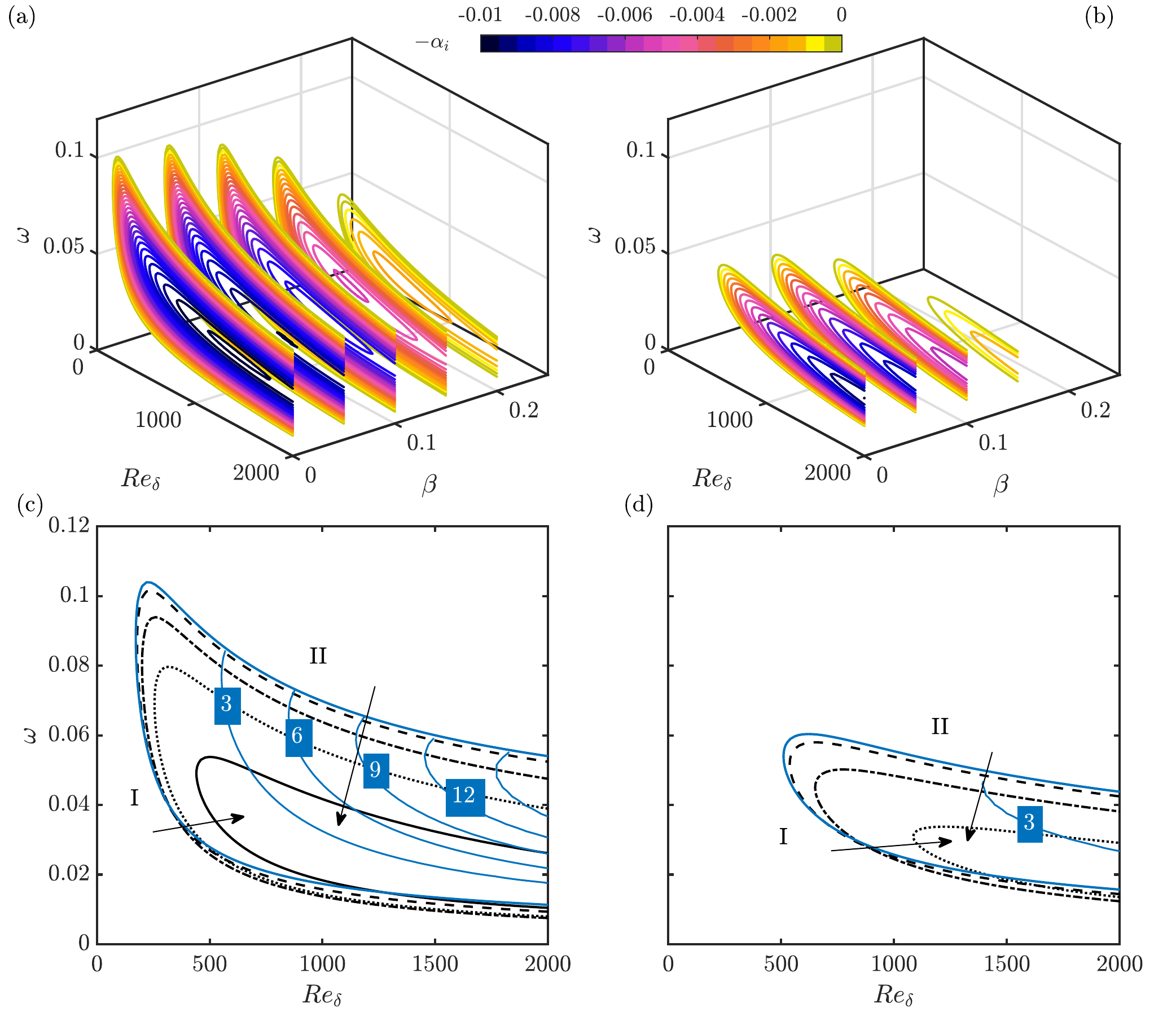}
\caption{\label{fig:a5}Slices of $-\alpha_i$-contours in the $Re_\delta$--$\omega$--$\beta$ (a,b) and neutral stability curves in the $Re_\delta$--$\omega$ (c,d) space: (a,c) T09w085, (b,d) T11w105. Arrows in (c,d) indicate a $\beta$-increase ($\Delta \beta=0.05$) of branch I and II (from solid blue to solid black line). Maximum $N$-factors are indicated in blue and are located always at $\beta=\beta_{max}=0$.}
\end{figure}

\begin{figure}[!tb]
\centering
\includegraphics[angle=-0,trim=0 0 0 0, clip,width=0.7\textwidth]{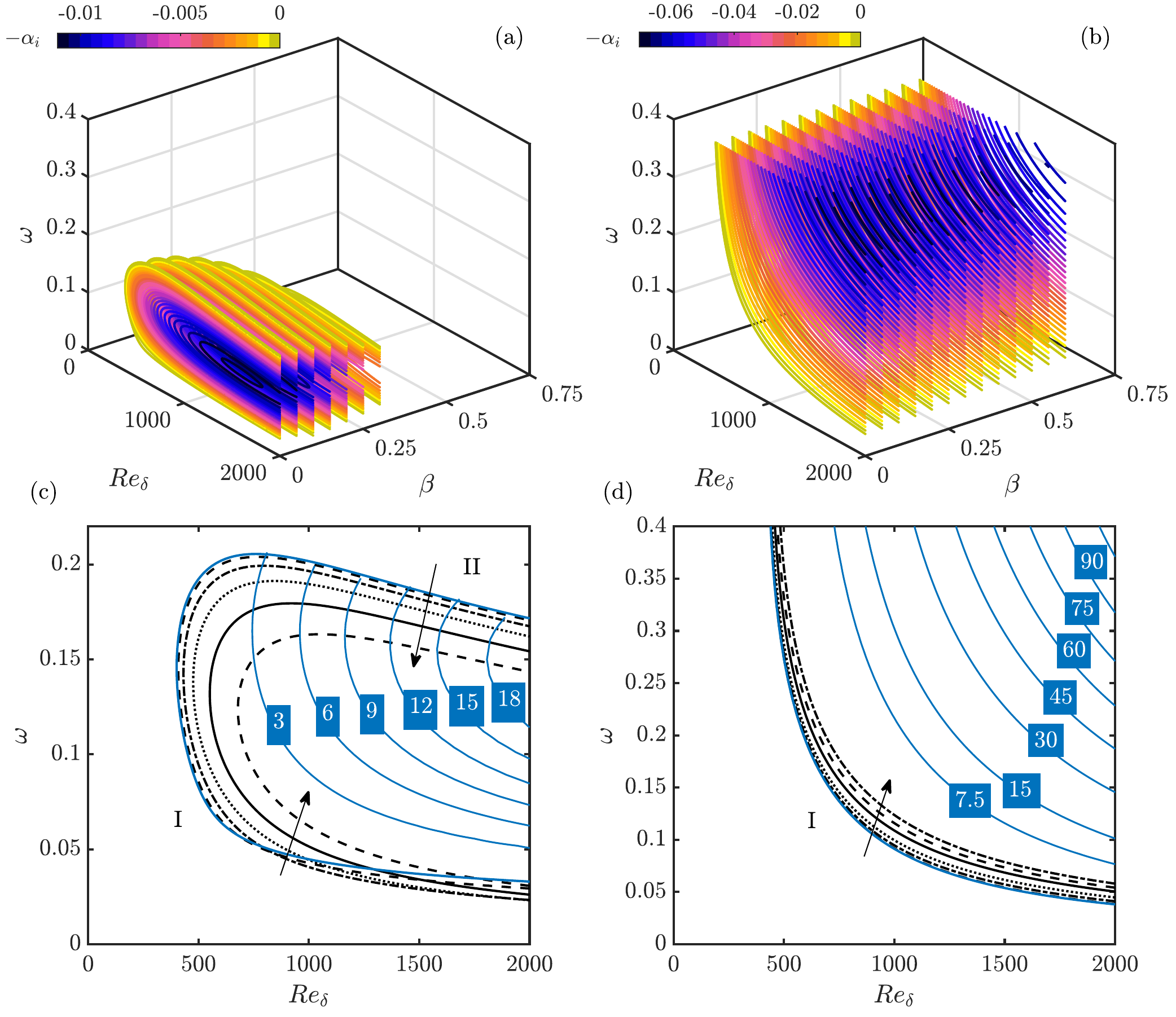}
\caption{\label{fig:a6}Case T09w105. Slices of $-\alpha_i$-contours in the $Re_\delta$--$\omega$--$\beta$ (a,b) and neutral stability curves in the $Re_\delta$--$\omega$ (c,d) space: (a,c) Mode I, (b,d) Mode II. Arrows in (c,d) indicate a $\beta$-increase ($\Delta \beta=0.05$) of branch I and II (from solid blue to dashed line in (c) and to dash-dotted line in (d)). Maximum $N$-factors are indicated in blue and are located always at $\beta=\beta_{max}=0$.}
\end{figure}

\begin{figure}[!tb]
\centering
\includegraphics[angle=-0,trim=0 0 0 0, clip,width=0.7\textwidth]{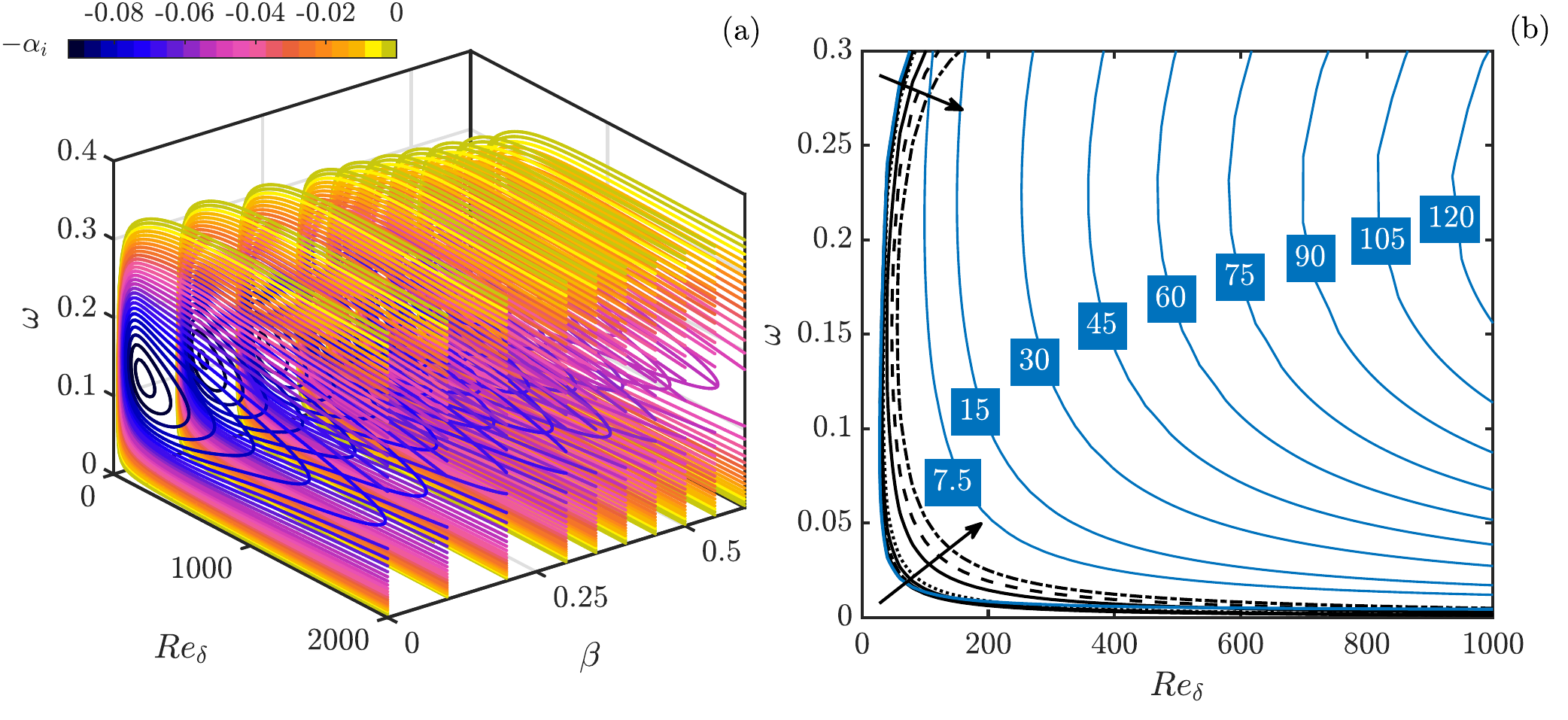}
\caption{\label{fig:a7}Case T11w095. Slices of $-\alpha_i$-contours in the $Re_\delta$--$\omega$--$\beta$ (a) and neutral stability curves in the $Re_\delta$--$\omega$ (b) space. Arrows in (b) indicate a $\beta$-increase ($\Delta \beta=0.05$) of branch I (from solid blue to dash-dotted line). Maximum $N$-factors are indicated in blue and are located always at $\beta=\beta_{max}=0$.}
\end{figure}

The influence of 3-D perturbations on modal growth needs to be investigated for the $N$-factor comparison in Sec.~\ref{sec:44}. In Ref.~\cite{Ren2}, the maximum local growth ratio, defined as $\max_{\forall \omega } \{\alpha_{i,\beta=\beta_{max}}/\alpha_{i,\beta=0}\}$, was calculated for an adiabatic wall at finite Eckert numbers. 2-D perturbations were found to be most unstable in the subcritical regime, whereas a maximum local-growth ratio larger than one was obtained for oblique disturbances in the supercritical regime. In the transcritical regime, Mode I and Mode II were most locally amplified in 3-D and 2-D, respectively. In this study, a similar investigation is undertaken with a special focus on the integral amplification rather than the local one. Stability diagrams are first obtained in the $Re_\delta$--$\omega$--$\beta$ space up to $Re_\delta=2000$. Cuts of $Re_\delta$--$\omega$ are then showcased at different $\beta$ with iso-contours of the largest $N$-factor over all spanwise wavenumbers. Figs.~\ref{fig:a5}(a,c) and \ref{fig:a5}(b,d) exemplify non-transcritical cases T09w085 and T11w105, respectively. Their largest growth rate in the stability diagram and their largest integral amplification are detected at $\beta=\beta_{max}=0$, with larger values for the subcritical wall-cooling regime. For instance, the largest eigenvalues correspond to $(c_r,\alpha_i,Re_\delta,\beta)=(0.301,-0.010,1400,0)$ and $(c_r,\alpha_i,Re_\delta,\beta)=(0.252,-0.0053,2200,0)$ for T09w085 and T110w105, respectively. The real part of the phase velocity $c_r$ is computed as $\omega/\alpha_r$. Note that, as illustrated in Figs.~\ref{fig:a5}(c,d), there are spanwise wavenumbers at which 3-D modes are locally more amplified than 2-D modes. Differently than at finite Eckert numbers \cite{Ren2}, all non-transcritical cases at $M_\infty=10^{-3}$ have their maximum local-growth ratio and largest integral amplification for 2-D perturbations.
The effect of an increasing spanwise wavenumber is displayed in Fig.~\ref{fig:a6} for the transcritical case T09w105. Mode II exhibits its largest growth rate over all frequencies at $\beta=\beta_m=0$, while Mode I can also be most unstable for 3-D disturbances. The largest eigenvalues are found at $(c_r,\alpha_i,Re_\delta,\beta)=(0.457,-0.011,2000,0)$ and $(c_r,\alpha_i,Re_\delta,\beta)=(0.312,-0.072,2060,0)$ for Mode I and Mode II, respectively. Nevertheless, Mode II is almost unaffected by a $\beta$-increase, being extremely unstable in the considered parameter space. Hence, $N$-factors up to $90$ are achieved at $Re_\delta=2000$. For Mode I, an increase in $\beta$ leads to a stabilization analogous to the non-transcritical cases.
For case T11w095, in Fig.~\ref{fig:a7}, only one highly unstable mode is detected in agreement with Ref.~\cite{Ren3}. It is inviscid, given the GIP in Fig.~\ref{fig:fig2}(c). Similar to Mode II in the transcritical wall-heating case, this mode is barely unaffected by a $\beta$-increase. Yet, it extends up to the flat-plate leading edge and to relatively high frequencies. Interestingly, the largest growth rate, i.e., $\alpha_{i,max}=-0.0946$, is found at a very low local Reynolds number of $Re_\delta=220$ with a phase speed of $c_r=0.217$. With respect to the integral amplification, we observe an $N$-factor of 15 already at $Re_\delta=200$. We can conclude that, for all cases in Tab.~\ref{tab:tableBF}, the current modal analysis has revealed the "dominance" of 2-D disturbances at $M=10^{-3}$. Both the local amplification (over most of the frequency spectrum) as well as the maximum local-growth ratio, and the $N$-factor are found largest for 2-D modes. Thus, these results clearly indicate that reducing the Mach number has the effect of shifting the maximum (local and integral) amplification from a 3-D to a 2-D mode independently of the considered thermodynamic regime at supercritical pressure.

\section{\label{sec:app6}Influence of the reduced pressure on the transient growth}
\begin{figure}[!t]
\centering
\includegraphics[angle=-0,trim=0 0 0 0, clip,width=0.85\textwidth]{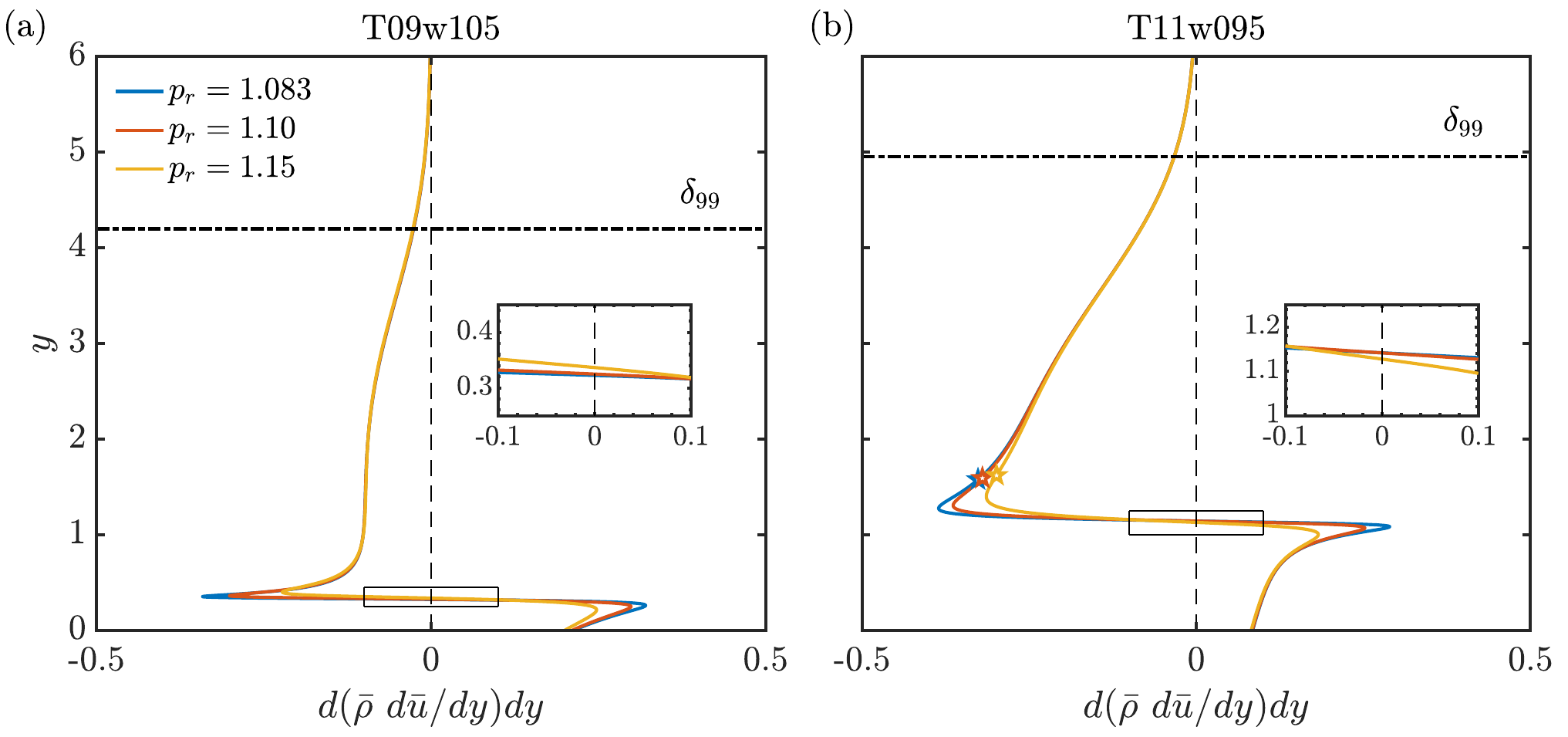}
\caption{\label{fig:a8}Generalized inflection point of the boundary-layer profile at different reduced pressures ($p_r=1.083, 1.10, 1.15$): (a) case T09w105, (b) case T11w095. Note that $\delta_{99}$ is not altered by a pressure change.}
\end{figure}

\begin{figure}[!t]
\centering
\includegraphics[angle=-0,trim=0 0 0 0, clip,width=0.9\textwidth]{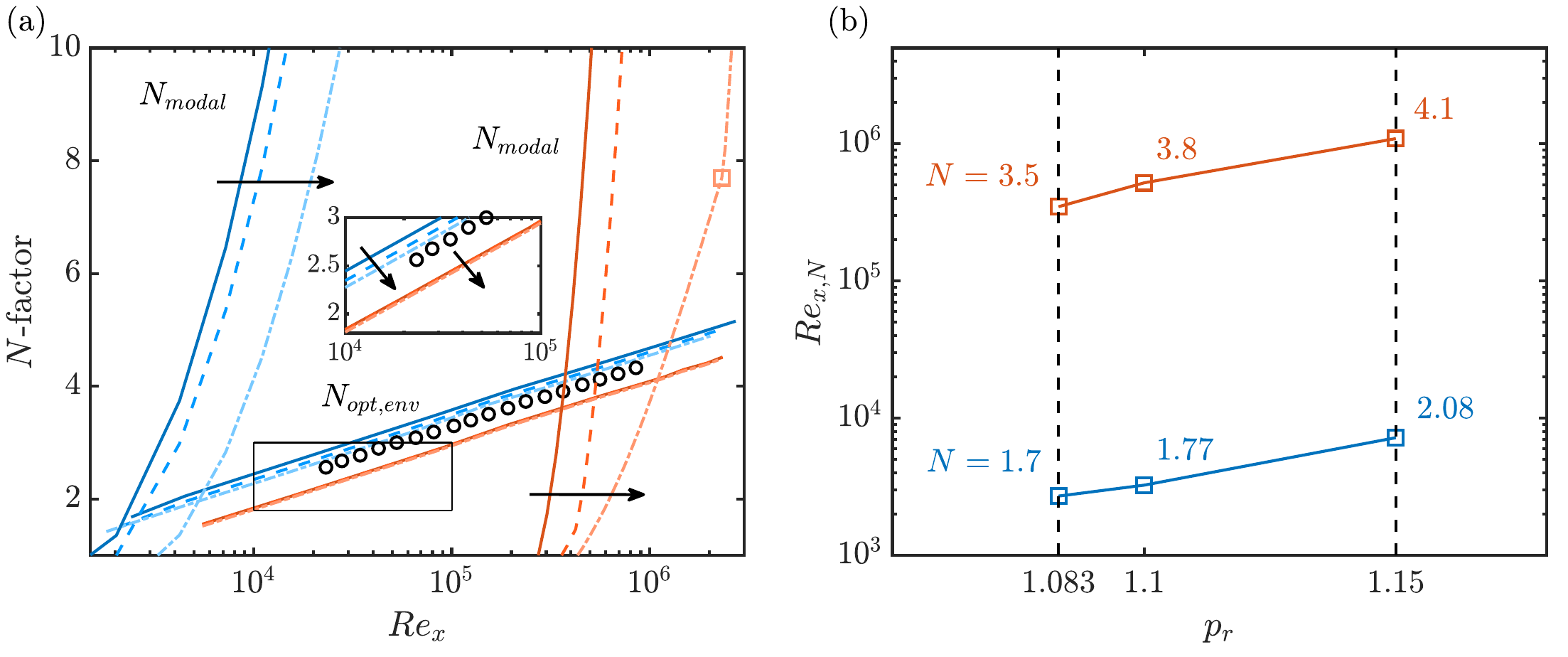}
\caption{\label{fig:a9}Effect of $p_r$-increase on the envelope curves of optimal non-modal and modal growth: case T09w105 (\textcolor{mycolor2}{\rule[0.5ex]{0.3cm}{1pt}}), case T11w095 (\textcolor{mycolor1}{\rule[0.5ex]{0.3cm}{1pt}}). In (a), $p_r=1.083$: continuous lines, $p_r=1.10$: dashed lines, and $p_r=1.15$: dash-dotted lines. In (a), black arrows indicate an increase in reduced pressure. The black circle (\textcolor{black}{$\circ$}) symbol refers to the adverse pressure case of Ref.~\cite{Levin1}. In (b), $N$-factors are indicated at the Reynolds number $Re_{x,N}$ where $N_{modal}=N_{opt}$.}
\end{figure}

The influence of the reduced pressure, $p_r=p^*/p^*_c$, on transient growth is investigated. As one approaches the critical point, the gradients of thermo-physical properties are more pronounced near the Widom line (e.g., $c^*_p$ in Fig.~\ref{fig:a1}), making the base flow becomes more inflectional and enhancing modal growth \cite{Ren2}, thus reducing the likelihood of transition below the critical Reynolds number. To assess whether transient growth could be the critical transition mechanism, the supercritical pressure is increased from a constant pressure of $p^*=\SI{80}{bar}$ (i.e., $p_r=1.083$) to $\SI{81.15}{bar}$ ($p_r=1.10$) and $\SI{84.84}{bar}$ ($p_r=1.15$) for cases T09w105 and T11w095. Other base-flow parameters in Tab.~\ref{tab:tableBF} are marginally influenced by this pressure variation due to the Mach number of $M_\infty=10^{-3}$. In Fig.~\ref{fig:a8}, the base-flow profiles of $d(\bar{\rho}\, d\bar{u}/dy)dy$ decrease with increasing reduced pressure for both transcritical cases. In this regard, the location of the generalized IP is barely modified. For case T11w095 in Fig.~\ref{fig:a8}(b) the IP location (colored star (\textcolor{black}{$\smallstar$}) symbol) shifts slightly away both from the wall and the Widom line, where $y_{WL}\approx y_{GIP}$.
For each reduced pressure, the same analysis conducted in Fig.~\ref{fig:24} is performed and displayed in Fig.~\ref{fig:a9}. Firstly, regardless of the selected supercritical pressure, the same optimal growth mechanisms for the transcritical regime previously analyzed in Sec.~\ref{sec:41} are observed: lift-up and Orr mechanism for T09w105, and only lift-up effect for T11w095.
Fig.~\ref{fig:a9}(a) presents the envelope curves for modal and non-modal growth, verifying the robustness of transient growth with respect to a thermodynamic variation. $N_{opt,env}$ is slightly affected by the pressure change, negligible for the transcritical wall-heating case T09w105. Regarding modal growth in case T09w105 (Fig.~\ref{fig:a8}(a)), the base-flow profile becomes less inflectional with increased reduced pressure, stabilizing Mode II while Mode I remains nearly unchanged. At $p_r=1.15$, Mode I dominates until $Re_x \approx 2.2 \times 10^6$, marked by a kink in the $N$-factor curve (colored ($\square$) symbol). Subsequently, the Reynolds number $Re_{x,N}$ at which $N_{modal} = N_{opt}$ is displayed in in Fig.~\ref{fig:a9}(b). $Re_{x,N}$ grows linearly by a factor of $3$ from $p_r=1.084$ to $1.15$ (red line), but non-modal amplification remains moderate. In case T11w095, a similar behavior is witnessed (Fig.~\ref{fig:a8}(b)). The increase of the inflectional location, i.e., $d^2 \bar{u}/dy^2=0$, weakly stabilizes modal instability, while non-modal amplification is almost unaffected by the pressure change. Hence, $Re_{x,N}$ more than doubles from $p_r=1.083$ to $1.15$ (blue line), but near the flat-plate leading edge, $N_{opt}$ remains still too low for transition below the critical Reynolds number. In conclusion, for a transcritical wall-cooled boundary layer, modal growth dominates near the critical point. For transcritical wall-heating, with large free-stream disturbances as in \cite{Levin1}, bypassing the exponential growth of Mode I and II is more likely when the gradients of thermo-physical properties are no longer abrupt.

\clearpage
\bibliography{main}% Produces the bibliography via BibTeX.

\providecommand{\noopsort}[1]{}\providecommand{\singleletter}[1]{#1}%
\begin{thebibliography}{57}%
\makeatletter
\providecommand \@ifxundefined [1]{%
 \@ifx{#1\undefined}
}%
\providecommand \@ifnum [1]{%
 \ifnum #1\expandafter \@firstoftwo
 \else \expandafter \@secondoftwo
 \fi
}%
\providecommand \@ifx [1]{%
 \ifx #1\expandafter \@firstoftwo
 \else \expandafter \@secondoftwo
 \fi
}%
\providecommand \natexlab [1]{#1}%
\providecommand \enquote  [1]{``#1''}%
\providecommand \bibnamefont  [1]{#1}%
\providecommand \bibfnamefont [1]{#1}%
\providecommand \citenamefont [1]{#1}%
\providecommand \href@noop [0]{\@secondoftwo}%
\providecommand \href [0]{\begingroup \@sanitize@url \@href}%
\providecommand \@href[1]{\@@startlink{#1}\@@href}%
\providecommand \@@href[1]{\endgroup#1\@@endlink}%
\providecommand \@sanitize@url [0]{\catcode `\\12\catcode `\$12\catcode `\&12\catcode `\#12\catcode `\^12\catcode `\_12\catcode `\%12\relax}%
\providecommand \@@startlink[1]{}%
\providecommand \@@endlink[0]{}%
\providecommand \url  [0]{\begingroup\@sanitize@url \@url }%
\providecommand \@url [1]{\endgroup\@href {#1}{\urlprefix }}%
\providecommand \urlprefix  [0]{URL }%
\providecommand \Eprint [0]{\href }%
\providecommand \doibase [0]{https://doi.org/}%
\providecommand \selectlanguage [0]{\@gobble}%
\providecommand \bibinfo  [0]{\@secondoftwo}%
\providecommand \bibfield  [0]{\@secondoftwo}%
\providecommand \translation [1]{[#1]}%
\providecommand \BibitemOpen [0]{}%
\providecommand \bibitemStop [0]{}%
\providecommand \bibitemNoStop [0]{.\EOS\space}%
\providecommand \EOS [0]{\spacefactor3000\relax}%
\providecommand \BibitemShut  [1]{\csname bibitem#1\endcsname}%
\let\auto@bib@innerbib\@empty
%</preamble>
\bibitem [{\citenamefont {Liao}\ \emph {et~al.}(2019)\citenamefont {Liao}, \citenamefont {Liu}, \citenamefont {Jiaqiang}, \citenamefont {Zhang}, \citenamefont {Chen}, \citenamefont {Deng},\ and\ \citenamefont {Zhu}}]{Liao1}%
  \BibitemOpen
  \bibfield  {author} {\bibinfo {author} {\bibfnamefont {G.}~\bibnamefont {Liao}}, \bibinfo {author} {\bibfnamefont {L.}~\bibnamefont {Liu}}, \bibinfo {author} {\bibfnamefont {E.}~\bibnamefont {Jiaqiang}}, \bibinfo {author} {\bibfnamefont {F.}~\bibnamefont {Zhang}}, \bibinfo {author} {\bibfnamefont {J.}~\bibnamefont {Chen}}, \bibinfo {author} {\bibfnamefont {Y.}~\bibnamefont {Deng}},\ and\ \bibinfo {author} {\bibfnamefont {H.}~\bibnamefont {Zhu}},\ }\bibfield  {title} {\bibinfo {title} {Effects of technical progress on performance and application of supercritical carbon dioxide power cycle:~a review},\ }\href@noop {} {\bibfield  {journal} {\bibinfo  {journal} {Energy Convers.~Manag.}\ }\textbf {\bibinfo {volume} {199}},\ \bibinfo {pages} {111986} (\bibinfo {year} {2019})}\BibitemShut {NoStop}%
\bibitem [{\citenamefont {Kawai}(2019)}]{Kawai1}%
  \BibitemOpen
  \bibfield  {author} {\bibinfo {author} {\bibfnamefont {S.}~\bibnamefont {Kawai}},\ }\bibfield  {title} {\bibinfo {title} {Heated transcritical and unheated non-transcritical turbulent boundary layers at supercritical pressures},\ }\href@noop {} {\bibfield  {journal} {\bibinfo  {journal} {J.~Fluid Mech.}\ }\textbf {\bibinfo {volume} {865}},\ \bibinfo {pages} {563–601} (\bibinfo {year} {2019})}\BibitemShut {NoStop}%
\bibitem [{\citenamefont {Peeters}\ \emph {et~al.}(2016)\citenamefont {Peeters}, \citenamefont {Pecnik}, \citenamefont {Rohde}, \citenamefont {van~der Hagen},\ and\ \citenamefont {Boersma}}]{Peeters1}%
  \BibitemOpen
  \bibfield  {author} {\bibinfo {author} {\bibfnamefont {J.~W.~R.}\ \bibnamefont {Peeters}}, \bibinfo {author} {\bibfnamefont {R.}~\bibnamefont {Pecnik}}, \bibinfo {author} {\bibfnamefont {M.}~\bibnamefont {Rohde}}, \bibinfo {author} {\bibfnamefont {T.~H.~J.~J.}\ \bibnamefont {van~der Hagen}},\ and\ \bibinfo {author} {\bibfnamefont {B.~J.}\ \bibnamefont {Boersma}},\ }\bibfield  {title} {\bibinfo {title} {Turbulence attenuation in simultaneously heated and cooled annular flows at supercritical pressure},\ }\href@noop {} {\bibfield  {journal} {\bibinfo  {journal} {J.~Fluid Mech.}\ }\textbf {\bibinfo {volume} {799}},\ \bibinfo {pages} {505–540} (\bibinfo {year} {2016})}\BibitemShut {NoStop}%
\bibitem [{\citenamefont {Yoo}(2013)}]{Yoo1}%
  \BibitemOpen
  \bibfield  {author} {\bibinfo {author} {\bibfnamefont {J.~Y.}\ \bibnamefont {Yoo}},\ }\bibfield  {title} {\bibinfo {title} {The turbulent flows of supercritical fluids with heat transfer},\ }\href@noop {} {\bibfield  {journal} {\bibinfo  {journal} {Annu.~Rev.~Fluid Mech.}\ }\textbf {\bibinfo {volume} {45}},\ \bibinfo {pages} {495–525} (\bibinfo {year} {2013})}\BibitemShut {NoStop}%
\bibitem [{\citenamefont {Nemati}\ \emph {et~al.}(2016)\citenamefont {Nemati}, \citenamefont {Patel}, \citenamefont {Boersma},\ and\ \citenamefont {Pecnik}}]{Nemati1}%
  \BibitemOpen
  \bibfield  {author} {\bibinfo {author} {\bibfnamefont {H.}~\bibnamefont {Nemati}}, \bibinfo {author} {\bibfnamefont {A.}~\bibnamefont {Patel}}, \bibinfo {author} {\bibfnamefont {B.~J.}\ \bibnamefont {Boersma}},\ and\ \bibinfo {author} {\bibfnamefont {R.}~\bibnamefont {Pecnik}},\ }\bibfield  {title} {\bibinfo {title} {The effect of thermal boundary conditions on forced convection heat transfer to fluids at supercritical pressure},\ }\href@noop {} {\bibfield  {journal} {\bibinfo  {journal} {J.~Fluid Mech.}\ }\textbf {\bibinfo {volume} {800}},\ \bibinfo {pages} {531–556} (\bibinfo {year} {2016})}\BibitemShut {NoStop}%
\bibitem [{\citenamefont {Gloerfelt}\ \emph {et~al.}(2020)\citenamefont {Gloerfelt}, \citenamefont {Robinet}, \citenamefont {Sciacovelli}, \citenamefont {Cinnella},\ and\ \citenamefont {Grasso}}]{Gloerfelt1}%
  \BibitemOpen
  \bibfield  {author} {\bibinfo {author} {\bibfnamefont {X.}~\bibnamefont {Gloerfelt}}, \bibinfo {author} {\bibfnamefont {J.-C.}\ \bibnamefont {Robinet}}, \bibinfo {author} {\bibfnamefont {L.}~\bibnamefont {Sciacovelli}}, \bibinfo {author} {\bibfnamefont {P.}~\bibnamefont {Cinnella}},\ and\ \bibinfo {author} {\bibfnamefont {F.}~\bibnamefont {Grasso}},\ }\bibfield  {title} {\bibinfo {title} {Dense-gas effects on compressible boundary-layer stability},\ }\href@noop {} {\bibfield  {journal} {\bibinfo  {journal} {J.~Fluid Mech.}\ }\textbf {\bibinfo {volume} {893}},\ \bibinfo {pages} {A19} (\bibinfo {year} {2020})}\BibitemShut {NoStop}%
\bibitem [{\citenamefont {Ren}\ \emph {et~al.}(2019{\natexlab{a}})\citenamefont {Ren}, \citenamefont {Fu},\ and\ \citenamefont {Pecnik}}]{Ren1}%
  \BibitemOpen
  \bibfield  {author} {\bibinfo {author} {\bibfnamefont {J.}~\bibnamefont {Ren}}, \bibinfo {author} {\bibfnamefont {S.}~\bibnamefont {Fu}},\ and\ \bibinfo {author} {\bibfnamefont {R.}~\bibnamefont {Pecnik}},\ }\bibfield  {title} {\bibinfo {title} {Linear instability of {P}oiseuille flows with highly non-ideal fluid},\ }\href@noop {} {\bibfield  {journal} {\bibinfo  {journal} {J.~Fluid Mech.}\ }\textbf {\bibinfo {volume} {859}},\ \bibinfo {pages} {89–125} (\bibinfo {year} {2019}{\natexlab{a}})}\BibitemShut {NoStop}%
\bibitem [{\citenamefont {Ren}\ \emph {et~al.}(2019{\natexlab{b}})\citenamefont {Ren}, \citenamefont {Marxen},\ and\ \citenamefont {Pecnik}}]{Ren2}%
  \BibitemOpen
  \bibfield  {author} {\bibinfo {author} {\bibfnamefont {J.}~\bibnamefont {Ren}}, \bibinfo {author} {\bibfnamefont {O.}~\bibnamefont {Marxen}},\ and\ \bibinfo {author} {\bibfnamefont {R.}~\bibnamefont {Pecnik}},\ }\bibfield  {title} {\bibinfo {title} {Boundary-layer stability of supercritical fluids in the vicinity of the {W}idom line},\ }\href@noop {} {\bibfield  {journal} {\bibinfo  {journal} {J.~Fluid Mech.}\ }\textbf {\bibinfo {volume} {871}},\ \bibinfo {pages} {831–864} (\bibinfo {year} {2019}{\natexlab{b}})}\BibitemShut {NoStop}%
\bibitem [{\citenamefont {Ly}\ and\ \citenamefont {Ihme}(2022)}]{Ly1}%
  \BibitemOpen
  \bibfield  {author} {\bibinfo {author} {\bibfnamefont {N.}~\bibnamefont {Ly}}\ and\ \bibinfo {author} {\bibfnamefont {M.}~\bibnamefont {Ihme}},\ }\bibfield  {title} {\bibinfo {title} {Destabilization of binary mixing layer in supercritical conditions},\ }\href@noop {} {\bibfield  {journal} {\bibinfo  {journal} {J.~Fluid Mech.}\ }\textbf {\bibinfo {volume} {945}},\ \bibinfo {pages} {R2} (\bibinfo {year} {2022})}\BibitemShut {NoStop}%
\bibitem [{\citenamefont {Bugeat}\ \emph {et~al.}(2022)\citenamefont {Bugeat}, \citenamefont {Boldini},\ and\ \citenamefont {Pecnik}}]{Bugeat1}%
  \BibitemOpen
  \bibfield  {author} {\bibinfo {author} {\bibfnamefont {B.}~\bibnamefont {Bugeat}}, \bibinfo {author} {\bibfnamefont {P.~C.}\ \bibnamefont {Boldini}},\ and\ \bibinfo {author} {\bibfnamefont {R.}~\bibnamefont {Pecnik}},\ }\bibfield  {title} {\bibinfo {title} {On the new unstable mode in the boundary layer flow of supercritical fluids},\ }in\ \href@noop {} {\emph {\bibinfo {booktitle} {Proceedings of the 12th International Symposium on Turbulence and Shear Flow Phenomena (TSFP-12)}}}\ (\bibinfo {year} {2022})\BibitemShut {NoStop}%
\bibitem [{\citenamefont {Bugeat}\ \emph {et~al.}(2024)\citenamefont {Bugeat}, \citenamefont {Boldini}, \citenamefont {Hasan},\ and\ \citenamefont {Pecnik}}]{Bugeat2}%
  \BibitemOpen
  \bibfield  {author} {\bibinfo {author} {\bibfnamefont {B.}~\bibnamefont {Bugeat}}, \bibinfo {author} {\bibfnamefont {P.~C.}\ \bibnamefont {Boldini}}, \bibinfo {author} {\bibfnamefont {A.~M.}\ \bibnamefont {Hasan}},\ and\ \bibinfo {author} {\bibfnamefont {R.}~\bibnamefont {Pecnik}},\ }\bibfield  {title} {\bibinfo {title} {Instability in strongly stratified plane {C}ouette flow with application to supercritical fluids},\ }\href@noop {} {\bibfield  {journal} {\bibinfo  {journal} {J.~Fluid Mech.}\ }\textbf {\bibinfo {volume} {984}},\ \bibinfo {pages} {A31} (\bibinfo {year} {2024})}\BibitemShut {NoStop}%
\bibitem [{\citenamefont {Ren}\ and\ \citenamefont {Kloker}(2022)}]{Ren3}%
  \BibitemOpen
  \bibfield  {author} {\bibinfo {author} {\bibfnamefont {J.}~\bibnamefont {Ren}}\ and\ \bibinfo {author} {\bibfnamefont {M.}~\bibnamefont {Kloker}},\ }\bibfield  {title} {\bibinfo {title} {Instabilities in three-dimensional boundary-layer flows with a highly non-ideal fluid},\ }\href@noop {} {\bibfield  {journal} {\bibinfo  {journal} {J.~Fluid Mech.}\ }\textbf {\bibinfo {volume} {951}},\ \bibinfo {pages} {A9} (\bibinfo {year} {2022})}\BibitemShut {NoStop}%
\bibitem [{\citenamefont {Banuti}(2015)}]{Banuti1}%
  \BibitemOpen
  \bibfield  {author} {\bibinfo {author} {\bibfnamefont {D.~T.}\ \bibnamefont {Banuti}},\ }\bibfield  {title} {\bibinfo {title} {Crossing the {W}idom-line -- supercritical pseudo-boiling},\ }\href@noop {} {\bibfield  {journal} {\bibinfo  {journal} {J.~Supercrit.~Fluids}\ }\textbf {\bibinfo {volume} {98}},\ \bibinfo {pages} {12–16} (\bibinfo {year} {2015})}\BibitemShut {NoStop}%
\bibitem [{\citenamefont {Schmid}(2007)}]{Schmid2}%
  \BibitemOpen
  \bibfield  {author} {\bibinfo {author} {\bibfnamefont {P.~J.}\ \bibnamefont {Schmid}},\ }\bibfield  {title} {\bibinfo {title} {Nonmodal stability theory},\ }\href@noop {} {\bibfield  {journal} {\bibinfo  {journal} {Annu.~Rev.~Fluid Mech.}\ }\textbf {\bibinfo {volume} {39}},\ \bibinfo {pages} {129} (\bibinfo {year} {2007})}\BibitemShut {NoStop}%
\bibitem [{\citenamefont {Reshotko}(2001)}]{Reshotko1}%
  \BibitemOpen
  \bibfield  {author} {\bibinfo {author} {\bibfnamefont {E.}~\bibnamefont {Reshotko}},\ }\bibfield  {title} {\bibinfo {title} {Transient growth: A factor in bypass transition},\ }\href@noop {} {\bibfield  {journal} {\bibinfo  {journal} {Phys.~Fluids}\ }\textbf {\bibinfo {volume} {5}},\ \bibinfo {pages} {1067–1075} (\bibinfo {year} {2001})}\BibitemShut {NoStop}%
\bibitem [{\citenamefont {Reshotko}\ and\ \citenamefont {Tumin}(2000)}]{Reshotko3}%
  \BibitemOpen
  \bibfield  {author} {\bibinfo {author} {\bibfnamefont {E.}~\bibnamefont {Reshotko}}\ and\ \bibinfo {author} {\bibfnamefont {A.}~\bibnamefont {Tumin}},\ }\bibfield  {title} {\bibinfo {title} {The blunt body paradox -- a case for transient growth},\ }in\ \href@noop {} {\emph {\bibinfo {booktitle} {Laminar-Turbulent Transition}}}\ (\bibinfo  {publisher} {Springer},\ \bibinfo {address} {Berlin},\ \bibinfo {year} {2000})\ p.\ \bibinfo {pages} {403–408}\BibitemShut {NoStop}%
\bibitem [{\citenamefont {Hanifi}\ \emph {et~al.}(1996)\citenamefont {Hanifi}, \citenamefont {Schmid},\ and\ \citenamefont {Henningson}}]{Hanifi1}%
  \BibitemOpen
  \bibfield  {author} {\bibinfo {author} {\bibfnamefont {A.}~\bibnamefont {Hanifi}}, \bibinfo {author} {\bibfnamefont {P.~J.}\ \bibnamefont {Schmid}},\ and\ \bibinfo {author} {\bibfnamefont {D.~S.}\ \bibnamefont {Henningson}},\ }\bibfield  {title} {\bibinfo {title} {Transient growth in compressible boundary layer flow},\ }\href@noop {} {\bibfield  {journal} {\bibinfo  {journal} {Phys.~Fluids}\ }\textbf {\bibinfo {volume} {8}},\ \bibinfo {pages} {826–837} (\bibinfo {year} {1996})}\BibitemShut {NoStop}%
\bibitem [{\citenamefont {Andersson}\ \emph {et~al.}(1999)\citenamefont {Andersson}, \citenamefont {Berggren},\ and\ \citenamefont {Henningson}}]{Andersson1}%
  \BibitemOpen
  \bibfield  {author} {\bibinfo {author} {\bibfnamefont {P.}~\bibnamefont {Andersson}}, \bibinfo {author} {\bibfnamefont {M.}~\bibnamefont {Berggren}},\ and\ \bibinfo {author} {\bibfnamefont {D.~S.}\ \bibnamefont {Henningson}},\ }\bibfield  {title} {\bibinfo {title} {Optimal disturbances and bypass transition in boundary layers},\ }\href@noop {} {\bibfield  {journal} {\bibinfo  {journal} {J.~Fluid Mech.}\ }\textbf {\bibinfo {volume} {11}},\ \bibinfo {pages} {134–150} (\bibinfo {year} {1999})}\BibitemShut {NoStop}%
\bibitem [{\citenamefont {Landahl}(1980)}]{Landahl1}%
  \BibitemOpen
  \bibfield  {author} {\bibinfo {author} {\bibfnamefont {M.~T.}\ \bibnamefont {Landahl}},\ }\bibfield  {title} {\bibinfo {title} {A note on an algebraic instability of inviscid parallel shear flows},\ }\href@noop {} {\bibfield  {journal} {\bibinfo  {journal} {J.~Fluid Mech.}\ }\textbf {\bibinfo {volume} {98}},\ \bibinfo {pages} {243–251} (\bibinfo {year} {1980})}\BibitemShut {NoStop}%
\bibitem [{\citenamefont {Luchini}(2000)}]{Luchini1}%
  \BibitemOpen
  \bibfield  {author} {\bibinfo {author} {\bibfnamefont {P.}~\bibnamefont {Luchini}},\ }\bibfield  {title} {\bibinfo {title} {Reynolds number independent instability of the boundary layer over a flat surface:~optimal perturbations},\ }\href@noop {} {\bibfield  {journal} {\bibinfo  {journal} {J.~Fluid Mech.}\ }\textbf {\bibinfo {volume} {404}},\ \bibinfo {pages} {289–309} (\bibinfo {year} {2000})}\BibitemShut {NoStop}%
\bibitem [{\citenamefont {Hanifi}\ and\ \citenamefont {Henningson}(1998)}]{Hanifi2}%
  \BibitemOpen
  \bibfield  {author} {\bibinfo {author} {\bibfnamefont {A.}~\bibnamefont {Hanifi}}\ and\ \bibinfo {author} {\bibfnamefont {D.~S.}\ \bibnamefont {Henningson}},\ }\bibfield  {title} {\bibinfo {title} {The compressible inviscid algebraic instability for streamwise independent disturbances},\ }\href@noop {} {\bibfield  {journal} {\bibinfo  {journal} {Phys.~Fluids}\ }\textbf {\bibinfo {volume} {10}} (\bibinfo {year} {1998})}\BibitemShut {NoStop}%
\bibitem [{\citenamefont {Tumin}\ and\ \citenamefont {Reshotko}(2001)}]{Tumin1}%
  \BibitemOpen
  \bibfield  {author} {\bibinfo {author} {\bibfnamefont {A.}~\bibnamefont {Tumin}}\ and\ \bibinfo {author} {\bibfnamefont {E.}~\bibnamefont {Reshotko}},\ }\bibfield  {title} {\bibinfo {title} {Spatial theory of optimal disturbances in boundary layers},\ }\href@noop {} {\bibfield  {journal} {\bibinfo  {journal} {Phys.~Fluids}\ }\textbf {\bibinfo {volume} {13}},\ \bibinfo {pages} {2097–2104} (\bibinfo {year} {2001})}\BibitemShut {NoStop}%
\bibitem [{\citenamefont {Bitter}(2015)}]{Bitter1}%
  \BibitemOpen
  \bibfield  {author} {\bibinfo {author} {\bibfnamefont {N.~P.}\ \bibnamefont {Bitter}},\ }\emph {\bibinfo {title} {Stability of hypervelocity boundary layers}},\ \href@noop {} {Ph.D. thesis},\ \bibinfo  {school} {California Institute of Technology} (\bibinfo {year} {2015})\BibitemShut {NoStop}%
\bibitem [{\citenamefont {Paredes}\ \emph {et~al.}(2016)\citenamefont {Paredes}, \citenamefont {Choudhari}, \citenamefont {Li},\ and\ \citenamefont {Chang}}]{Paredes1}%
  \BibitemOpen
  \bibfield  {author} {\bibinfo {author} {\bibfnamefont {P.}~\bibnamefont {Paredes}}, \bibinfo {author} {\bibfnamefont {M.~M.}\ \bibnamefont {Choudhari}}, \bibinfo {author} {\bibfnamefont {F.}~\bibnamefont {Li}},\ and\ \bibinfo {author} {\bibfnamefont {C.~L.}\ \bibnamefont {Chang}},\ }\bibfield  {title} {\bibinfo {title} {Optimal growth in hypersonic boundary layers},\ }\href@noop {} {\bibfield  {journal} {\bibinfo  {journal} {AIAA J.}\ }\textbf {\bibinfo {volume} {54}},\ \bibinfo {pages} {3050–3061} (\bibinfo {year} {2016})}\BibitemShut {NoStop}%
\bibitem [{\citenamefont {Govindarajan}\ and\ \citenamefont {Sahu}(2014)}]{Govindarajan1}%
  \BibitemOpen
  \bibfield  {author} {\bibinfo {author} {\bibfnamefont {R.}~\bibnamefont {Govindarajan}}\ and\ \bibinfo {author} {\bibfnamefont {K.~C.}\ \bibnamefont {Sahu}},\ }\bibfield  {title} {\bibinfo {title} {Instabilities in viscosity-stratified flow},\ }\href@noop {} {\bibfield  {journal} {\bibinfo  {journal} {Annu.~Rev.~Fluid Mech.}\ }\textbf {\bibinfo {volume} {46}},\ \bibinfo {pages} {331–353} (\bibinfo {year} {2014})}\BibitemShut {NoStop}%
\bibitem [{\citenamefont {Malik}\ \emph {et~al.}(2008)\citenamefont {Malik}, \citenamefont {Dey},\ and\ \citenamefont {Alam}}]{Malik1}%
  \BibitemOpen
  \bibfield  {author} {\bibinfo {author} {\bibfnamefont {M.}~\bibnamefont {Malik}}, \bibinfo {author} {\bibfnamefont {J.}~\bibnamefont {Dey}},\ and\ \bibinfo {author} {\bibfnamefont {M.}~\bibnamefont {Alam}},\ }\bibfield  {title} {\bibinfo {title} {Linear stability, transient energy growth, and the role of viscosity stratification in compressible plane {C}ouette flow},\ }\href@noop {} {\bibfield  {journal} {\bibinfo  {journal} {Phys.~Rev.~E}\ }\textbf {\bibinfo {volume} {77}},\ \bibinfo {pages} {036322} (\bibinfo {year} {2008})}\BibitemShut {NoStop}%
\bibitem [{\citenamefont {Saikia}\ \emph {et~al.}(2017)\citenamefont {Saikia}, \citenamefont {Ramachandran}, \citenamefont {Sinha},\ and\ \citenamefont {Govindarajan}}]{Saikia1}%
  \BibitemOpen
  \bibfield  {author} {\bibinfo {author} {\bibfnamefont {B.}~\bibnamefont {Saikia}}, \bibinfo {author} {\bibfnamefont {A.}~\bibnamefont {Ramachandran}}, \bibinfo {author} {\bibfnamefont {K.}~\bibnamefont {Sinha}},\ and\ \bibinfo {author} {\bibfnamefont {R.}~\bibnamefont {Govindarajan}},\ }\bibfield  {title} {\bibinfo {title} {Effects of viscosity and conductivity stratification on the linear stability and transient growth within compressible {C}ouette flow},\ }\href@noop {} {\bibfield  {journal} {\bibinfo  {journal} {Phys.~Fluids}\ }\textbf {\bibinfo {volume} {29}},\ \bibinfo {pages} {1–20} (\bibinfo {year} {2017})}\BibitemShut {NoStop}%
\bibitem [{\citenamefont {Chikkadi}\ \emph {et~al.}(2005)\citenamefont {Chikkadi}, \citenamefont {Sameen},\ and\ \citenamefont {Govindarajan}}]{Chikkadi1}%
  \BibitemOpen
  \bibfield  {author} {\bibinfo {author} {\bibfnamefont {V.}~\bibnamefont {Chikkadi}}, \bibinfo {author} {\bibfnamefont {A.}~\bibnamefont {Sameen}},\ and\ \bibinfo {author} {\bibfnamefont {R.}~\bibnamefont {Govindarajan}},\ }\bibfield  {title} {\bibinfo {title} {Preventing transition to turbulence:~a viscosity stratification does not always help},\ }\href@noop {} {\bibfield  {journal} {\bibinfo  {journal} {Phys.~Rev.~Lett.}\ }\textbf {\bibinfo {volume} {95}},\ \bibinfo {pages} {264504} (\bibinfo {year} {2005})}\BibitemShut {NoStop}%
\bibitem [{\citenamefont {Sameen}\ and\ \citenamefont {Govindarajan}(2007)}]{Sameen1}%
  \BibitemOpen
  \bibfield  {author} {\bibinfo {author} {\bibfnamefont {A.}~\bibnamefont {Sameen}}\ and\ \bibinfo {author} {\bibfnamefont {R.}~\bibnamefont {Govindarajan}},\ }\bibfield  {title} {\bibinfo {title} {The effect of wall heating on instability of channel flow},\ }\href@noop {} {\bibfield  {journal} {\bibinfo  {journal} {J.~Fluid Mech.}\ }\textbf {\bibinfo {volume} {577}},\ \bibinfo {pages} {417–442} (\bibinfo {year} {2007})}\BibitemShut {NoStop}%
\bibitem [{\citenamefont {Sameen}\ \emph {et~al.}(2011)\citenamefont {Sameen}, \citenamefont {Bale},\ and\ \citenamefont {Govindarajan}}]{Sameen2}%
  \BibitemOpen
  \bibfield  {author} {\bibinfo {author} {\bibfnamefont {A.}~\bibnamefont {Sameen}}, \bibinfo {author} {\bibfnamefont {R.}~\bibnamefont {Bale}},\ and\ \bibinfo {author} {\bibfnamefont {R.}~\bibnamefont {Govindarajan}},\ }\bibfield  {title} {\bibinfo {title} {The effect of wall heating on instability of channel flow – {CORRIGENDUM}},\ }\href@noop {} {\bibfield  {journal} {\bibinfo  {journal} {J.~Fluid Mech.}\ }\textbf {\bibinfo {volume} {673}},\ \bibinfo {pages} {603–605} (\bibinfo {year} {2011})}\BibitemShut {NoStop}%
\bibitem [{\citenamefont {Jose}\ \emph {et~al.}(2020)\citenamefont {Jose}, \citenamefont {Brandt},\ and\ \citenamefont {Govindarajan}}]{Jose1}%
  \BibitemOpen
  \bibfield  {author} {\bibinfo {author} {\bibfnamefont {S.}~\bibnamefont {Jose}}, \bibinfo {author} {\bibfnamefont {L.}~\bibnamefont {Brandt}},\ and\ \bibinfo {author} {\bibfnamefont {R.}~\bibnamefont {Govindarajan}},\ }\bibfield  {title} {\bibinfo {title} {Localisation of optimal perturbations in variable viscosity channel flow},\ }\href@noop {} {\bibfield  {journal} {\bibinfo  {journal} {Int.~J.~Heat Fluid Flow}\ }\textbf {\bibinfo {volume} {85}},\ \bibinfo {pages} {108588} (\bibinfo {year} {2020})}\BibitemShut {NoStop}%
\bibitem [{\citenamefont {Parente}\ \emph {et~al.}(2020)\citenamefont {Parente}, \citenamefont {Robinet}, \citenamefont {{De Palma}},\ and\ \citenamefont {Cherubini}}]{Parente1}%
  \BibitemOpen
  \bibfield  {author} {\bibinfo {author} {\bibfnamefont {E.}~\bibnamefont {Parente}}, \bibinfo {author} {\bibfnamefont {J.~C.}\ \bibnamefont {Robinet}}, \bibinfo {author} {\bibfnamefont {P.}~\bibnamefont {{De Palma}}},\ and\ \bibinfo {author} {\bibfnamefont {S.}~\bibnamefont {Cherubini}},\ }\bibfield  {title} {\bibinfo {title} {Modal and nonmodal stability of a stably stratified boundary layer flow},\ }\href@noop {} {\bibfield  {journal} {\bibinfo  {journal} {Phys.~Rev.~Fluids}\ }\textbf {\bibinfo {volume} {5}},\ \bibinfo {pages} {113901} (\bibinfo {year} {2020})}\BibitemShut {NoStop}%
\bibitem [{\citenamefont {Robinet}\ and\ \citenamefont {Gloerfelt}(2019)}]{Robinet1}%
  \BibitemOpen
  \bibfield  {author} {\bibinfo {author} {\bibfnamefont {J.-C.}\ \bibnamefont {Robinet}}\ and\ \bibinfo {author} {\bibfnamefont {X.}~\bibnamefont {Gloerfelt}},\ }\bibfield  {title} {\bibinfo {title} {Instabilities in non-ideal fluids},\ }\href@noop {} {\bibfield  {journal} {\bibinfo  {journal} {J.~Fluid Mech.}\ }\textbf {\bibinfo {volume} {880}},\ \bibinfo {pages} {1–4} (\bibinfo {year} {2019})}\BibitemShut {NoStop}%
\bibitem [{\citenamefont {Levin}\ and\ \citenamefont {Henningson}(2003)}]{Levin1}%
  \BibitemOpen
  \bibfield  {author} {\bibinfo {author} {\bibfnamefont {O.}~\bibnamefont {Levin}}\ and\ \bibinfo {author} {\bibfnamefont {D.~S.}\ \bibnamefont {Henningson}},\ }\bibfield  {title} {\bibinfo {title} {Exponential vs algebraic growth and transition prediction in boundary layer flow},\ }\href@noop {} {\bibfield  {journal} {\bibinfo  {journal} {Flow Turbul.~Combust.}\ }\textbf {\bibinfo {volume} {70}},\ \bibinfo {pages} {183–210} (\bibinfo {year} {2003})}\BibitemShut {NoStop}%
\bibitem [{\citenamefont {Tempelmann}\ \emph {et~al.}(2012)\citenamefont {Tempelmann}, \citenamefont {Hanifi},\ and\ \citenamefont {Henningson}}]{Tempelmann1}%
  \BibitemOpen
  \bibfield  {author} {\bibinfo {author} {\bibfnamefont {D.}~\bibnamefont {Tempelmann}}, \bibinfo {author} {\bibfnamefont {A.}~\bibnamefont {Hanifi}},\ and\ \bibinfo {author} {\bibfnamefont {D.~S.}\ \bibnamefont {Henningson}},\ }\bibfield  {title} {\bibinfo {title} {Spatial optimal growth in three-dimensional compressible boundary layers},\ }\href@noop {} {\bibfield  {journal} {\bibinfo  {journal} {J.~Fluid Mech.}\ }\textbf {\bibinfo {volume} {704}},\ \bibinfo {pages} {251–279} (\bibinfo {year} {2012})}\BibitemShut {NoStop}%
\bibitem [{\citenamefont {Schmid}\ and\ \citenamefont {Henningson}(2001)}]{Schmid3}%
  \BibitemOpen
  \bibfield  {author} {\bibinfo {author} {\bibfnamefont {P.~J.}\ \bibnamefont {Schmid}}\ and\ \bibinfo {author} {\bibfnamefont {D.~S.}\ \bibnamefont {Henningson}},\ }\href@noop {} {\emph {\bibinfo {title} {Stability and transition in shear flows}}}\ (\bibinfo  {publisher} {Springer},\ \bibinfo {year} {2001})\BibitemShut {NoStop}%
\bibitem [{\citenamefont {Lemmon}\ \emph {et~al.}(2013)\citenamefont {Lemmon}, \citenamefont {Huber},\ and\ \citenamefont {Mclinden}}]{Lemmon1}%
  \BibitemOpen
  \bibfield  {author} {\bibinfo {author} {\bibfnamefont {E.~W.}\ \bibnamefont {Lemmon}}, \bibinfo {author} {\bibfnamefont {M.~L.}\ \bibnamefont {Huber}},\ and\ \bibinfo {author} {\bibfnamefont {M.~O.}\ \bibnamefont {Mclinden}},\ }\href@noop {} {\bibinfo {title} {{NIST Standard Reference Database 23:~Reference Fluid Thermodynamic and Transport Properties - REFPROP, Version 9.1}}} (\bibinfo {year} {2013}),\ \bibinfo {note} {{Available at:~\url{http://www.nist.gov/srd/nist23.cfm}}}\BibitemShut {NoStop}%
\bibitem [{\citenamefont {White}(2006)}]{White1}%
  \BibitemOpen
  \bibfield  {author} {\bibinfo {author} {\bibfnamefont {F.~M.}\ \bibnamefont {White}},\ }\href@noop {} {\emph {\bibinfo {title} {Viscous Fluid Flow, 3rd Edition}}}\ (\bibinfo  {publisher} {McGraw-Hill},\ \bibinfo {address} {Boston},\ \bibinfo {year} {2006})\BibitemShut {NoStop}%
\bibitem [{\citenamefont {Nemati}\ \emph {et~al.}(2015)\citenamefont {Nemati}, \citenamefont {Patel}, \citenamefont {Boersma},\ and\ \citenamefont {Pecnik}}]{Nemati2}%
  \BibitemOpen
  \bibfield  {author} {\bibinfo {author} {\bibfnamefont {H.}~\bibnamefont {Nemati}}, \bibinfo {author} {\bibfnamefont {A.}~\bibnamefont {Patel}}, \bibinfo {author} {\bibfnamefont {B.~J.}\ \bibnamefont {Boersma}},\ and\ \bibinfo {author} {\bibfnamefont {R.}~\bibnamefont {Pecnik}},\ }\bibfield  {title} {\bibinfo {title} {Mean statistics of a heated turbulent pipe flow at supercritical pressure},\ }\href@noop {} {\bibfield  {journal} {\bibinfo  {journal} {Int.~J.~Heat Mass Transf.}\ }\textbf {\bibinfo {volume} {83}},\ \bibinfo {pages} {741} (\bibinfo {year} {2015})}\BibitemShut {NoStop}%
\bibitem [{\citenamefont {Malik}(1900)}]{Malik2}%
  \BibitemOpen
  \bibfield  {author} {\bibinfo {author} {\bibfnamefont {M.~R.}\ \bibnamefont {Malik}},\ }\bibfield  {title} {\bibinfo {title} {Numerical methods for hypersonic boundary layer stability},\ }\href@noop {} {\bibfield  {journal} {\bibinfo  {journal} {J.~Comput.~Phys.}\ }\textbf {\bibinfo {volume} {86}},\ \bibinfo {pages} {376} (\bibinfo {year} {1900})}\BibitemShut {NoStop}%
\bibitem [{\citenamefont {Chu}(1965)}]{Chu1}%
  \BibitemOpen
  \bibfield  {author} {\bibinfo {author} {\bibfnamefont {T.~B.}\ \bibnamefont {Chu}},\ }\bibfield  {title} {\bibinfo {title} {On the energy transfer to small disturbances in fluid flow (part {I})},\ }\href@noop {} {\bibfield  {journal} {\bibinfo  {journal} {Acta Mech.}\ }\textbf {\bibinfo {volume} {1}},\ \bibinfo {pages} {215–234} (\bibinfo {year} {1965})}\BibitemShut {NoStop}%
\bibitem [{\citenamefont {Mack}(1969)}]{Mack1}%
  \BibitemOpen
  \bibfield  {author} {\bibinfo {author} {\bibfnamefont {L.~M.}\ \bibnamefont {Mack}},\ }\href@noop {} {\bibinfo {title} {{Boundary layer stability theory}}},\ \bibinfo {howpublished} {{JPL Rep.~900-227.~Jet Propulsion Laboratory.}} (\bibinfo {year} {1969})\BibitemShut {NoStop}%
\bibitem [{\citenamefont {Chen}\ \emph {et~al.}(2022)\citenamefont {Chen}, \citenamefont {Wang},\ and\ \citenamefont {Fu}}]{Chen1}%
  \BibitemOpen
  \bibfield  {author} {\bibinfo {author} {\bibfnamefont {X.}~\bibnamefont {Chen}}, \bibinfo {author} {\bibfnamefont {L.}~\bibnamefont {Wang}},\ and\ \bibinfo {author} {\bibfnamefont {S.}~\bibnamefont {Fu}},\ }\bibfield  {title} {\bibinfo {title} {Energy transfer of hypersonic and high-enthalpy boundary layer instabilities and transition},\ }\href@noop {} {\bibfield  {journal} {\bibinfo  {journal} {Phys.~Rev.~Fluids}\ }\textbf {\bibinfo {volume} {7}},\ \bibinfo {pages} {033901} (\bibinfo {year} {2022})}\BibitemShut {NoStop}%
\bibitem [{\citenamefont {Schmid}\ and\ \citenamefont {Henningson}(1994)}]{Schmid1}%
  \BibitemOpen
  \bibfield  {author} {\bibinfo {author} {\bibfnamefont {P.~J.}\ \bibnamefont {Schmid}}\ and\ \bibinfo {author} {\bibfnamefont {D.~S.}\ \bibnamefont {Henningson}},\ }\bibfield  {title} {\bibinfo {title} {Optimal energy density growth in {H}agen–{P}oiseuille flow},\ }\href@noop {} {\bibfield  {journal} {\bibinfo  {journal} {J.~Fluid Mech.}\ }\textbf {\bibinfo {volume} {277}},\ \bibinfo {pages} {197–225} (\bibinfo {year} {1994})}\BibitemShut {NoStop}%
\bibitem [{\citenamefont {Corbett}\ and\ \citenamefont {Bottaro}(2000)}]{Corbett1}%
  \BibitemOpen
  \bibfield  {author} {\bibinfo {author} {\bibfnamefont {P.}~\bibnamefont {Corbett}}\ and\ \bibinfo {author} {\bibfnamefont {A.}~\bibnamefont {Bottaro}},\ }\bibfield  {title} {\bibinfo {title} {Optimal perturbations for boundary layers subject to stream-wise pressure gradient},\ }\href@noop {} {\bibfield  {journal} {\bibinfo  {journal} {Phys.~Fluids}\ }\textbf {\bibinfo {volume} {12}},\ \bibinfo {pages} {120–130} (\bibinfo {year} {2000})}\BibitemShut {NoStop}%
\bibitem [{\citenamefont {Butler}\ and\ \citenamefont {Farrell}(1992)}]{Butler1}%
  \BibitemOpen
  \bibfield  {author} {\bibinfo {author} {\bibfnamefont {K.~M.}\ \bibnamefont {Butler}}\ and\ \bibinfo {author} {\bibfnamefont {B.~F.}\ \bibnamefont {Farrell}},\ }\bibfield  {title} {\bibinfo {title} {Three-dimensional optimal perturbations in viscous shear flow},\ }\href@noop {} {\bibfield  {journal} {\bibinfo  {journal} {Phys.~Fluids A}\ }\textbf {\bibinfo {volume} {4}},\ \bibinfo {pages} {1637–1650} (\bibinfo {year} {1992})}\BibitemShut {NoStop}%
\bibitem [{\citenamefont {Brandt}(2014)}]{Brandt1}%
  \BibitemOpen
  \bibfield  {author} {\bibinfo {author} {\bibfnamefont {L.}~\bibnamefont {Brandt}},\ }\bibfield  {title} {\bibinfo {title} {The lift-up effect: the linear mechanism behind transition and turbulence in shear flows},\ }\href@noop {} {\bibfield  {journal} {\bibinfo  {journal} {Eur.~J.~Mech.~(B/Fluids)}\ }\textbf {\bibinfo {volume} {47}},\ \bibinfo {pages} {80–96} (\bibinfo {year} {2014})}\BibitemShut {NoStop}%
\bibitem [{\citenamefont {Farrell}\ and\ \citenamefont {Ioannou}(1993)}]{Farrell1}%
  \BibitemOpen
  \bibfield  {author} {\bibinfo {author} {\bibfnamefont {B.~F.}\ \bibnamefont {Farrell}}\ and\ \bibinfo {author} {\bibfnamefont {P.~J.}\ \bibnamefont {Ioannou}},\ }\bibfield  {title} {\bibinfo {title} {Optimal excitation of three-dimensional perturbations in viscous constant shear flow},\ }\href@noop {} {\bibfield  {journal} {\bibinfo  {journal} {Phys.~Fluids A}\ }\textbf {\bibinfo {volume} {5}},\ \bibinfo {pages} {1390–1400} (\bibinfo {year} {1993})}\BibitemShut {NoStop}%
\bibitem [{\citenamefont {Hack}\ and\ \citenamefont {Moin}(2017)}]{Hack1}%
  \BibitemOpen
  \bibfield  {author} {\bibinfo {author} {\bibfnamefont {M.~J.~P.}\ \bibnamefont {Hack}}\ and\ \bibinfo {author} {\bibfnamefont {P.}~\bibnamefont {Moin}},\ }\bibfield  {title} {\bibinfo {title} {Algebraic disturbance growth by interaction of {O}rr and lift-up mechanisms},\ }\href@noop {} {\bibfield  {journal} {\bibinfo  {journal} {J.~Fluid Mech.}\ }\textbf {\bibinfo {volume} {829}},\ \bibinfo {pages} {112–126} (\bibinfo {year} {2017})}\BibitemShut {NoStop}%
\bibitem [{\citenamefont {Orr}(1907)}]{Orr1}%
  \BibitemOpen
  \bibfield  {author} {\bibinfo {author} {\bibfnamefont {W.~M.~F.}\ \bibnamefont {Orr}},\ }\bibfield  {title} {\bibinfo {title} {The stability or instability of the steady motions of a perfect liquid and of a viscous liquid},\ }\href@noop {} {\bibfield  {journal} {\bibinfo  {journal} {Proc.~R.~Irish Acad.~A}\ }\textbf {\bibinfo {volume} {27}},\ \bibinfo {pages} {9–138} (\bibinfo {year} {1907})}\BibitemShut {NoStop}%
\bibitem [{\citenamefont {Berlin}\ \emph {et~al.}(1994)\citenamefont {Berlin}, \citenamefont {Lundbladh},\ and\ \citenamefont {Henningson}}]{Berlin1}%
  \BibitemOpen
  \bibfield  {author} {\bibinfo {author} {\bibfnamefont {S.}~\bibnamefont {Berlin}}, \bibinfo {author} {\bibfnamefont {A.}~\bibnamefont {Lundbladh}},\ and\ \bibinfo {author} {\bibfnamefont {D.}~\bibnamefont {Henningson}},\ }\bibfield  {title} {\bibinfo {title} {Spatial simulations of oblique transition in a boundary layer},\ }\href@noop {} {\bibfield  {journal} {\bibinfo  {journal} {Phys.~Fluids}\ }\textbf {\bibinfo {volume} {6}},\ \bibinfo {pages} {1949–1951} (\bibinfo {year} {1994})}\BibitemShut {NoStop}%
\bibitem [{\citenamefont {Gustavsson}(1991)}]{Gustavsson1}%
  \BibitemOpen
  \bibfield  {author} {\bibinfo {author} {\bibfnamefont {L.~H.}\ \bibnamefont {Gustavsson}},\ }\bibfield  {title} {\bibinfo {title} {Energy growth of three-dimensional disturbances in plane {P}oiseuille flow},\ }\href@noop {} {\bibfield  {journal} {\bibinfo  {journal} {J.~Fluid Mech.}\ }\textbf {\bibinfo {volume} {224}},\ \bibinfo {pages} {241–260} (\bibinfo {year} {1991})}\BibitemShut {NoStop}%
\bibitem [{\citenamefont {Jiao}\ \emph {et~al.}(2021)\citenamefont {Jiao}, \citenamefont {Hwang},\ and\ \citenamefont {Chernyshenko}}]{Jiao1}%
  \BibitemOpen
  \bibfield  {author} {\bibinfo {author} {\bibfnamefont {Y.}~\bibnamefont {Jiao}}, \bibinfo {author} {\bibfnamefont {Y.}~\bibnamefont {Hwang}},\ and\ \bibinfo {author} {\bibfnamefont {S.~I.}\ \bibnamefont {Chernyshenko}},\ }\bibfield  {title} {\bibinfo {title} {Orr mechanism in transition of parallel shear flow},\ }\href@noop {} {\bibfield  {journal} {\bibinfo  {journal} {Phys.~Rev.~Fluids}\ }\textbf {\bibinfo {volume} {6}},\ \bibinfo {pages} {023902} (\bibinfo {year} {2021})}\BibitemShut {NoStop}%
\bibitem [{\citenamefont {Tumin}(2007)}]{Tumin4}%
  \BibitemOpen
  \bibfield  {author} {\bibinfo {author} {\bibfnamefont {A.}~\bibnamefont {Tumin}},\ }\bibfield  {title} {\bibinfo {title} {Three-dimensional spatial normal modes in compressible boundary layers},\ }\href@noop {} {\bibfield  {journal} {\bibinfo  {journal} {J.~Fluid Mech.}\ }\textbf {\bibinfo {volume} {586}},\ \bibinfo {pages} {295–322} (\bibinfo {year} {2007})}\BibitemShut {NoStop}%
\bibitem [{\citenamefont {Kaminski}\ \emph {et~al.}(2014)\citenamefont {Kaminski}, \citenamefont {Caulfield},\ and\ \citenamefont {Taylor}}]{Kaminski1}%
  \BibitemOpen
  \bibfield  {author} {\bibinfo {author} {\bibfnamefont {A.~K.}\ \bibnamefont {Kaminski}}, \bibinfo {author} {\bibfnamefont {C.~P.}\ \bibnamefont {Caulfield}},\ and\ \bibinfo {author} {\bibfnamefont {J.~R.}\ \bibnamefont {Taylor}},\ }\bibfield  {title} {\bibinfo {title} {Transient growth in strongly stratified shear layers},\ }\href@noop {} {\bibfield  {journal} {\bibinfo  {journal} {J.~Fluid Mech.}\ }\textbf {\bibinfo {volume} {758}},\ \bibinfo {pages} {R4} (\bibinfo {year} {2014})}\BibitemShut {NoStop}%
\bibitem [{\citenamefont {Tumin}\ and\ \citenamefont {Reshotko}(2003)}]{Tumin2}%
  \BibitemOpen
  \bibfield  {author} {\bibinfo {author} {\bibfnamefont {A.}~\bibnamefont {Tumin}}\ and\ \bibinfo {author} {\bibfnamefont {E.}~\bibnamefont {Reshotko}},\ }\bibfield  {title} {\bibinfo {title} {Optimal disturbances in compressible boundary layers},\ }\href@noop {} {\bibfield  {journal} {\bibinfo  {journal} {AIAA J.}\ }\textbf {\bibinfo {volume} {41}},\ \bibinfo {pages} {2357–2363} (\bibinfo {year} {2003})}\BibitemShut {NoStop}%
\bibitem [{\citenamefont {Reshotko}\ and\ \citenamefont {Tumin}(2004)}]{Reshotko2}%
  \BibitemOpen
  \bibfield  {author} {\bibinfo {author} {\bibfnamefont {E.}~\bibnamefont {Reshotko}}\ and\ \bibinfo {author} {\bibfnamefont {A.}~\bibnamefont {Tumin}},\ }\bibfield  {title} {\bibinfo {title} {Role of transient growth in roughness-induced transition},\ }\href@noop {} {\bibfield  {journal} {\bibinfo  {journal} {AIAA J.}\ }\textbf {\bibinfo {volume} {42 (4)}},\ \bibinfo {pages} {766–770} (\bibinfo {year} {2004})}\BibitemShut {NoStop}%
\end{thebibliography}%

\end{document}